\documentclass[trackchanges, twocolumn]{aastex701}
\usepackage{amsmath}
\begin{document}

\title{Mind the Information Gap: Unveiling Detailed Morphologies of z\,$\boldsymbol{\sim}$\,0.5--1.0 Galaxies with SLACS Strong Lenses and Data-Driven Analysis}

\author[orcid=0000-0001-9459-6316,gname=Ronan,sname=Legin]{Ronan Legin}
\affiliation{Ciela - Montreal Institute for Astrophysical Data Analysis and Machine Learning, Montr{\'e}al, Canada}
\affiliation{Universit{\'e} de Montr{\'e}al, Montr{\'e}al, Canada}
\affiliation{Mila - Quebec Artificial Intelligence Institute, Montr{\'e}al, Canada}
\email{ronan.legin@umontreal.ca}

\author[orcid=0000-0002-9086-6398,gname=Connor,sname=Stone]{Connor Stone}
\affiliation{Ciela - Montreal Institute for Astrophysical Data Analysis and Machine Learning, Montr{\'e}al, Canada}
\affiliation{Universit{\'e} de Montr{\'e}al, Montr{\'e}al, Canada}
\affiliation{Mila - Quebec Artificial Intelligence Institute, Montr{\'e}al, Canada}
\email{}

\author[orcid=0000-0001-8806-7936,gname=Alexandre,sname=Adam]{Alexandre Adam}
\affiliation{Ciela - Montreal Institute for Astrophysical Data Analysis and Machine Learning, Montr{\'e}al, Canada}
\affiliation{Universit{\'e} de Montr{\'e}al, Montr{\'e}al, Canada}
\affiliation{Mila - Quebec Artificial Intelligence Institute, Montr{\'e}al, Canada}
\email{}

\author[orcid=0009-0008-5839-5937,gname=Gabriel,sname=Missael Barco]{Gabriel Missael Barco}
\affiliation{Ciela - Montreal Institute for Astrophysical Data Analysis and Machine Learning, Montr{\'e}al, Canada}
\affiliation{Universit{\'e} de Montr{\'e}al, Montr{\'e}al, Canada}
\affiliation{Mila - Quebec Artificial Intelligence Institute, Montr{\'e}al, Canada}
\email{}

\author[orcid=0000-0002-0055-1780,gname=Adam,sname=Coogan]{Adam Coogan}
\affiliation{Ciela - Montreal Institute for Astrophysical Data Analysis and Machine Learning, Montr{\'e}al, Canada}
\affiliation{Universit{\'e} de Montr{\'e}al, Montr{\'e}al, Canada}
\affiliation{Mila - Quebec Artificial Intelligence Institute, Montr{\'e}al, Canada}
\email{}

\author[orcid=0009-0009-5797-0300,gname=Nikolay,sname=Malkin]{Nikolay Malkin}
\affiliation{Universit{\'e} de Montr{\'e}al, Montr{\'e}al, Canada}
\affiliation{Mila - Quebec Artificial Intelligence Institute, Montr{\'e}al, Canada}
\affiliation{University of Edinburgh, Edinburgh, Scotland}
\email{}

\author[orcid=0000-0003-3544-3939,gname=Laurence,sname=Perreault-Levasseur]{Laurence Perreault-Levasseur}
\affiliation{Ciela - Montreal Institute for Astrophysical Data Analysis and Machine Learning, Montr{\'e}al, Canada}
\affiliation{Universit{\'e} de Montr{\'e}al, Montr{\'e}al, Canada}
\affiliation{Mila - Quebec Artificial Intelligence Institute, Montr{\'e}al, Canada}
\affiliation{Center for Computational Astrophysics, Flatiron Institute, 162 5th Avenue, 10010, New York, NY, USA}
\affiliation{Trottier Space Institute, Montr{\'e}al, Canada}
\affiliation{Perimeter Institute, Waterloo, Canada}
\email{}

\author[orcid=0000-0002-8669-5733,gname=Yashar,sname=Hezaveh]{Yashar Hezaveh}
\affiliation{Ciela - Montreal Institute for Astrophysical Data Analysis and Machine Learning, Montr{\'e}al, Canada}
\affiliation{Universit{\'e} de Montr{\'e}al, Montr{\'e}al, Canada}
\affiliation{Mila - Quebec Artificial Intelligence Institute, Montr{\'e}al, Canada}
\affiliation{Center for Computational Astrophysics, Flatiron Institute, 162 5th Avenue, 10010, New York, NY, USA}
\affiliation{Trottier Space Institute, Montr{\'e}al, Canada}
\email{}

\begin{abstract}
We present new state-of-the-art lens models for strong gravitational lensing systems from the Sloan Lens ACS (SLACS) survey, developed within a Bayesian framework that employs high-dimensional (pixellated), data-driven priors for the background source, foreground lens light, and point-spread function (PSF). Unlike conventional methods, our approach delivers high-resolution reconstructions of all major physical components of the lensing system and substantially reduces model–data residuals compared to previous work. For the majority of 30 lensing systems analyzed, we also provide posterior samples capturing the full uncertainty of each physical model parameter. The reconstructions of the background sources reveal high significance morphological structures as small as $\sim$ 200 parsecs in galaxies at redshifts of $z\sim0.5-1.0$, demonstrating the power of strong lensing and the analysis method to be used as a cosmic telescope to study the high redshift universe.  This study marks the first application of data-driven generative priors to modeling real strong-lensing data and establishes a new benchmark for strong lensing precision modeling in the era of large-scale imaging surveys. 

\end{abstract}



\section{Introduction} 

Strong gravitational lensing occurs when the gravitational potential of a foreground object, such as a galaxy or galaxy cluster, bends the trajectory of light from distant background sources into multiple magnified and distorted images. The ability to produce multiple resolved images makes strong lensing a powerful and versatile astrophysical probe, enabling precision measurements of the Hubble constant \citep[e.g.,][]{Wong2019, Shajib2020, Qi2021, Cola_2025}, the detection of dark matter substructure \citep[e.g.,][]{Mao1998, Vegetti2010, Xu2015, Hezaveh2016, Gilman2019, Quinn2021, Powell2023, BayerI2023, BayerII2023, Ballard2024, Keeley2024, Lange2025}, and detailed studies of galaxy formation at high redshift through the magnification of otherwise unresolved sources \citep[e.g.,][]{Zitrin2012, Coe2013, Fan2019, Jacobs2019, Shu2022, Shajib_2023, Nightingale2025}. With upcoming surveys such as LSST and Euclid expected to discover tens of thousands of new lensing systems \citep{Collett_2015, Shajib2025} --- for example, Euclid’s Q1 release has already discovered $\sim500$ galaxy–galaxy strong lenses \citep{Euclid2025Q1A, Euclid2025Q1B, Euclid2025Q1C, Euclid2025Q1D, Euclid2025Q1E} --- strong lensing is poised to deliver some of the most stringent constraints yet on key questions in cosmology and galaxy evolution.

Realizing the scientific potential of strong lensing requires accurate lens modeling, the process of inferring the physical and instrumental components that contribute to a strong lensing observation. These components typically include the foreground lens mass distribution, the lens light profile, the light profile of the background source, and instrumental effects such as the telescope’s Point Spread Function \citep[PSF, e.g.,][]{Tessore2015, Pearson2021, Meneghetti2022, Mishrasharma2022,  Adam_2022, Adam2023, Shajib2024}. Lens modeling is typically performed within a Bayesian framework, where each lens component is represented by an analytic or pixelated model, and prior distributions are assigned to constrain their values to physically plausible regions. The accuracy of these modeling choices is critical: overly simplistic assumptions or poorly constrained priors can lead to biased inferences, ultimately compromising the robustness of key scientific conclusions drawn from strong lensing data \citep{Xu2015, Sonnenfeld2018, Galan2024, Ballard2024, Barco2025}.

Despite the rich information content of strong lensing data, accurate modeling has often been limited by computational cost and the complexity of the inference. Consequently, model parametrization and priors are often chosen for computational tractability rather than physical realism. Background source and lens light profiles are commonly modeled with simple analytic functions that are fast to simulate and easy to regularize, but fail to capture the true morphological complexity of galaxies, leading to noticeable residuals between the model and data \citep[e.g.,][]{Bolton2008, Savary2022, Knabel2023, Huang2025, Cao2025}. More flexible representations, such as pixelated images, are widely assumed to follow analytic prior distributions chosen for computational convenience \citep[e.g., Gaussian or wavelet-based, ][]{Suyu2006, Birrer_2015, Galan2021, Karchev2022, Etherington2022, Tan2024, Sheu2025}. However, such priors impose artificial smoothness scales that can limit reconstruction fidelity, and, again, fail to capture the complex morphological features of real galaxies. \citet{Galan2024} demonstrated that the inferred mass distribution of the lens can exhibit systematic biases depending on the choice of source reconstruction priors and modeling assumptions. Their results highlight that seemingly reasonable yet distinct modeling choices can lead to inconsistent inferences, underscoring the sensitivity of lens model parameters to the choice of background source prior distribution. \citet{Ballard2024} and \citet{Stacey2025} showed that the inferred significance of a dark matter subhalo in the system SDSSJ0946+1006 and SDP.81 is highly sensitive to the choice of source model and regularization—specifically, whether curvature or gradient regularization is used in the case of \citet{Ballard2024} and whether fixed or adaptive pixelated grid is used for source reconstruction in \citet{Stacey2025}—and stress the importance of carefully selecting the background source model prior distribution before attempting detailed lens modeling. Moreover, models of instrumental parameters, such as pixelated PSF models, are often fixed rather than marginalized over \citep{Enzi2025}, due to the computational intractability of high-dimensional inference and the lack of explicit, tractable priors.

Recent advances in generative modeling offer a powerful alternative to traditional lens modeling approaches. Data-driven generative models, such as diffusion models \citep{Song_2019, Ho_2020, Song_2020, Song_2021, Ling_2022}, can learn complex, high-dimensional priors directly from data, enabling modeling with realistic physical features. 
They have been used for Bayesian analysis of high dimensional problems in several scientific fields, including astrophysics \citep[e.g., ][]{Dia2023, Adam2023, Remy2023, Legin2023, Sun2023, Bourdin2024, Cuesta2024, Wu2024, Rozet2024, Sampson2024, Rozet2025, Adam_2025, Andry2025, Barco_2025, Supranta2025}. These models can potentially enable more accurate and physically motivated lens modeling.

In this paper, we revisit high-quality strong lensing observations from the SLACS survey and achieve state-of-the-art lens reconstructions by incorporating high-dimensional, data-driven priors into the lens modeling process. We focus on galaxy-galaxy strong lensing systems, jointly inferring the mass distribution of the foreground lens and the light profiles of both the lens and background source galaxies. We train diffusion-based generative models to learn expressive score-based priors for the foreground lens light, source galaxy, and PSF. Our lensing analysis marginalizes over the PSF uncertainty using generated PSF posterior samples from \citet[][hereafter CS25]{Stone2025} for each lensing system.

The paper is organized as follows. In Section~\ref{sec:methods}, we describe the data, our modeling choices, and provide a brief overview of score-based generative models together with details of the architectures adopted in this work. Section~\ref{sec:results} presents our main results, including reconstructed lens models and comparisons with previous studies. In Section~\ref{sec:discussion}, we discuss the significance of these findings in the broader context of strong lensing analyses and compare them with results in the literature. Finally, Section~\ref{sec:conclusion} summarizes our conclusions.

\section{Methods}
\label{sec:methods}

\subsection{Overview}
\subsubsection{Data}
We perform lens modeling on 30 high-quality strong gravitational lensing systems from the SLACS survey \citep{Bolton_2006, Treu_2006, Koopmans_2006, Bolton2008}. Specifically, we use flat-field-corrected (FLT) \textit{Hubble Space Telescope} (\textit{HST}) imaging data in the F814W filter available from the Mikulski Archive for Space Telescopes (MAST)\footnote{\url{https://mast.stsci.edu}}, under program IDs 10886, 10174 and 10587. For each system, our data consists of a set of four dithered FLT exposures, each probing slightly different sky coordinates of the lensing system. From each frame, we extract a $128 \times 128$ pixel cutout, denoted as $\mathbf{d}_i$, centered on the foreground lens galaxy. We denote the full set of cutouts as $\mathbf{D} = \{\mathbf{d}_{i}\}_{i=1}^4$, where $i$ denotes the frame number.

\subsubsection{Analysis Framework}
For each strong lensing system, we infer a set of astrophysical parameters $\boldsymbol{\Theta} = \{\mathbf{L}, \mathbf{M}, \mathbf{S}\}$, representing the foreground lens light profile $\mathbf{L}$, the lens mass model parameters $\mathbf{M}$, and the background source $\mathbf{S}$. Following the methodology established in CS25, we use a distinct PSF model, $\mathbf{P}_{i}$, per FLT frame, with the full set of models denoted as $\mathbf{P} = \{\mathbf{P}_{i} \}_{i=1}^4$. 
The PSF is inferred from isolated star cutouts within the same FLT frames (but excluding the lensing system) and is thus independent of the lensing analysis. We therefore use posterior samples obtained in CS25 to marginalize over PSF uncertainty when performing lensing analysis as detailed in section \ref{sec:gibbs_strategy}. Finally, we also infer a relative $x$–$y$ positional shift $\boldsymbol{\delta}_{i} = \{\delta_x, \delta_y \}$ in the frame's central sky coordinate  (CRVAL) for each individual frame, denoted $\boldsymbol{\delta} = \{\boldsymbol{\delta}_{i}\}_{i=1}^4$, to ensure that the lens model, evaluated at the four sets of dithered sky coordinates, properly aligns with each data cutout.

From Bayes' theorem, the posterior distribution $p(\boldsymbol{\delta}, \mathbf{P}, \boldsymbol{\Theta} | \mathbf{D})$ is proportional to the prior $p(\boldsymbol{\delta}, \mathbf{P}, \boldsymbol{\Theta})$ distribution weighted by the likelihood $p(\mathbf{D} | \boldsymbol{\delta}, \mathbf{P}, \boldsymbol{\Theta})$. Assuming the four dithered data images $\mathbf{D} = \{\mathbf{d}_i\}_{i=1}^4$ are conditionally independent given model parameters $\boldsymbol{\Theta}$, $\boldsymbol{\delta}$ and $\mathbf{P}$, and that the prior distribution for each model parameter ($\boldsymbol{\delta}$, $\mathbf{P}$, $\boldsymbol{\Theta}$) is independent, the posterior factorizes as:
{\small
\begin{equation}
    p(\boldsymbol{\delta}, \mathbf{P}, \boldsymbol{\Theta} \mid \mathbf{D}) \propto p(\boldsymbol{\Theta}) \prod_{i=1}^4 p(\mathbf{d}_i \mid \boldsymbol{\delta}_i, \mathbf{P}_i, \boldsymbol{\Theta}) \, p(\boldsymbol{\delta}_i) p(\mathbf{P}_i).
    \label{eq:posterior}
\end{equation}
}
We assume that the frame-specific components $\boldsymbol{\delta}_i$ and $\mathbf{P}_i$ for $i \in \{1, 2, 3, 4\}$ are drawn independently from common prior distributions $p(\boldsymbol{\delta})$ and $p(\mathbf{P})$. Furthermore, we assume the lensing parameters $\boldsymbol{\Theta} = \{\mathbf{L}, \mathbf{M}, \mathbf{S}\}$ have independent prior distributions, i.e., $p(\boldsymbol{\Theta}) = p(\mathbf{L}) p(\mathbf{M}) p(\mathbf{S})$ and that $\boldsymbol{\Theta}$ is shared across data cutouts $\mathbf{d}_i$. In this work, we assume the individual likelihood distributions $p(\mathbf{d}_i \mid \boldsymbol{\delta}_i, \mathbf{P}_i, \boldsymbol{\Theta})$ are Gaussian with diagonal covariance matrices. We use the square of the error  \texttt{ERR} image array provided in the FITS file for each frame $i$ as an estimate of the noise variance in each data pixel $j$. For each frame, we construct a mask $\mathbf{m}$ by combining the Data Quality (\texttt{DQ}) image array from the FITS file—which flags features such as cosmic rays and bad pixels—with manually generated masks for external bright sources not associated with the lensing system. Given these choices, the likelihood distribution for data frame $i$ is summarized as

{\footnotesize \begin{equation}
p(\mathbf{d}_i \mid \boldsymbol{\delta}_i, \mathbf{P}_i, \boldsymbol{\Theta}) \propto \exp{(-\frac{1}{2} \sum_{j} m_{i,j} \left( \frac{d_{i,j} - f_{i,j}(\boldsymbol{\delta}_i, \mathbf{P}_i, \boldsymbol{\Theta})}{\mathrm{ERR}_{i,j}} \right)^2)},
\label{eq:likelihood}
\end{equation}}
where $j$ indexes the pixels in each data frame $d_i$, $m_{i,j}\in \{0,1\}$ is a mask for cosmic rays and external sources, and $\mathbf{f}_i(\boldsymbol{\delta}_i, \mathbf{P}_i, \boldsymbol{\Theta})$ represents our forward model of the lensing system evaluated at the sky coordinates of data frame $i$.

Our forward model $\mathbf{f}_i(\boldsymbol{\delta}_i, \mathbf{P}_i, \boldsymbol{\Theta})$ makes use of the single-lens-plane approximation \citep[e.g., see][]{Meneghetti2022}, in which the lens galaxy lies in an  image plane defined by angular coordinates $\boldsymbol{\theta} = \{\theta_x, \theta_y \}$, and the background source lies in the source plane defined with coordinates $\boldsymbol{\beta} = \{\beta_x, \beta_y \}$. Lensing involves mapping image-plane coordinates to the source plane via the lens equation:
\begin{equation}
\boldsymbol{\beta} = \boldsymbol{\theta} - \boldsymbol{\alpha}(\boldsymbol{\theta};\mathbf{M}),
\end{equation}
where $\boldsymbol{\alpha}(\boldsymbol{\theta};\mathbf{M})$ is the reduced deflection angle determined by the lens mass distribution $\mathbf{M}$ and related to the convergence $\kappa(\boldsymbol{\theta};\mathbf{M})$: 

\begin{equation}
\boldsymbol{\alpha}(\boldsymbol{\theta};\mathbf{M}) 
= \frac{1}{\pi} \int \mathrm{d}^2 \theta' \, 
\kappa(\boldsymbol{\theta}';\mathbf{M}) \,
\frac{\boldsymbol{\theta} - \boldsymbol{\theta}'}{|\boldsymbol{\theta} - \boldsymbol{\theta}'|^2}.
\end{equation}

To simulate a lensed image, we ray-trace each coordinate $\boldsymbol{\theta}$ to its source-plane counterpart $\boldsymbol{\beta}$ using the lens equation and evaluate the source surface brightness at $\boldsymbol{\beta}$ given by our source brightness model $\mathbf{S}$. The resulting lens image is combined with the lens light $\mathbf{L}$, convolved with the frame-specific PSF $\mathbf{P}_i$ and downsampled by four to match the data resolution. These steps define a linear operator in $\mathbf{S}$ and $\mathbf{L}$ (for fixed $\mathbf{M}, \mathbf{P}_i, \boldsymbol{\delta}_i$):

\begin{equation}
    \mathbf{f}_i(\boldsymbol{\delta}_i, \mathbf{P}_i, \boldsymbol{\Theta}) = R_4(\mathbf{P}_i \ast [\mathcal{A}(\mathbf{M}) \mathbf{S} + \mathbf{L}]_{\boldsymbol{\delta}_i}) + \mathrm{sky}_i,
    \label{eq:forward_model}
\end{equation}
where $\mathcal{A}(\mathbf{M})$ is the lens distortion matrix performing the lensing of the background source light $\mathbf{S}$ and $\mathbf{P}_i \ast$ is the convolution of the lensed image by the PSF. Here, $\mathrm{sky}_i$ is a constant scalar representing a background sky level estimated by taking the median of the full data frame $i$, and $[...]_{\boldsymbol{\delta}_i}$ denotes the lensing model centered with a relative sky coordinate shift $\boldsymbol{\delta}_i$ for frame $i$. The forward model is evaluated at four times the pixel resolution of the data frame $\mathbf{d}_i$ and then downsampled to match the data resolution, as represented by the $R_4$ operator. 
Note that we have a single ``on-sky'' model for $\mathbf{L}$, $\mathbf{M}$, and $\mathbf{S}$ which gets sampled in the four FLT frames to produce four $\mathbf{f}_i$ images.

Unlike using the drizzled data product, modeling on the raw FLT data requires correcting for geometric distortions in the sky coordinates \citep{Stark2024, Adam_2025}. We account for them using the fifth-order polynomial corrections that relate pixel coordinates to sky coordinates as provided in the FITS header of each file \citep{Shupe2005}.

The likelihood and forward model are evaluated using \texttt{caustics} \citep{Stone2024}, a strong lensing simulation and inference package built on PyTorch \citep{Pytorch2019}, which supports GPU acceleration and end-to-end autodifferentiation.

\subsection{Modeling choices and Priors}

For the foreground mass model, we adopt an Elliptical Power–Law (EPL) profile augmented by external shear and by third- and fourth-order multipole perturbations \citep{Tessore2015, Xu2015}. Using the notation of \citet{Tessore2015} for the EPL and \citet{Xu2015} for the multipoles, the convergences are

\begin{equation}
\kappa_{\mathrm{EPL}}(\theta_x, \theta_y)
= \frac{2-t}{2}\,
\left[
\frac{R_{\rm E}\,\sqrt{q}}{\sqrt{q^2 \theta_x^2 + \theta_y^2}}
\right]^t,
\end{equation}
and
\begin{equation}
\kappa_{\mathrm{multi}}(\theta_x, \theta_y)
= \sum_{m\in\{3,4\}}
\frac{a_m}{2\,r}\,
\cos\!\big[m\,(\psi-\phi_m)\big],
\end{equation}
where $r=\sqrt{\theta_x^2+\theta_y^2}$ and $\psi=\arctan(\theta_y,\theta_x)$ are multipole angular coordinates about the lens center. The EPL component is parameterized by the Einstein radius $R_{\rm E}$, axis ratio $q$, and density slope $t$. In addition, the profile has an angle parameter $\phi_{\mathrm{EPL}}$, which applies a rotation to the EPL profile. The multipoles are specified by amplitudes $a_m$ and orientations $\phi_m$ for $m \in \{3,4\}$. The external shear has zero convergence and is included through the reduced deflection angle as
\begin{equation}
\boldsymbol{\alpha}_{\mathrm{ext}}(\theta_x,\theta_y)
=\gamma_{\mathrm{ext}}
\begin{pmatrix}
\theta_x \cos(2\phi_{\mathrm{ext}}) + \theta_y \sin(2\phi_{\mathrm{ext}})\\
\theta_x \sin(2\phi_{\mathrm{ext}})  - \theta_y \cos(2\phi_{\mathrm{ext}})
\end{pmatrix}
\end{equation}
specified by external shear magnitude $\gamma_{\rm ext}$ and orientation angle $\phi_{\rm ext}$.

We adopt uniform priors for most lens model parameters: $R_{\rm E} \in [0.1, 3.0]$, $q \in [0.2, 1.0]$ and $\gamma_{\mathrm{ext}} \in [0, 0.5]$. Angular parameters $\phi_{\mathrm{EPL}}$, $\phi_{\mathrm{ext}}$ and $\phi_m$ are treated as unconstrained, with uniform priors that permit values outside the $[0, 2\pi]$ interval during inference. We place the following Gaussian priors $ \mathcal{N}(\mu, \sigma^2)$ on the EPL slope and multipole amplitudes: $t \sim \mathcal{N}(1.0, 0.2^2)$ and $a_3, a_4 \sim \mathcal{N}(0, 0.01^2)$, effectively centering the prior distribution on the singular isothermal ellipsoid (SIE) model \citep{Kormann1994}.

We assume a uniform prior for the alignment parameters $\boldsymbol{\delta}$ with bounds within the range of the data cutout field-of-view. The lens light $\mathbf{L}$, background source $\mathbf{S}$, and PSF models $\mathbf{P} = \{\mathbf{P}_i\}_{i=1}^4$ are represented as pixelated images of size $256^2$, $256^2$, and $128^2$ pixels, respectively. For $\mathbf{L}$ and $\mathbf{P}$, the pixel resolutions are chosen to be two and four times finer than those of the FLT data images. The source pixel resolution, however, is selected independently for each lensing system to ensure optimal reconstruction of the lensed galaxy image. During lensing simulations, we use bilinear interpolation to evaluate the pixelated source and lenslight models at the relevant coordinates.

The combined dimensionality of the lens light $\mathbf{L}$, PSF $\mathbf{P}$, and background source $\mathbf{S}$ parameter space well exceeds the dimensionality of the data. Inferring $\mathbf{L}$, $\mathbf{P}$, and $\mathbf{S}$ from the data therefore constitutes a severely ill-posed inverse problem. Solving this problem requires high-dimensional priors that can constrain the solution to physically plausible configurations without imposing overly restrictive models. To this end, we employ score-based diffusion models \citep{Song_2019, Ho_2020, Song_2020} as expressive high-dimensional generative priors for the source $\mathbf{S}$, lens light $\mathbf{L}$, and PSF $\mathbf{P}$.

\subsection{Data-driven priors and score-posterior sampling}
\label{sec:score_models_intro}

Score-based generative models enable flexible modeling and sampling of high-dimensional probability distributions $p(\mathbf{x})$ by approximating the score function of a time-dependent convolved distribution, $\nabla_{\mathbf{x}_t} \log p_t(\mathbf{x}_t)$ \citep{Song_2019, Ho_2020, Song_2020, Song_2021, Ling_2022}. 
Following the framework introduced in \citet{Song_2020}, these models define a forward diffusion process via a stochastic differential equation (SDE), in which samples $\mathbf{x}\sim p(\mathbf{x})$ are progressively perturbed by Gaussian noise whose amplitude increases according to a predefined schedule $\sigma(t)$ over a continuous time variable $t \in [0, 1]$. 
To generate new samples from the original data distribution $p(\mathbf{x})$, a reverse-time SDE is solved, effectively denoising samples from a Gaussian distribution with variance $\sigma(t=1)^2$ by following the gradient of the convolved distribution $\nabla_{\mathbf{x}_t} \log p_t(\mathbf{x}_t)$. 
Since $\nabla_{\mathbf{x}_t} \log p_t(\mathbf{x}_t)$ is typically unknown for complex and high-dimensional data distributions $p(\mathbf{x})$, a neural network is trained to approximate it using samples drawn from $p(\mathbf{x})$ or some close approximation \citep{Hyvarinen_2005, Vincent_2011, Ho_2020, Song_2020, Ling_2022}.

A key advantage of score-based models is their ability to sample from conditional distributions, such as the posterior distribution, enabling Bayesian inference in high-dimensional spaces (see, e.g., the review by \citet{Daras2024} and references therein). Given model parameter $\mathbf{x}$ (e.g., $\mathbf{x} = \{\mathbf{S}, \mathbf{L}, \mathbf{P} \}$) and data $\mathbf{y}$ (e.g., FLT data cutouts $\mathbf{D}$), conditional sampling from $p(\mathbf{x} \mid \mathbf{y})$ requires solving the reverse-time SDE with the convolved posterior score function $\nabla_{\mathbf{x}_t} \log p_t(\mathbf{x}_t \mid \mathbf{y})$, which, following \citet{Song_2020}, can be decomposed as

\begin{equation}
\nabla_{\mathbf{x}_t} \log p_t(\mathbf{x}_t \mid \mathbf{y}) \approx \nabla_{\mathbf{x}_t} \log p_t(\mathbf{x}_t) + \nabla_{\mathbf{x}_t} \log p_t(\mathbf{y}_t \mid \mathbf{x}_t),
\end{equation}
where $\nabla_{\mathbf{x}_t} \log p_t(\mathbf{y}_t \mid \mathbf{x}_t)$ is the convolved likelihood score with $\mathbf{y}_t$ representing data generated by diffusion-noised model parameters $\mathbf{x}_t$.
This approximation is exact for $t=0$ (specifically $\sigma_t = 0$) and reasonably accurate for $t>0$.
Following \citet{Adam_2022}, if the base likelihood $p(\mathbf{y} \mid \mathbf{x})$ is modeled as $\mathbf{y} \sim \mathcal{N}(\mathbf{A}\mathbf{x}, \mathbf{\Sigma}_{\mathbf{y}})$, where $\mathbf{A}$ is the forward model operator (e.g., lensing distortion matrix $\mathcal{A}$, PSF convolution $\mathbf{P} \ast$, downsampling operator $R_4$, see Eq. \ref{eq:forward_model}), the convolved likelihood score $\nabla_{\mathbf{x}_t} \log p_t(\mathbf{y}_t \mid \mathbf{x}_t)$ can be approximated by
\begin{equation}
\nabla_{\mathbf{x}_t} \log p_t(\mathbf{y}_t \mid \mathbf{x}_t) 
= \mathbf{A}^\top \mathbf{\Sigma}^{-1} \big(\mathbf{y} - \mathbf{A} \mathbf{x}_t\big),
\label{eq:convolved_likelihood}
\end{equation}
where the effective covariance is
\begin{equation}
\mathbf{\Sigma} \;=\; \mathbf{\Sigma_y} + \mathbf{\Sigma_{\mathbf{A}}}(t),
\qquad 
\mathbf{\Sigma_{\mathbf{A}}}(t) = \sigma(t)^2 \mathbf{A} \mathbf{A}^\top.
\end{equation}
Here $\mathbf{\Sigma_y}$ denotes the observational (e.g., \textit{HST} data) noise covariance, while $\mathbf{\Sigma}_{\mathbf{A}}(t)$ reflects the diffusion noise at time $t$ transformed by the forward model operator $\mathbf{A}$. The latter can be evaluated exactly by differentiating the forward model with respect to $\mathbf{x}$, or more memory-efficiently (when $\mathbf{A}$ is large) by taking the sample covariance of repeated noise realizations transformed by $\mathbf{A}$.

We train neural networks to learn score-based priors for the PSF, $\nabla_{\mathbf{P}_{t}} \log p_t(\mathbf{P}_{t})$; lens light, $\nabla_{\mathbf{L}_{t}} \log p_t(\mathbf{L}_{t})$; and background source, $\nabla_{\mathbf{S}_{t}} \log p_t(\mathbf{S}_{t})$, using the denoising objective from \citet{Vincent_2011, Song_2020}. To sample from the posterior defined in Eq.~\ref{eq:posterior}, we solve the reverse-time SDE using the convolved Gaussian likelihood approximation from Eq. \ref{eq:convolved_likelihood}. This approximation is justified given the linearity of our forward model (Eq.~\ref{eq:forward_model}) and the assumption that the likelihood for each individual FLT frame in Eq.~\ref{eq:likelihood} is Gaussian. We adopt the Variance-Exploding (VE) stochastic differential equation (SDE) introduced by \citet{Song_2020}, where the noise schedule $\sigma(t)$ follows an exponential schedule $\sigma(t) = \sigma_{\mathrm{min}}  (\frac{\sigma_{\mathrm{max}}}{\sigma_{\mathrm{min}}})^t$, smoothly interpolating between minimum and maximum noise levels $\sigma_{\mathrm{min}}$ and $\sigma_{\mathrm{max}}$.

\subsection{Score Models}
\label{sec:score_models}
We utilize the \texttt{scoremodels} Python package \citep{Adam_score_models} to learn score-based priors for the background source $\mathbf{S}$, foreground lens light $\mathbf{L}$, and PSF $\mathbf{P}$. Each score prior is modeled using the NCSN++ variant of the U-Net neural network architecture proposed in \citet{Song_2021}, with different choices of hyperparameters as detailed below.

For the background source score model, we train on high-resolution galaxy images from the Dark Energy Spectroscopic Instrument (DESI) Photometry and Rotation curve OBservations from Extragalactic Surveys
(PROBES) dataset \citep{Stone_2019, Stone_2021} following the procedure outlined in \citet{Adam_2022}. The training set consists of approximately 2,000 \textit{g}-band galaxy images of 256 x 256 pixels, normalized following \citet{Smith2022}. The minimum and maximum noise levels, $\sigma_{\mathrm{min}}$ and $\sigma_{\mathrm{max}}$, were set to $10^{-5}$ and $500$, respectively. The score neural network consists of five downsampling stages with convolutional layers producing $[128,\,128,\,256,\,256,\,256,\, 512]$ feature maps at each resolution level. Training was conducted for approximately three days on four NVIDIA A100 GPUs using a batch size of 4 and the Adam optimizer \citep{Kingma2014} with a learning rate of $2 \times 10^{-5}$.

We train the lens light score prior using simulated parametric galaxy brightness profiles generated with the \texttt{AstroPhot} Python package \citep{Stone_2023}. Each profile is simulated as a sum of $N$ Sérsic models \citep{Sersic1963}, with $N$ drawn uniformly from 1 to 5. Each Sérsic model is defined by the effective radius $R_\mathrm{eff}$, Sérsic index $n$, central intensity $I_0$, axis ratio $q_L$, and position angle (PA). Simulations are generated on-the-fly during training, with each set of Sérsic parameters drawn from broad uniform priors: $n \in [2.0, 6.0]$, $R_\mathrm{eff} \in [1.0, 10.0]$, $q_L \in [0.5, 1.0]$ and $\mathrm{PA} \in [0, 2\pi]$. The peak surface brightness of generated lens light profiles are uniformly resampled between 0.4 and 5.  All Sérsic components in a given training example share the same centroid and PA. For each Sérsic, we include a 10\% probability of sampling a boxy Sérsic profile, with strength parameter $C$ defined in \texttt{AstroPhot} uniformly sampled in the range $C \in [2, 3]$, with $C = 2$ corresponding to a standard ellipse. We also allow a 10\% probability of sampling a Sérsic profile with a smooth peak defined by a softening parameter $s \in [0, 0.2]$, introducing a smooth core with varying strength. We set diffusion noise levels $\sigma_{\mathrm{min}} = 10^{-5}$ and $\sigma_{\mathrm{max}} = 200$. The score network architecture comprises seven downsampling levels, with convolutional layers producing $[16, 32, 64, 64, 128, 128, 256, 256]$ feature maps across resolution levels. The model was trained for one day on a single NVIDIA A100 GPU using a batch size of 16 and the Adam optimizer \citep{Kingma2014} with a learning rate of $10^{-4}$.

For the PSF model, posterior samples were generated using the score generative prior model introduced in CS25, which was trained on four-times upsampled empirical PSF models estimated from ACS/WFC HST data ~\citep{Anderson2006}. Further details on the training procedure, dataset, and network architecture hyperparameters are provided in CS25.

\subsection{Joint posterior sampling strategy}
\label{sec:gibbs_strategy}
We use a Gibbs sampling approach to draw samples from the joint posterior distribution of the background source $\mathbf{S}$, foreground mass distribution $\mathbf{M}$, foreground light $\mathbf{L}$, alignment parameters $\boldsymbol{\delta}$, and PSF models $\mathbf{P}$. In addition, we jointly sample source transformation parameters $\boldsymbol{\Theta}_{\mathrm{src}} = [\,\mathbf{t}_s,\; s_s\,]$, where $\mathbf{t}_s = (t_x, t_y)$ is a 2D translation of the source centroid and $s_s$ is an isotropic scale factor that sets the pixel scale of the source image model $\mathbf{S}$, effectively controlling the physical size of the source. Although the score-based prior over sources $\mathbf{S}$ could, in principle, absorb these transformations, introducing $\mathbf{t}_s$ and $s_s$ explicitly and sampling them jointly with $\mathbf{M}$ improves exploration of foreground $\mathbf{M}$ configurations. We assume broad independent uniform priors for $t_x, t_y$ and $s_s$.

At each iteration, we sample one component conditioned on the current values of all others, cycling sequentially through all components. Specifically, a single Gibbs step consists of a single set of updates in the following order: $\mathbf{S},\, (\mathbf{M}, \boldsymbol{\delta}, \boldsymbol{\Theta}_{\mathrm{src}}),\, \mathbf{L},\, \mathbf{P}$, where $(\mathbf{M}, \boldsymbol{\delta}, \boldsymbol{\Theta}_{\mathrm{src}})$ are jointly sampled. The set of proposed samples over $N$ Gibbs steps forms an empirical approximation of the joint posterior distribution.

We generate posterior samples for $\mathbf{S}$, $\mathbf{L}$, and $\mathbf{P}$ using the score-based priors described in Section~\ref{sec:score_models} and following the methodology outlined in Section~\ref{sec:score_models_intro}. To improve sampling efficiency, we do not solve the full reverse SDE from $t = 1$ to $t = 0$. Instead, we start from a smaller noise level of $t = 0.6$, adding Gaussian noise with standard deviation $\sigma(t = 0.6)$ to the current Gibbs sample and running the reverse SDE from there to $t = 0$ as our proposal distribution. To initialize the Gibbs chain, however, we run the full reverse process from $t = 1$. For the PSF models $\mathbf{P}$, we draw from pre-generated posterior samples obtained in CS25, which were modeled from star cutouts within the FLT frames, excluding the data cutout containing the strong lensing system. As a result, the PSF samples are drawn from a marginal posterior independent of the parameters $\mathbf{M}$, $\boldsymbol{\delta}$, $\mathbf{L}$, and $\mathbf{S}$.  

We perform joint updates of \(\mathbf{M}\), \(\boldsymbol{\delta}\), and \(\boldsymbol{\Theta}_{\mathrm{src}}\) using the Metropolis-adjusted Langevin algorithm (MALA; \citealt{MALA}). At each Gibbs step, we first optimize \(\mathbf{M}\), \(\boldsymbol{\delta}\) and \(\boldsymbol{\Theta}_{\mathrm{src}}\) to the maximum a posteriori (MAP) point, from which the MALA chain is  initialized at. The chain comprises a burn-in phase—during which the step size and mass matrix (the inverse covariance matrix used in the MALA sampling procedure) are adapted—followed by a sampling phase in which these are held fixed and posterior samples are collected.

Our full Gibbs sampling procedure is divided into three stages: (1) an initial optimization phase to reach a high-posterior-density solution and avoid poor local minima; (2) an annealed sampling phase for posterior exploration and production of uncorrelated samples; and (3) a final non-annealed sampling phase for accurate posterior estimation.

\begin{enumerate}
    \item \textbf{Initialization and optimization.} For each lensing system, we first align the FLT frames by optimizing the sky coordinate alignment parameters $\boldsymbol{\delta}$ using a centered Gaussian profile in place of the pixelated lens light model. We then jointly optimize $\boldsymbol{\delta}$ and sample lens light models from the score-based posterior, applying custom masks to exclude the lensed arcs from the likelihood. The background source $\mathbf{S}$ is initialized from the score-based posterior, while the mass model $\mathbf{M}$ is initialized with parameter values from \citet{Bolton2008}, fixing the EPL slope $t$ to one and multipole moments $a_3$ and $a_4$ to zero. We then jointly update $\mathbf{S}$, $\mathbf{M}$ (excluding multipole parameters) under an annealed likelihood, starting with noise variance ten times the base \texttt{ERR} variance and gradually reducing it, before further updating $\mathbf{L}$, $\boldsymbol{\delta}$ and the multipoles parameters. The PSF $\mathbf{P}$ remains fixed to an arbitrary pre-generated posterior sample from CS25.

    \item \textbf{Annealed sampling.} Starting from the final model obtained from Stage~1, we perform 100 Gibbs iterations under an annealed likelihood (\texttt{ERR} noise variance scaled by a factor of 8), jointly sampling $\mathbf{S}$, $\mathbf{L}$, $\mathbf{M}$ (including multipole parameters), and $\boldsymbol{\delta}$, while continuing to draw a new $\mathbf{P}$ from its marginal posterior at each sampling step. An exception in the number of iterations is SDSSJ1430+4105 (see Fig.~\ref{fig:corner_maintext}), for which we ran 320 annealed Gibbs iterations. For all lenses, the first 20 iterations are treated as burn-in and discarded.

    \item \textbf{Final non-annealed sampling.} For each of the retained annealed samples, we run 5 additional Gibbs steps under the full likelihood to fine-tune them to the non-annealed target distribution. This produces our final set of posterior samples for $\mathbf{S}$, $\mathbf{L}$, $\mathbf{M}$, and $\boldsymbol{\delta}$ for which we obtain an empirical estimation of the true posterior distribution.
    
\end{enumerate}

We also generate a higher-resolution $512 \times 512$ source model $\mathbf{S}$ based on the parameter values of $\mathbf{M}$, $\mathbf{L}$, and $\boldsymbol{\delta}$ obtained at the end of the initialization and optimization stage. This high-resolution fit is presented as an example of a lens reconstruction model in Figure~\ref{fig:SDSSJ1430+4105_bestfit} and in Appendix~\ref{sec:appendix_results}. The third Gibbs stage is run in parallel on multiple NVIDIA A100 GPUs. The combined runtime of all three stages for a single lensing system is approximately 200 GPU hours (600 GPU hours for SDSSJ1430+4105).

\section{Results}
\label{sec:results}
\begin{figure*}[!t]
    \centering
    \includegraphics[width=0.95\textwidth]{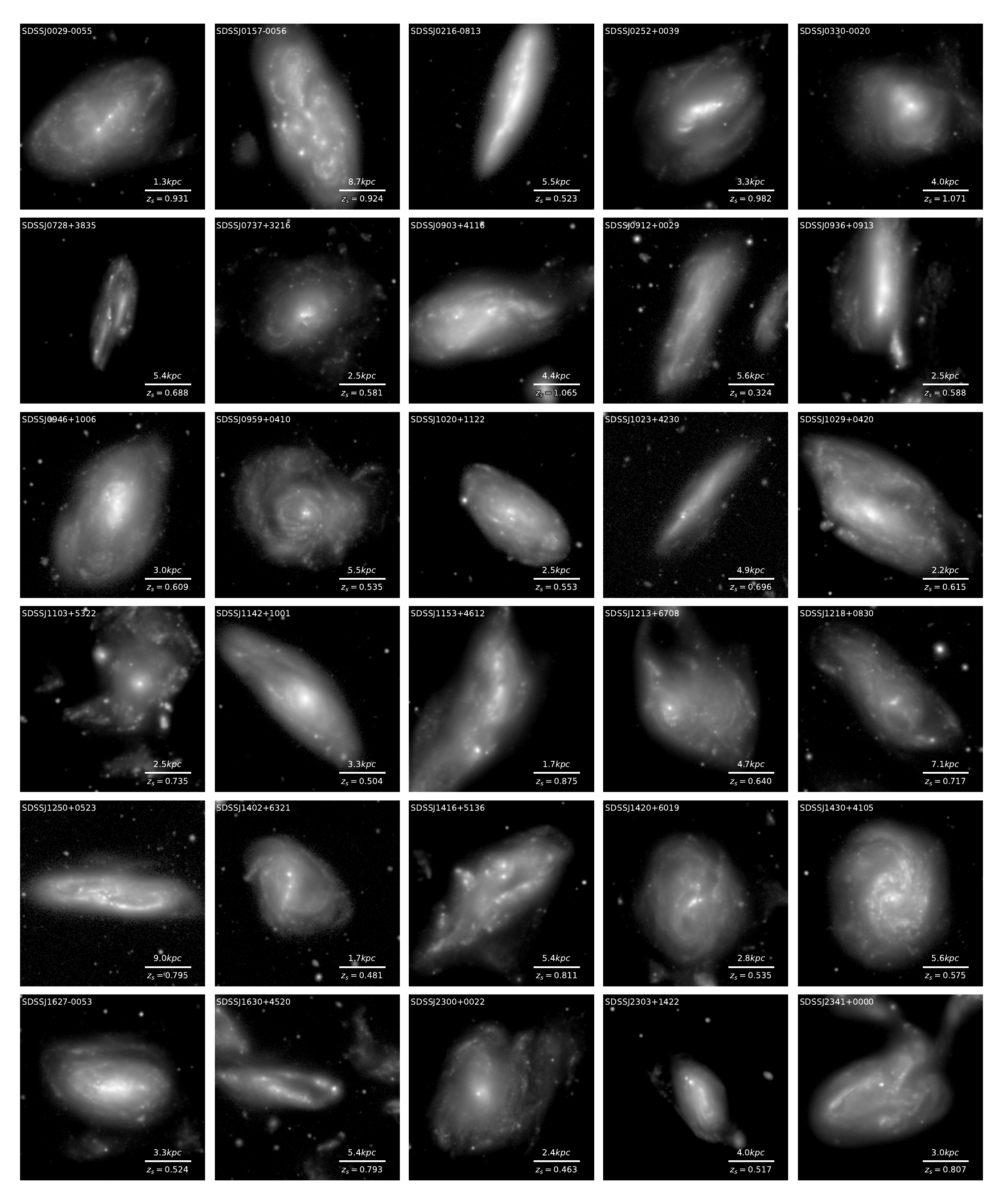}
    \caption{Example high-resolution ($512 \times 512$) sample of the background source for representative lensing systems. Each panel shows the reconstructed source with the lensing caustic overlaid, along with the name of the lensing system, source redshift and a physical reference scale in kiloparsecs (kpc). The physical scale is computed assuming a flat $\Lambda$CDM cosmology with $\Omega_{\mathrm{m}} = 0.3$, $\Omega_{\Lambda} = 0.7$, and $H_{0} = 70~\mathrm{km,s^{-1},Mpc^{-1}}$. Full lens model reconstructions and data–model residuals based on these source models are presented in Appendix~\ref{sec:appendix_results}.}
    \label{fig:sources_good}
\end{figure*}

For each lens system, we present a best-fit model of the lensing data that shows a sample of inferred source and lens light models alongside data-model residuals. An example for SDSSJ1430+4105 is given in Figure~\ref{fig:SDSSJ1430+4105_bestfit}, with analogous reconstructions for all other systems provided in Appendix~\ref{sec:appendix_results}. For these figures, we use the high-resolution source model sample generated on a upsampled \(512\times512\) grid  using our source score model detailed in Section \ref{sec:score_models}. The full set of high-resolution source models is shown in Figure~\ref{fig:sources_good}, overlaid with source redshift \(z_s\) and physical scale reference in kiloparsecs.

\begin{figure*}[t]
    \centering
    \includegraphics[width=\textwidth,trim={10 0 10 0}, clip]{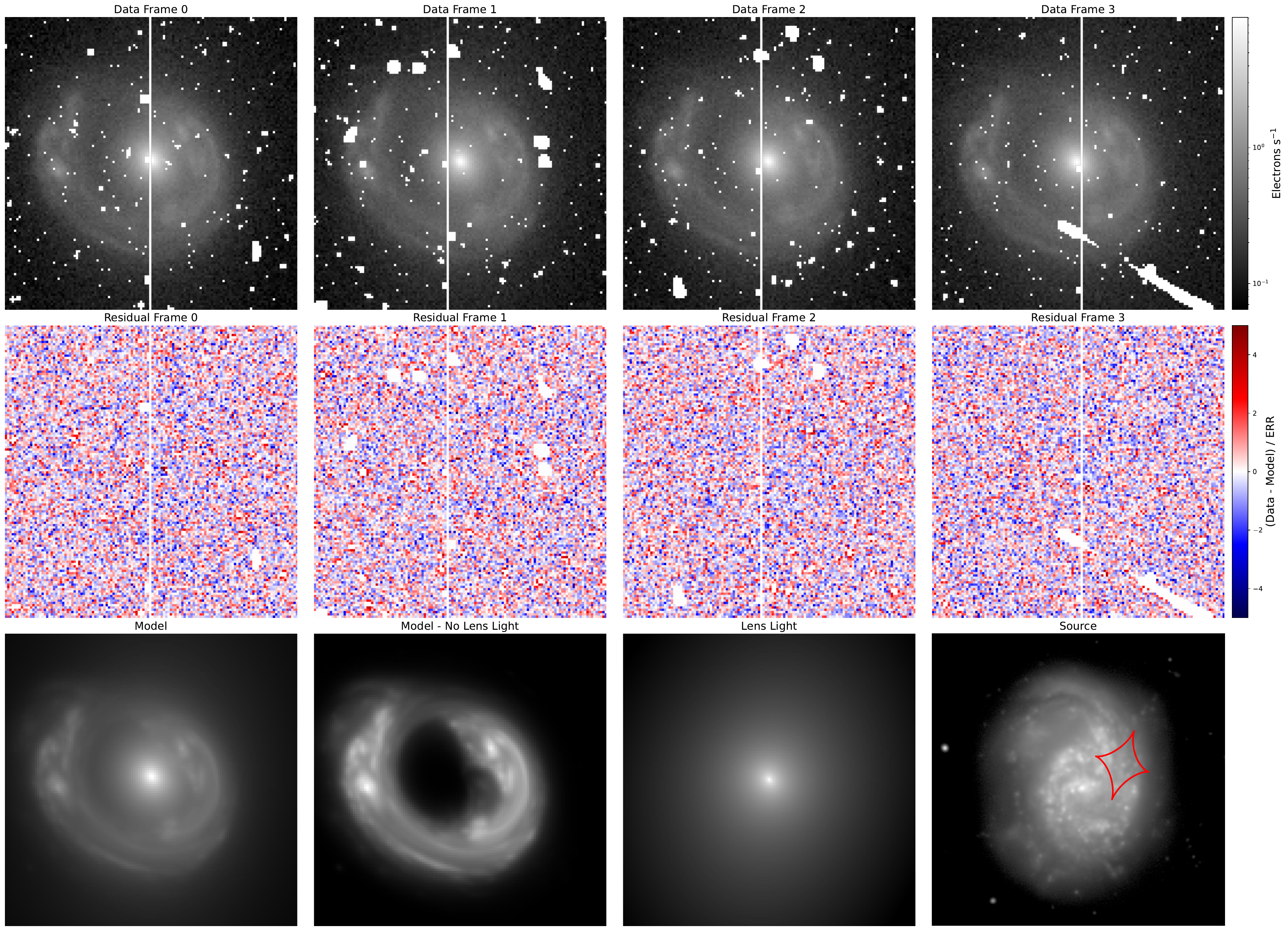}
    \caption{Lens model of SDSSJ1430+4105. From top to bottom: the four dithered FLT frame data cutouts (log-scaled) with the applied data-quality \texttt{DQ} mask; the residuals between the data and the lens model normalized by the pixel-wise likelihood standard deviation, $\sigma = \texttt{ERR}$, with colormap centered and clipped at $\pm 5$ ($5\sigma$); and a summary of the lens model. The bottom row shows, from left to right, the full lens model, the lens model without lens light, the isolated lens light model, and the reconstructed background source (all in log-scale). The red curve indicates the lensing caustic overlaid on the source model: source features inside the caustic are lensed into four observable images.}
    \label{fig:SDSSJ1430+4105_bestfit}
\end{figure*}

\begin{figure*}[!t]
    \centering
    \includegraphics[width=\textwidth]{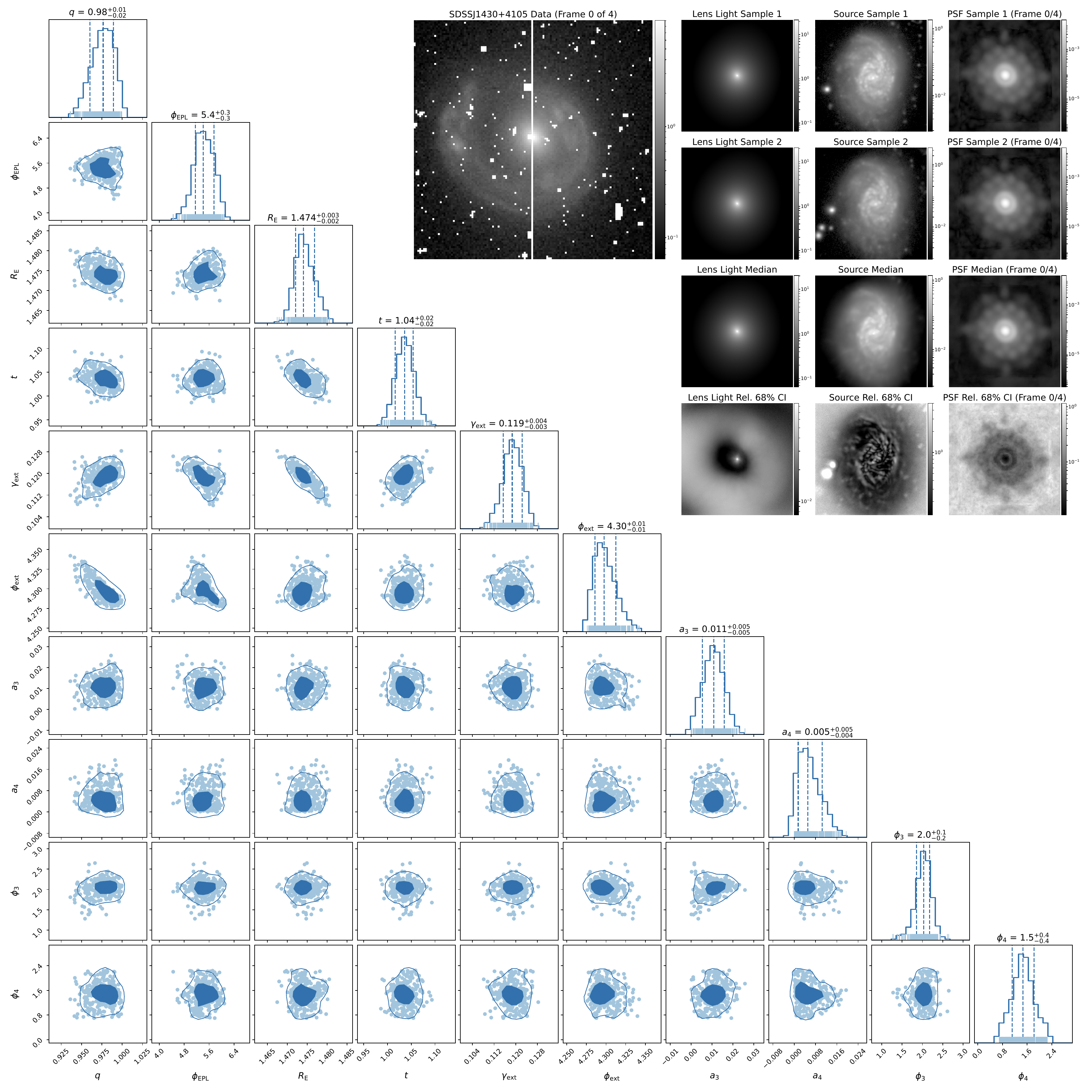}
    \caption{Joint posterior distribution for SDSSJ1430+4105. The corner plot shows posterior samples (light blue) of the foreground mass parameters $\mathbf{M}$, with contours (dark blue) marking the 1$\sigma$ and 2$\sigma$ regions. The 1-D corner plot marginals display the posterior median of the foreground mass parameters $\mathbf{M}$ with 16th–84th percentile uncertainties, indicated by dashed vertical lines and reported above each panel. Posterior samples and uncertainties for the source $\mathbf{S}$, lens light $\mathbf{L}$, and PSF model $\mathbf{P}_0$ (frame 0 of 4) are also shown. The upper-right panels display, from top to bottom, two posterior samples, the median model, and the relative 68\% credible interval (interval width divided by the median). All results are based on 300 posterior samples and are marginalized over alignment parameters $\boldsymbol{\delta}$.
}
    \label{fig:corner_maintext}
\end{figure*}

For most lenses in this work, we report posterior summary statistics obtained from performing all stages of the Gibbs sampling strategy outlined in Section~\ref{sec:gibbs_strategy} (i.e., optimization, annealed sampling, and final non-annealed sampling). Figure~\ref{fig:corner_maintext} shows the joint and marginal posteriors for all major model components for SDSSJ1430+4105; for conciseness, we omit the PSF models for frames 2, 3, and 4, as well as the alignment parameters $\boldsymbol{\delta}$ marginal posteriors. 
For a subset of lenses (SDSSJ0912+0029, SDSSJ0959+0410, SDSSJ1029+0420, SDSSJ1103+5322, SDSSJ1213+6708, SDSSJ1218+0830, SDSSJ1420+6019, SDSSJ2341+0000), we only report best-fit model parameters in Table \ref{tab:macro_params_other}, without associated uncertainties. In these cases, while the initial optimization stage converged to a high-probability region, posterior sampling showed signs of instability.

This instability most often arose from two contributing factors:
\begin{itemize}
\item The chosen temperature of the annealed likelihood, which in some cases was too high to confine the chains within a local minimum consistent with the data;
\item The lens light prior too restrictive to capture complex features in the lensing galaxy light profile, leading to leaking lens light into the background source reconstruction and consequently resulted in unrealistic background sources.
\end{itemize}
Section~\ref{sec:discussion} discusses the specific sources of posterior instability and bias for each of these systems. Further investigations will be needed in future work to diagnose and solve these problems on a case-by-case basis.

Although full posterior sampling was not successful for this subset of lensing systems, the reported best-fit parameters still correspond to high-density regions of parameter space with good residual agreement to the data. We include these best-fit results in Table~\ref{tab:macro_params_other}, along with source reconstructions in Appendix~\ref{sec:appendix_results}, where they may serve as valuable initializations for future modeling efforts.

For the lens systems where posterior sampling was stable, Table~\ref{tab:macro_params} reports the median and $68\%$ credible intervals of relevant foreground mass parameters $\mathbf{M}$.

\begin{figure*}[!t]
    \centering
    \includegraphics[width=\textwidth]{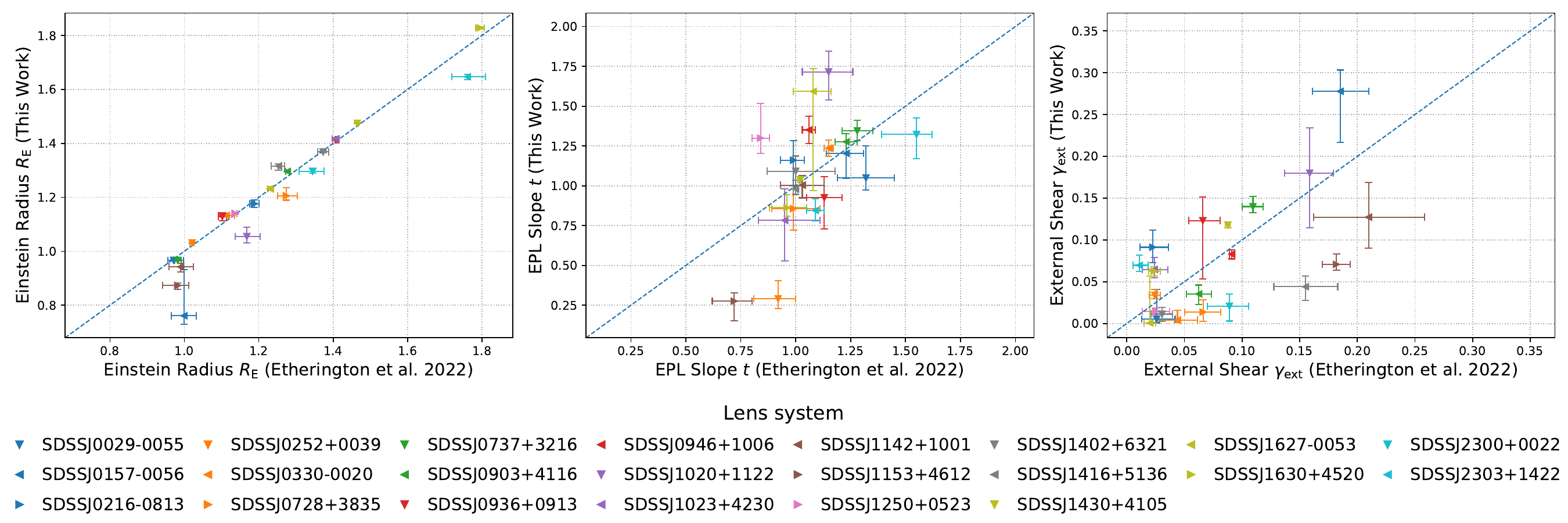}
    \caption{Comparison of inferred foreground mass parameter values between this work and \citet{Etherington2022} (obtained from Table B1 of their supplementary material) for the Einstein radius \(R_{\mathrm{E}}\), EPL slope \(t\) and external shear magnitude \(\gamma_{\mathrm{ext}}\) with error bars signifying the $68\%$ credible-interval widths. Only lenses modeled in both studies are shown; systems for which our inference did not converge or yield stable posterior sampling are excluded. Because \citet{Etherington2022} did not include multipole terms in the foreground mass model, exact agreement with our results—which do include third- and fourth-order multipoles—is not expected. Note that \citet[Table B1]{Etherington2022} report $68\%$ credible intervals on the Cartesian components of the external shear $(\gamma_{1,\mathrm{ext}}, \gamma_{2,\mathrm{ext}})$. The uncertainties we report on their magnitude $\gamma_{\mathrm{ext}}$ are approximate: we draw bootstrap samples of $(\gamma_{1,\mathrm{ext}}, \gamma_{2,\mathrm{ext}})$ from independent Gaussian distributions with $\sigma$ set by their reported $68\%$ credible widths, transform the samples to the shear magnitude, and take the resulting $68\%$ credible width.}
    \label{fig:etherington_milex}
\end{figure*}

To benchmark our inference on foreground-mass model parameters $\mathbf{M}$, we compare our inferred parameter values with those reported by \citet{Bolton2008}, \citet{Etherington2022}, and \citet{Tan2024}. Figures \ref{fig:etherington_milex}, \ref{fig:tan_milex} and \ref{fig:bolton_milex} present comparisons for the Einstein radius \(R_{\mathrm{E}}\), EPL slope \(t\) and external shear \(\gamma_{\mathrm{ext}}\). Figures~\ref{fig:milex_etherington_error_width} and \ref{fig:milex_tan_error_width} show the ratio of \(68\%\) credible-interval widths for the aforementioned mass parameters between our work and the works of \citealt{Etherington2022} and \citealt{Tan2024}. Only lenses modeled in both studies are included; systems with unstable posteriors (i.e, lenses featured in Table \ref{tab:macro_params_other}) are excluded.

\begin{table*}
\centering
\begin{tabular}{ccccccccc}
\hline
System Name & $q$ & $\phi_{\mathrm{EPL}}$ & $R_{\mathrm{E}}$ & $t$ & $\gamma_{\mathrm{ext}}$ & $\phi_{\mathrm{ext}}$ & $a_3$ & $a_4$ \\
\hline
SDSSJ0029$-$0055 & $0.877^{+0.063}_{-0.042}$ & $1.129^{+0.059}_{-0.090}$ & $0.967^{+0.003}_{-0.003}$ & $1.049^{+0.200}_{-0.076}$ & $0.005^{+0.035}_{-0.004}$ & $0.589^{+0.274}_{-0.946}$ & $0.012^{+0.004}_{-0.006}$ & $0.004^{+0.006}_{-0.002}$ \\
SDSSJ0157$-$0056 & $0.445^{+0.075}_{-0.071}$ & $5.328^{+0.108}_{-0.162}$ & $0.761^{+0.171}_{-0.031}$ & $1.203^{+0.125}_{-0.155}$ & $0.278^{+0.025}_{-0.061}$ & $5.192^{+0.164}_{-0.092}$ & $0.009^{+0.012}_{-0.006}$ & $0.008^{+0.010}_{-0.005}$ \\
SDSSJ0216$-$0813 & $0.864^{+0.075}_{-0.059}$ & $5.866^{+0.193}_{-0.172}$ & $1.176^{+0.014}_{-0.013}$ & $1.161^{+0.124}_{-0.158}$ & $0.091^{+0.021}_{-0.018}$ & $5.194^{+0.060}_{-0.088}$ & $0.013^{+0.008}_{-0.009}$ & $0.012^{+0.008}_{-0.008}$ \\
SDSSJ0252$+$0039 & $0.913^{+0.016}_{-0.014}$ & $5.781^{+0.049}_{-0.080}$ & $1.031^{+0.002}_{-0.002}$ & $0.291^{+0.114}_{-0.060}$ & $0.034^{+0.007}_{-0.004}$ & $5.724^{+0.055}_{-0.107}$ & $0.004^{+0.004}_{-0.002}$ & $0.008^{+0.003}_{-0.002}$ \\
SDSSJ0330$-$0020 & $0.896^{+0.024}_{-0.029}$ & $6.098^{+0.120}_{-0.049}$ & $1.133^{+0.003}_{-0.006}$ & $1.236^{+0.051}_{-0.051}$ & $0.004^{+0.012}_{-0.003}$ & $0.008^{+0.313}_{-0.226}$ & $0.007^{+0.009}_{-0.005}$ & $0.005^{+0.009}_{-0.003}$ \\
SDSSJ0728$+$3835 & $0.799^{+0.038}_{-0.061}$ & $0.220^{+0.046}_{-0.061}$ & $1.206^{+0.030}_{-0.016}$ & $0.856^{+0.141}_{-0.136}$ & $0.014^{+0.014}_{-0.011}$ & $4.987^{+0.186}_{-1.301}$ & $0.008^{+0.006}_{-0.005}$ & $0.009^{+0.011}_{-0.006}$ \\
SDSSJ0737$+$3216 & $0.851^{+0.019}_{-0.020}$ & $6.118^{+0.089}_{-0.100}$ & $0.967^{+0.002}_{-0.003}$ & $1.345^{+0.065}_{-0.063}$ & $0.140^{+0.012}_{-0.007}$ & $4.550^{+0.022}_{-0.018}$ & $0.013^{+0.005}_{-0.005}$ & $0.031^{+0.005}_{-0.005}$ \\
SDSSJ0903$+$4116 & $0.787^{+0.025}_{-0.022}$ & $4.757^{+0.093}_{-0.072}$ & $1.296^{+0.004}_{-0.003}$ & $1.276^{+0.052}_{-0.072}$ & $0.035^{+0.011}_{-0.012}$ & $4.428^{+0.154}_{-0.069}$ & $0.016^{+0.009}_{-0.008}$ & $0.009^{+0.007}_{-0.006}$ \\
SDSSJ0936$+$0913 & $0.626^{+0.197}_{-0.084}$ & $4.974^{+0.197}_{-0.168}$ & $1.129^{+0.014}_{-0.017}$ & $0.926^{+0.130}_{-0.197}$ & $0.123^{+0.028}_{-0.070}$ & $1.663^{+0.184}_{-0.216}$ & $0.012^{+0.015}_{-0.008}$ & $0.010^{+0.008}_{-0.006}$ \\
SDSSJ0946$+$1006 & $0.853^{+0.013}_{-0.013}$ & $4.339^{+0.033}_{-0.036}$ & $1.413^{+0.002}_{-0.002}$ & $1.351^{+0.086}_{-0.086}$ & $0.083^{+0.005}_{-0.006}$ & $3.742^{+0.022}_{-0.021}$ & $0.043^{+0.008}_{-0.007}$ & $0.021^{+0.008}_{-0.009}$ \\
SDSSJ1020$+$1122 & $0.628^{+0.226}_{-0.193}$ & $6.103^{+0.860}_{-2.400}$ & $1.055^{+0.035}_{-0.025}$ & $1.716^{+0.130}_{-0.178}$ & $0.180^{+0.054}_{-0.065}$ & $3.030^{+0.159}_{-0.220}$ & $0.012^{+0.013}_{-0.009}$ & $0.014^{+0.010}_{-0.010}$ \\
SDSSJ1023$+$4230 & $0.771^{+0.042}_{-0.035}$ & $4.698^{+0.069}_{-0.081}$ & $1.417^{+0.005}_{-0.004}$ & $0.783^{+0.199}_{-0.255}$ & $0.065^{+0.015}_{-0.010}$ & $4.583^{+0.055}_{-0.110}$ & $0.007^{+0.007}_{-0.005}$ & $0.049^{+0.010}_{-0.015}$ \\
SDSSJ1142$+$1001 & $0.752^{+0.106}_{-0.099}$ & $5.344^{+0.146}_{-0.117}$ & $0.942^{+0.013}_{-0.018}$ & $1.004^{+0.059}_{-0.080}$ & $0.127^{+0.042}_{-0.037}$ & $5.072^{+0.096}_{-0.108}$ & $0.006^{+0.006}_{-0.005}$ & $0.006^{+0.008}_{-0.004}$ \\
SDSSJ1153$+$4612 & $0.918^{+0.010}_{-0.015}$ & $0.683^{+0.179}_{-0.093}$ & $0.873^{+0.008}_{-0.015}$ & $0.276^{+0.053}_{-0.124}$ & $0.071^{+0.013}_{-0.007}$ & $1.020^{+0.038}_{-0.012}$ & $0.018^{+0.006}_{-0.005}$ & $0.014^{+0.005}_{-0.006}$ \\
SDSSJ1250$+$0523 & $0.984^{+0.013}_{-0.025}$ & $1.623^{+1.470}_{-2.033}$ & $1.140^{+0.003}_{-0.003}$ & $1.300^{+0.217}_{-0.098}$ & $0.014^{+0.004}_{-0.006}$ & $1.191^{+0.107}_{-0.158}$ & $0.006^{+0.006}_{-0.005}$ & $0.041^{+0.012}_{-0.008}$ \\
SDSSJ1402$+$6321 & $0.730^{+0.034}_{-0.035}$ & $0.465^{+0.043}_{-0.039}$ & $1.368^{+0.003}_{-0.004}$ & $1.091^{+0.095}_{-0.142}$ & $0.011^{+0.008}_{-0.009}$ & $0.793^{+0.802}_{-0.431}$ & $0.007^{+0.008}_{-0.004}$ & $0.010^{+0.007}_{-0.007}$ \\
SDSSJ1416$+$5136 & $0.828^{+0.056}_{-0.045}$ & $1.471^{+0.199}_{-0.163}$ & $1.316^{+0.007}_{-0.017}$ & $0.980^{+0.054}_{-0.034}$ & $0.044^{+0.013}_{-0.016}$ & $5.570^{+0.285}_{-0.185}$ & $0.008^{+0.011}_{-0.006}$ & $0.012^{+0.012}_{-0.007}$ \\
SDSSJ1430$+$4105 & $0.977^{+0.013}_{-0.016}$ & $5.416^{+0.345}_{-0.256}$ & $1.474^{+0.003}_{-0.002}$ & $1.036^{+0.018}_{-0.020}$ & $0.119^{+0.004}_{-0.003}$ & $4.298^{+0.015}_{-0.011}$ & $0.011^{+0.005}_{-0.005}$ & $0.005^{+0.005}_{-0.004}$ \\
SDSSJ1627$-$0053 & $0.747^{+0.046}_{-0.111}$ & $1.367^{+0.200}_{-0.029}$ & $1.232^{+0.009}_{-0.002}$ & $1.593^{+0.144}_{-0.623}$ & $0.001^{+0.056}_{-0.001}$ & $0.554^{+1.088}_{-0.545}$ & $0.010^{+0.013}_{-0.006}$ & $0.014^{+0.003}_{-0.010}$ \\
SDSSJ1630$+$4520 & $0.754^{+0.012}_{-0.014}$ & $0.199^{+0.034}_{-0.022}$ & $1.828^{+0.004}_{-0.005}$ & $0.863^{+0.082}_{-0.055}$ & $0.063^{+0.005}_{-0.006}$ & $0.149^{+0.058}_{-0.030}$ & $0.023^{+0.013}_{-0.015}$ & $0.038^{+0.004}_{-0.009}$ \\
SDSSJ2300$+$0022 & $0.652^{+0.043}_{-0.036}$ & $0.006^{+0.091}_{-0.063}$ & $1.297^{+0.009}_{-0.007}$ & $1.322^{+0.104}_{-0.151}$ & $0.021^{+0.015}_{-0.018}$ & $3.645^{+2.643}_{-1.834}$ & $0.011^{+0.009}_{-0.008}$ & $0.007^{+0.006}_{-0.005}$ \\
SDSSJ2303$+$1422 & $0.659^{+0.029}_{-0.033}$ & $1.200^{+0.051}_{-0.032}$ & $1.647^{+0.009}_{-0.010}$ & $0.845^{+0.074}_{-0.065}$ & $0.070^{+0.012}_{-0.007}$ & $1.719^{+0.121}_{-0.067}$ & $0.040^{+0.006}_{-0.007}$ & $0.016^{+0.009}_{-0.011}$ \\
\hline
\end{tabular}
\caption{Summary of foreground mass model parameters $\mathbf{M}$. Reported values are posterior medians with 68\% credible intervals marginalized over the source $\mathbf{S}$, lens light $\mathbf{L}$, PSF $\mathbf{P}$ and alignment parameters $\boldsymbol{\delta}$.}
\label{tab:macro_params}
\end{table*}

\section{Discussion}
\label{sec:discussion}

The lens models featured in Figure~\ref{fig:SDSSJ1430+4105_bestfit} and Appendix~\ref{sec:appendix_results} demonstrate that the vast majority of signal in \textit{HST}-like data can be effectively modeled by combining parametric lens-mass models (EPL, external shear, and third- and fourth-order multipoles) with high-dimensional, data-driven models for the source, lens light, and PSF. This holds across a representative sample of high-quality SLACS lenses with source galaxies spanning diverse morphologies, including irregular or clumpy galaxies such as SDSSJ1416+5136 (Figure~\ref{fig:stage1_SDSSJ1416+5136}), SDSSJ0903+4116 (Figure~\ref{fig:stage1_SDSSJ0903+4116}) and SDSSJ0157$-$0056 (Figure~\ref{fig:stage1_SDSSJ0157-0056}), redshifts ($z_s \in [0.324, 1.071]$), and physical sizes.

SDSSJ1430+4105 (Figure~\ref{fig:SDSSJ1430+4105_bestfit}) provides a clear showcase: the model achieves noise-consistent residuals and a source model with realistic galaxy morphology. By contrast, prior analyses \citep[e.g.,][]{Bolton2008, Eichner2012, Vegetti2014, Etherington2022, BayerI2023} have generally relied on parametric or pixelated sources with simple, non-physically motivated priors (e.g., smoothness regularization, Gaussian or wavelet-based priors) and have resulted in significant residuals or unrealistic source morphologies.

Beyond the reconstructions showcased in Figure \ref{fig:SDSSJ1430+4105_bestfit} and Appendix \ref{sec:appendix_results}, Figure \ref{fig:corner_maintext} illustrates that our framework is also capable of providing posterior uncertainty estimates for all components of the lens model: the background source $\mathbf{S}$, the lens light $\mathbf{L}$, the foreground mass distribution $\mathbf{M}$, and the four PSF models $\mathbf{P}$ (one PSF model per data frame). The $68\%$ credible interval of posterior samples for the source, lens light and PSF, exhibit features consistent with our expectations on where the highest amount of  information is available from the observed data. For example, the source model uncertainties are largest in the outskirts of the galaxy, which are regions with low brightness, and smallest near the lensing caustic, where features are lensed into multiple highly magnified images. Lens light uncertainties are highest near the lensed arcs, reflecting degeneracies between the lens light and source brightness, and at the center brightess peak, where information on the cuspiness of the lens light peak is limited due to PSF blurring, uncertainties in the PSF model and uncertainties in the relative alignment shift parameter $\boldsymbol{\delta}$, for which we marginalize over during inference. For the PSF models, beyond the diffraction spikes, the relative uncertainties are nearly uniform, indicating prior-dominated regions that are unconstrained by the star cutouts used for modeling. These patterns observed in the pixel-wise $68\%$ credible interval of $\mathbf{S}$, $\mathbf{L}$ and $\mathbf{P}$ are consistent with expectations and highlight where future observations or complementary constraints could most improve the precision of the lens model.

In Figures~\ref{fig:etherington_milex}, \ref{fig:tan_milex}, and \ref{fig:bolton_milex}, we compare our inferred values of the Einstein radius \(R_{\mathrm{E}}\), mass slope \(t\) and external shear \(\gamma_{\mathrm{ext}}\) with the values reported by \citet{Bolton2008}, \citet{Etherington2022} and \citet{Tan2024}. All studies adopt a power-law mass profile with external shear with the exception of \citet{Bolton2008}, where an SIE lens profile was assumed (equivalent to a mass slope of \(t=1\)). The models in this work additionally include third- and fourth-order multipoles, so perfect agreement across studies is not expected. Aside from a few systems, \(R_{\mathrm{E}}\) closely tracks the results of these previous works; by contrast, the correlations for \(t\) and \(\gamma_{\mathrm{ext}}\) are weaker as these parameters exhibit larger uncertainties. It is interesting to note, however, that our inferred \(t\) values appear to scatter symmetrically about \(1\), indicating no clear trend for profiles systematically steeper or shallower than the SIE profile given the population of lenses analyzed in this work. Figures~\ref{fig:milex_etherington_error_width} and \ref{fig:milex_tan_error_width} show, for each lens, the ratio of the \(68\%\) credible-interval widths on the foreground mass parameters between our analysis and those of \citet{Etherington2022} and \citet{Tan2024}. The most striking pattern is the comparison to \citet{Tan2024}, where our inferred uncertainties are systematically larger across all parameters. In contrast, the ratio of parameter uncertainties between our analysis and \citet{Etherington2022} as seen in Figure~\ref{fig:milex_etherington_error_width} are more closely distributed around unity, with the exception of the Einstein radius $R_{\mathrm{E}}$, where our uncertainties are overall lower than \citet{Etherington2022}.

Given that the studies differ primarily in their source and lens light prior distributions, differences in inferred statistics for \(R_{\mathrm{E}}\), \(t\)  and \(\gamma_{\mathrm{ext}}\), point towards systematics driven by the choice of prior distribution. Specifically, in \citet{Tan2024}, the source is represented by a S\'ersic component plus a shapelet basis with fixed number of shapelet components per lens, and the lens light is modeled with two S\'ersic components. These choices tightly constrain the lens light and background source degrees of freedom and specifically for the background source, limit the ability to capture high-frequency or complex morphological structure, as seen from figures of data-model residuals in \citet{Tan2024}. This parametric restriction on foreground and background light profiles can in turn lead to artificially narrow posteriors on the mass parameters as it limits the exploration of the full space of solutions of the lens model. In contrast to \citet{Tan2024}, \citet{Etherington2022} employed a more flexible source representation in which the source is reconstructed on an adaptive Voronoi mesh, with a smoothing regularization that acts as an effective prior by penalizing large differences between neighboring mesh pixels. This fully pixelated source representation is a plausible reason for the more conservative error bars seen on their inferred foreground mass parameters when compared to \citet{Tan2024}, closely matching the overall width of the \(68\%\) credible-interval with our results (Figure~\ref{fig:milex_etherington_error_width}). It is important to note that \citet{Etherington2022}, similar to \citet{Tan2024}, also adopts a double Sérsic lens light model for the majority of lenses, so differences stem primarily due to the chosen source-model representation. These results therefore seem to point towards the conclusion that adopting more realistic and flexible source representations could lead to less bias in inferred lens parameters.

In terms of the data-model residuals for the various lenses reported by \citet{Etherington2022} and \citet{Tan2024}, we highlight a few key points. For the majority of lenses in this work (with a few exceptions; e.g., Figures~\ref{fig:stage1_SDSSJ0959+0410}, \ref{fig:stage1_SDSSJ1402+6321}, \ref{fig:stage1_SDSSJ1416+5136}), we are able to fully reconstruct the observed foreground lens light, achieving noise-consistent residuals, especially near the central brightest peak. In contrast, \citet{Tan2024} report difficulty reconstructing the central peak of the lens light and mask the central region when plotting data-model residuals. Similarly, \citet{Etherington2022} show significant residuals near the center peak on the majority of SLACS lenses analyzed in their work; both works assumed a double S\'ersic lens light model (with a few exceptions in \citet{Etherington2022}, where a B-spline lens light model was used for a few lenses). By contrast, our lens light score prior learns the prior distribution directly in pixel space, and is not restricted to a specific parametric form (despite being trained on S\'ersic profiles, score models are known to learn distributions with support on their entire parameter space, enabling them to generalize well outside of their training set), meaning it can model more complicated lens light profiles deviating from the analytic S\'ersic form. This flexibility enables faithful reconstruction of the central lens light peaks without having to resort to masking. For the few lenses that exhibit central residuals in our work (e.g., SDSSJ1103+5322 in Figure~\ref{fig:stage1_SDSSJ1103+5322} and SDSSJ2341+0000 in Figure~\ref{fig:stage1_SDSSJ2341+0000}), the results do suggest that a lens light score prior trained solely on parametric profiles is too restrictive. To minimize residuals, future improvements could include training the score prior on samples drawn directly from observed galaxies or hydrodynamical simulations that span a broader, more realistic range of lens light morphologies.

In reconstructing the lensed background source, both our framework and the one employed in \citet{Etherington2022} recover the lensing signal down to observational noise for most systems. By contrast, the S\'ersic plus wavelet source model in \citet{Tan2024} shows difficulty reproducing high-frequency structure in the arcs, leaving more residual signal in their data--model comparisons. 

Although we generally achieve excellent agreement with the data, our fixed size pixelated grid for our source model is a limiting factor in reconstructing highly magnified features present in the lensed arcs. For example, in SDSSJ0029$-$0055, noticeable residuals persist in the upper-right lensed image of the source (see Figure~\ref{fig:stage1_SDSSJ0029-0055}). This corresponds to a compact point-like feature sitting near the cusp of the lensing caustic in our source model, a region of high magnification. 
Although it is difficult to judge definitively from the residual plots in \citet{Etherington2022}, their adaptive Voronoi mesh appears better able to capture this highly magnified feature, consistent with the resolution adaptivity of their source model. 
This suggests that higher resolution may be needed to capture the most highly magnified regions. 
This can be achieved by simply increasing the resolution of the source reconstruction (requiring training on higher resolution images of unlensed galaxies for the prior). Alternatively, recent advances in continuous-field diffusion generative models \citep{Zhuang2023, Lim2023} may alleviate the limitations associated with using a uniform pixel resolution grid. 

In addition to residuals seen in SDSSJ0029-0055, the reconstructions of SDSSJ0252+0039 (Figure \ref{fig:stage1_SDSSJ0252+0039}) and SDSSJ0946+1006 (Figure \ref{fig:stage1_SDSSJ0946+1006}) exhibit noticeable residuals near the lensing arcs that may indicate imperfect modeling of the foreground lens mass distribution. These discrepancies may stem from the absence of dark matter substructure in our modeling, as previous studies have reported evidence for substructure in SDSSJ0252+0039 \citep{BayerI2023,BayerII2023} and SDSSJ0946+1006 \citep{Vegetti2010, Quinn2021, Ballard2024, Enzi2025}. However, the flexible, high-resolution, data-driven priors used for the background source in this work may have reduced the significance of residuals previously attributed to substructure by more faithfully capturing the background source galaxy. In line with this argument, recent work has demonstrated how differences in assumed source model representation and prior distribution can alter the significance of dark matter substructure detection \citep{Ballard2024, Stacey2025}. We therefore argue that the use of realistic, data-driven source models warrants a reassessment of the significance of previously reported substructure detections.

A key limitation in our framework is the potential mismatch between the assumed and true prior and likelihood distributions. For the source galaxy, our prior is based on the PROBES dataset \citep{Stone_2019, Stone_2021}, which consists of high-resolution, low-redshift galaxy images taken from the Dark Energy Spectroscopic Instrument (DESI) Legacy Surveys. This can bias reconstructions of higher-redshift, irregular galaxies, imprinting features characteristic of low-redshift galaxies that are prior rather than likelihood driven. Analyzing the median of posterior samples helps distinguish likelihood-driven from prior-driven morphologies: for example, in SDSSJ1430+4105 (Figure \ref{fig:corner_maintext}), spiral arms persist in the median reconstruction, suggesting they are likelihood-driven, while individual features in the outskirts become washed out, likely indicating they are prior-driven. Note that changes in source size can also wash out features consistent between posterior samples. Also, note that the PROBES dataset features galaxies imaged at different wavelengths than the HST observations used in this work, which is another source of distributional shift that would need to be addressed in future work.

An analogous mismatch is the assumed data noise distribution. Our use of a Gaussian likelihood simplifies the treatment of observational noise but fails to capture the statistics of non-Gaussian noise commonly present in \textit{HST} data \citep{Stark2024}. Such deviations can bias the interpretation of residuals, with non-Gaussian noise potentially mimicking substructure or biasing the inference of the background source. More flexible, data-driven noise models such as SLIC \citep{Legin2023_slic}, which use score-based models to learn the noise distribution directly from data examples, offer a promising route to mitigate these potential biases.

\section{Conclusion}
\label{sec:conclusion}
In conclusion, we revisit 30 strong lensing systems from the SLACS survey and present state-of-the-art lens modeling results powered by high-dimensional, data-driven priors for the background source, foreground lens light, and point-spread function (PSF). For each system, we provide a high-resolution lens reconstruction, and for the majority of lenses, we additionally report posterior sample statistics over the foreground mass distribution, lens light, source, and PSF models. This work represents the first application of data-driven priors to real strong lensing data and demonstrates their potential to overcome limitations of traditional parametric models. By capturing the complex, realistic structure of lensing components, our framework enables more accurate and physically meaningful reconstructions. These advances pave the way for robust and high-fidelity strong lens modeling in preparation for the next generation of wide-field surveys.

\begin{acknowledgments}
R.L. and A.A. acknowledge funding from the NSERC CGS D scholarships. C.S. acknowledges the support of a NSERC Postdoctoral Fellowship and a CITA National Fellowship. G.M.B. acknowledges support from the Fonds de recherche du Québec – Nature et technologies (FRQNT) under a Doctoral Research Scholarship (https://doi.org/10.69777/368273). Y.H. and L.P. acknowledge support from the Canada Research Chairs Program, the National Sciences and Engineering Council of Canada through grants RGPIN-2020- 05073 and 05102, and the Fonds de recherche du Québec through grants 2022-NC-301305 and 300397.
\end{acknowledgments}

\software{
\texttt{caustics} \citep{Stone2024}, 
\texttt{astrophot} \citep{Stone_2023},
\texttt{scoremodels} \citep{Adam_score_models},
\texttt{pytorch} \citep{Pytorch2019}
}


\bibliography{bibliography}{}
\bibliographystyle{aasjournalv7}

\clearpage        
\appendix

\section{Notes for Each Lens}

In the following, we provide brief notes on individual lens systems. These comments highlight features of the reconstructed source and lens light models, residual structures identified in the data–model comparison plots (Appendix~\ref{sec:appendix_results}), foreground mass parameter values and, where relevant, insights drawn from the posterior samples generated from this work.

\begin{figure*}[t]
    \centering
    \includegraphics[width=\textwidth]{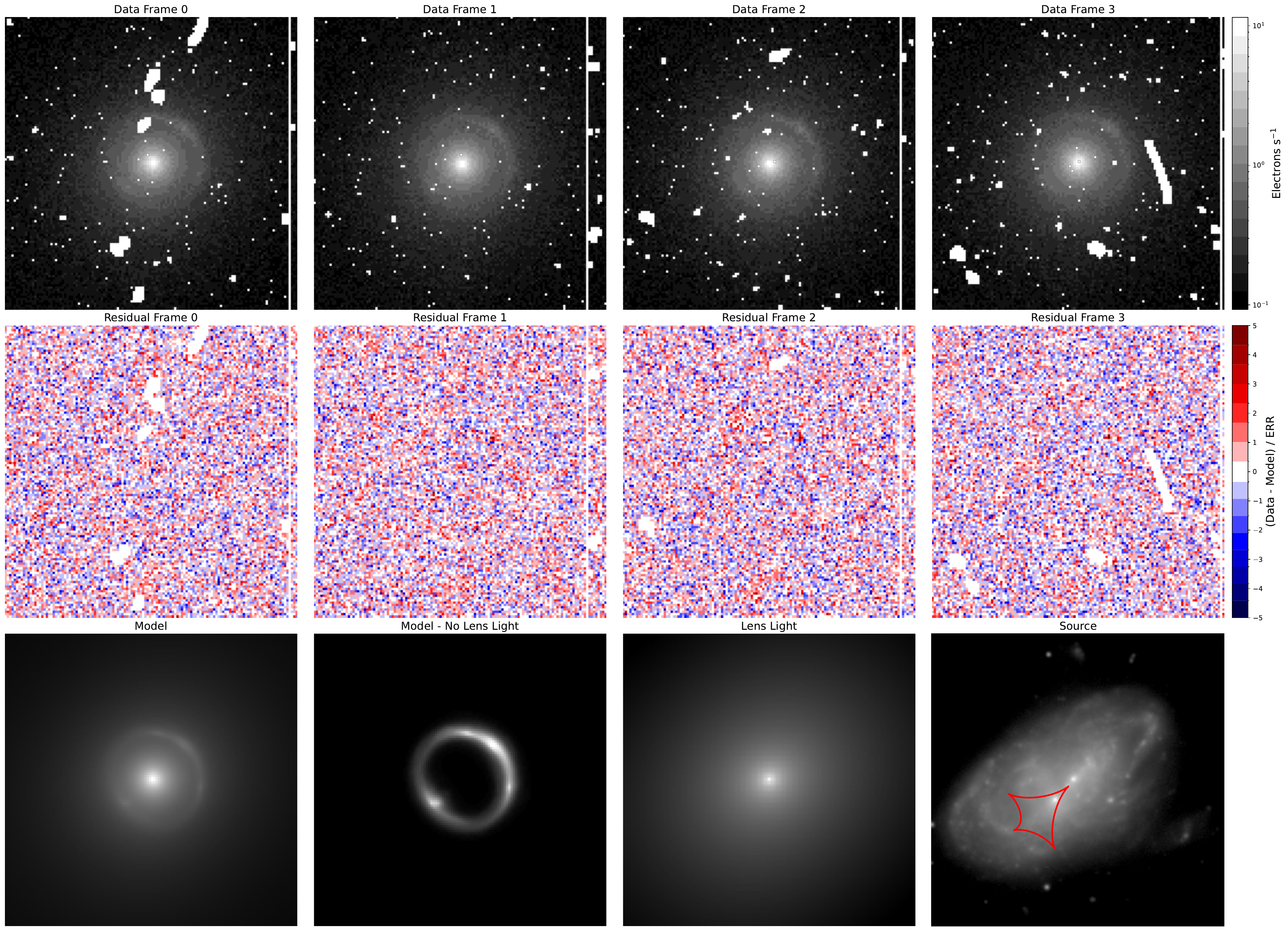}
    \caption{Same as Figure~\ref{fig:SDSSJ1430+4105_bestfit}, but for SDSSJ0029-0055.}
    \label{fig:stage1_SDSSJ0029-0055}
\end{figure*}

\textit{SDSSJ0029-0055} High-redshift background source galaxy ($z_s = 0.931$) featuring two compact bright points in the high-resolution ($512 \times 512$) sample seen in Figure \ref{fig:stage1_SDSSJ0029-0055}: one quadruply imaged within the inner caustic and one doubly imaged near the caustic cusp. Noticeable residuals persist near the upper-right brightest lensed counter image (see Figure ~\ref{fig:stage1_SDSSJ0029-0055}). Posterior samples reveal compact features of size of order $\sim$200 pc.

\textit{SDSSJ0157-0056} High-redshift background source galaxy ($z_s = 0.924$) with multiple compact regions of size of order $\sim$1000 pc, potentially signaling regions of star formation. This system exhibits one of the largest predicted uncertainties on the Einstein radius $R_{\rm E}$. It also shows the highest predicted external shear $\gamma_{\mathrm{ext}}$, which is consistent with the results of \citet{Bolton2008} and \citet{Etherington2022}. 

\textit{SDSSJ0216-0813} Elongated background source galaxy ($z_s = 0.523$), possibly viewed edge-on, with a slight curve located near the lensing caustic. Two external galaxies were masked in the cutout (to the right and bottom); a multi-deflector mass model should be considered to verify whether these external sources affect the lensed image. The value of $\gamma_{\mathrm{ext}}$ from \citet{Bolton2008} and \citet{Etherington2022} are predicted lower than the values we report, albeit still consistent within our uncertainties.

\begin{figure*}[t]
    \centering
    \includegraphics[width=\textwidth]{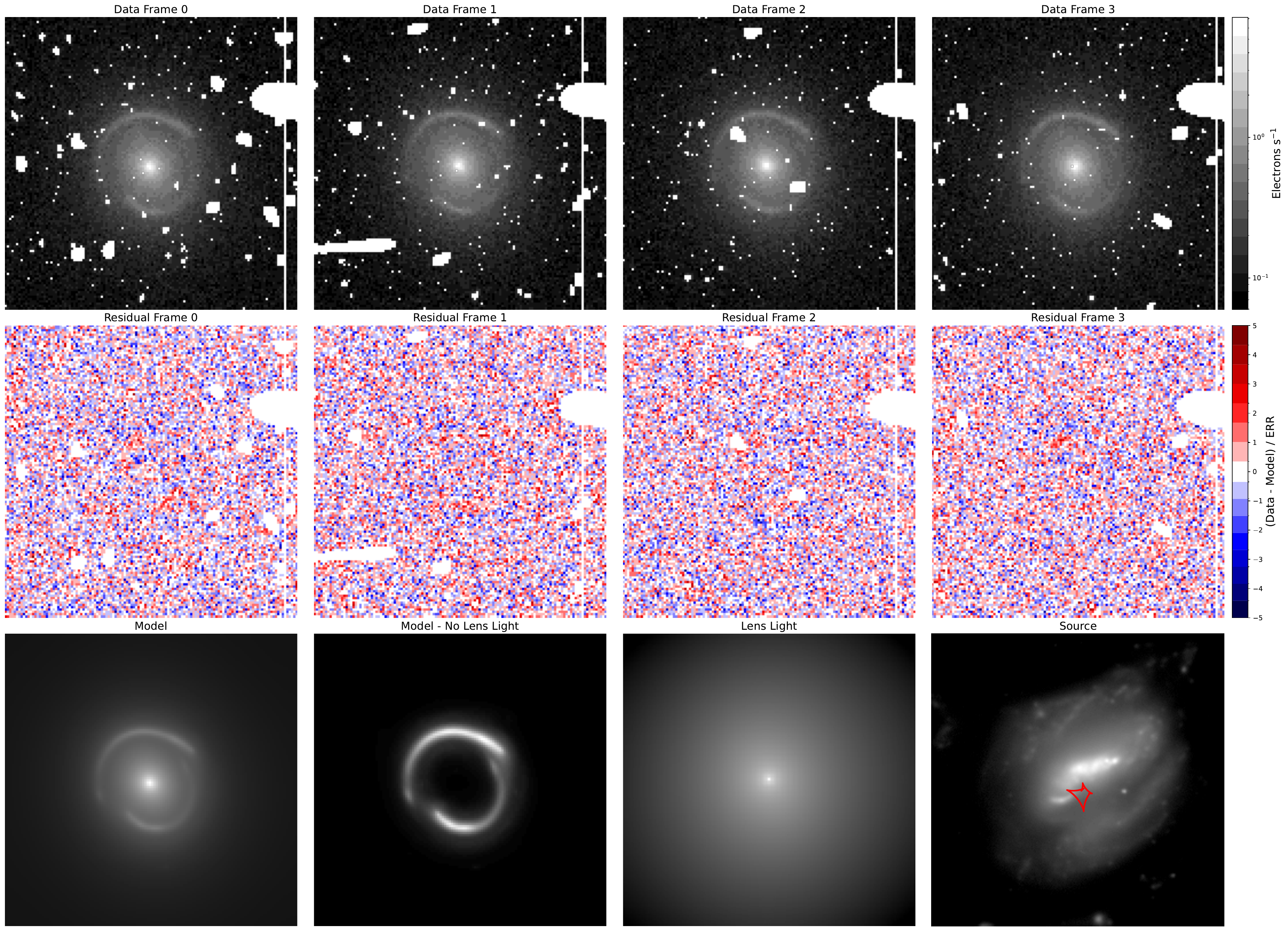}
    \caption{Same as Figure~\ref{fig:SDSSJ1430+4105_bestfit}, but for SDSSJ0252+0039.}
    \label{fig:stage1_SDSSJ0252+0039}
\end{figure*}

\textit{SDSSJ0252+0039} High-redshift background source galaxy ($z_s = 0.982$). Figure ~\ref{fig:stage1_SDSSJ0252+0039} shows noticeable residuals near the arcs, suggestive of model mismatch, potentially due to dark-matter substructure \citep{BayerI2023, BayerII2023}. Foreground mass parameter estimates are generally consistent with \citet{Bolton2008}, \citet{Etherington2022}, and \citet{Tan2024}, although a noticeable discrepancy is observed in the mass slope $t$ compared to \citet{Etherington2022}.

\begin{figure*}[t]
    \centering
    \includegraphics[width=\textwidth]{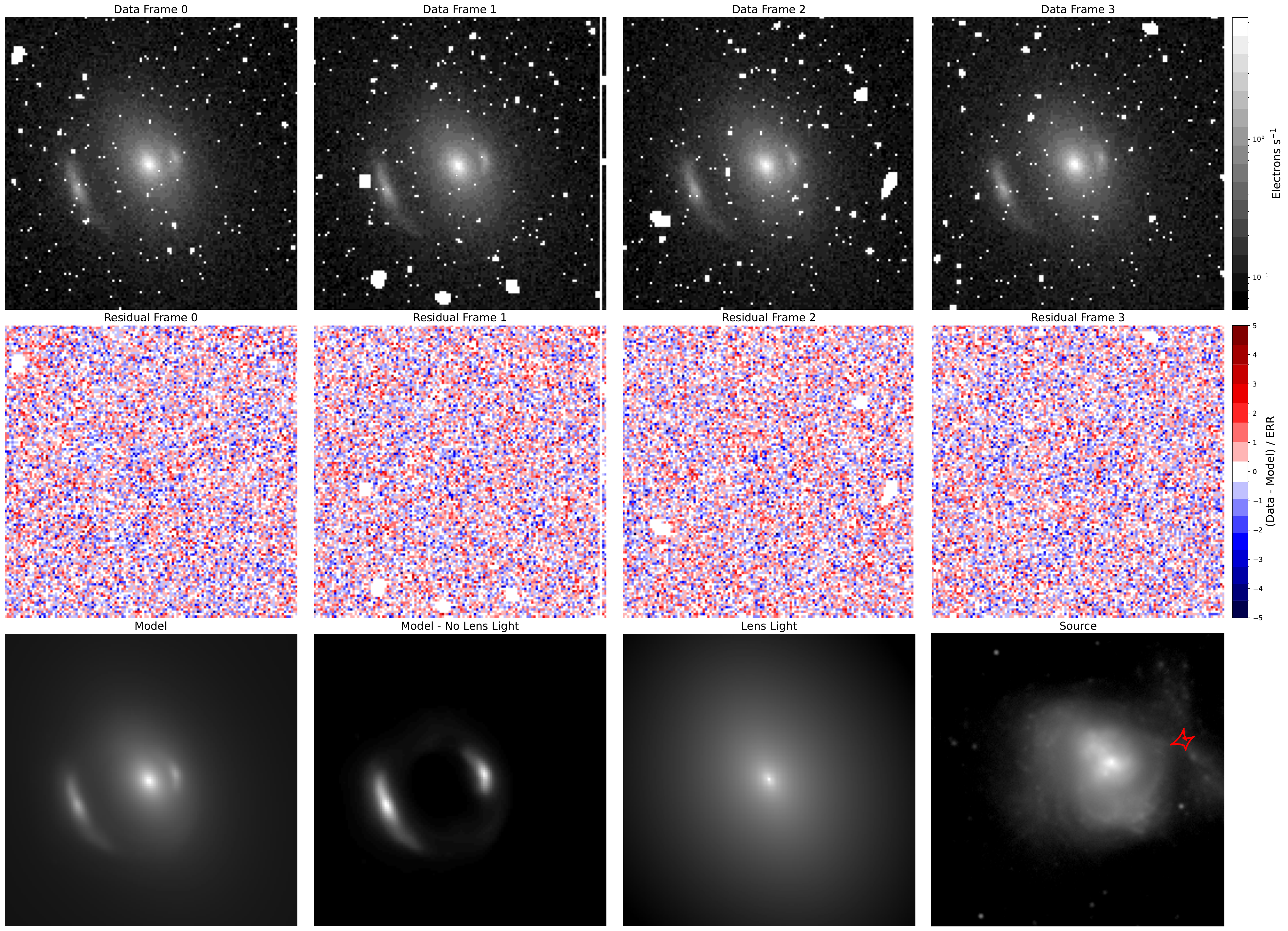}
    \caption{Same as Figure~\ref{fig:SDSSJ1430+4105_bestfit}, but for SDSSJ0330-0020.}
    \label{fig:stage1_SDSSJ0330-0020}
\end{figure*}

\textit{SDSSJ0330-0020} High-redshift background source galaxy ($z_s = 1.071$) with a bright core and smooth outer edge as seen from Figure \ref{fig:stage1_SDSSJ0330-0020} and posterior samples. This is the highest redshift source in our sample of lenses.

\begin{figure*}[t]
    \centering
    \includegraphics[width=\textwidth]{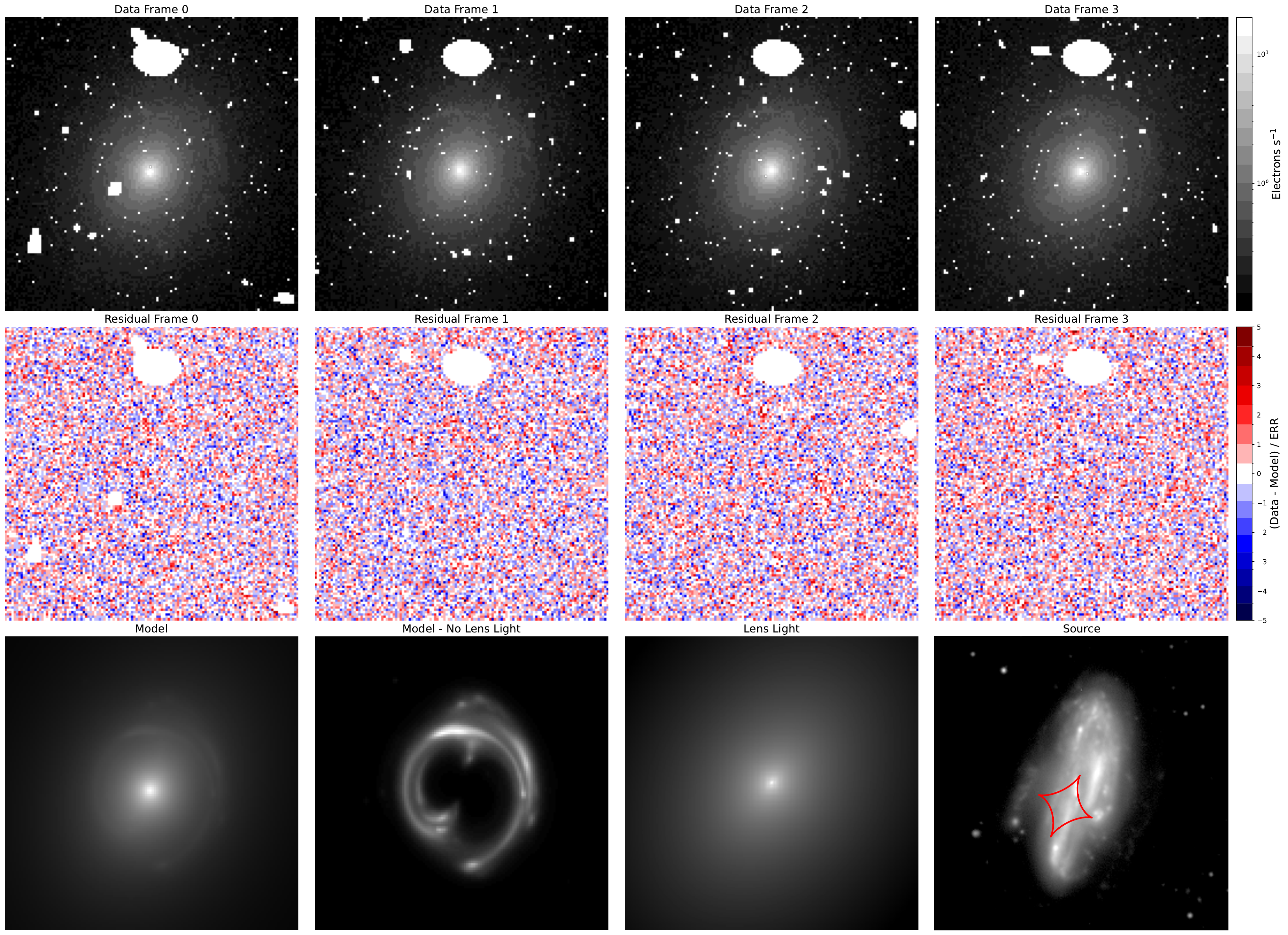}
    \caption{Same as Figure~\ref{fig:SDSSJ1430+4105_bestfit}, but for SDSSJ0728+3835.}
    \label{fig:stage1_SDSSJ0728+3835}
\end{figure*}

\textit{SDSSJ0728+3835} Background source galaxy ($z_s = 0.688$) whose reconstruction resembles two galaxies aligned  parallel to each other, as seen in Figure \ref{fig:stage1_SDSSJ0728+3835} and posterior samples. A bright object was masked in the data to avoid potential modeling bias. Foreground mass parameters are generally consistent with \citet{Bolton2008}, \citet{Etherington2022}, and \citet{Tan2024}.

\begin{figure*}[t]
    \centering
    \includegraphics[width=\textwidth]{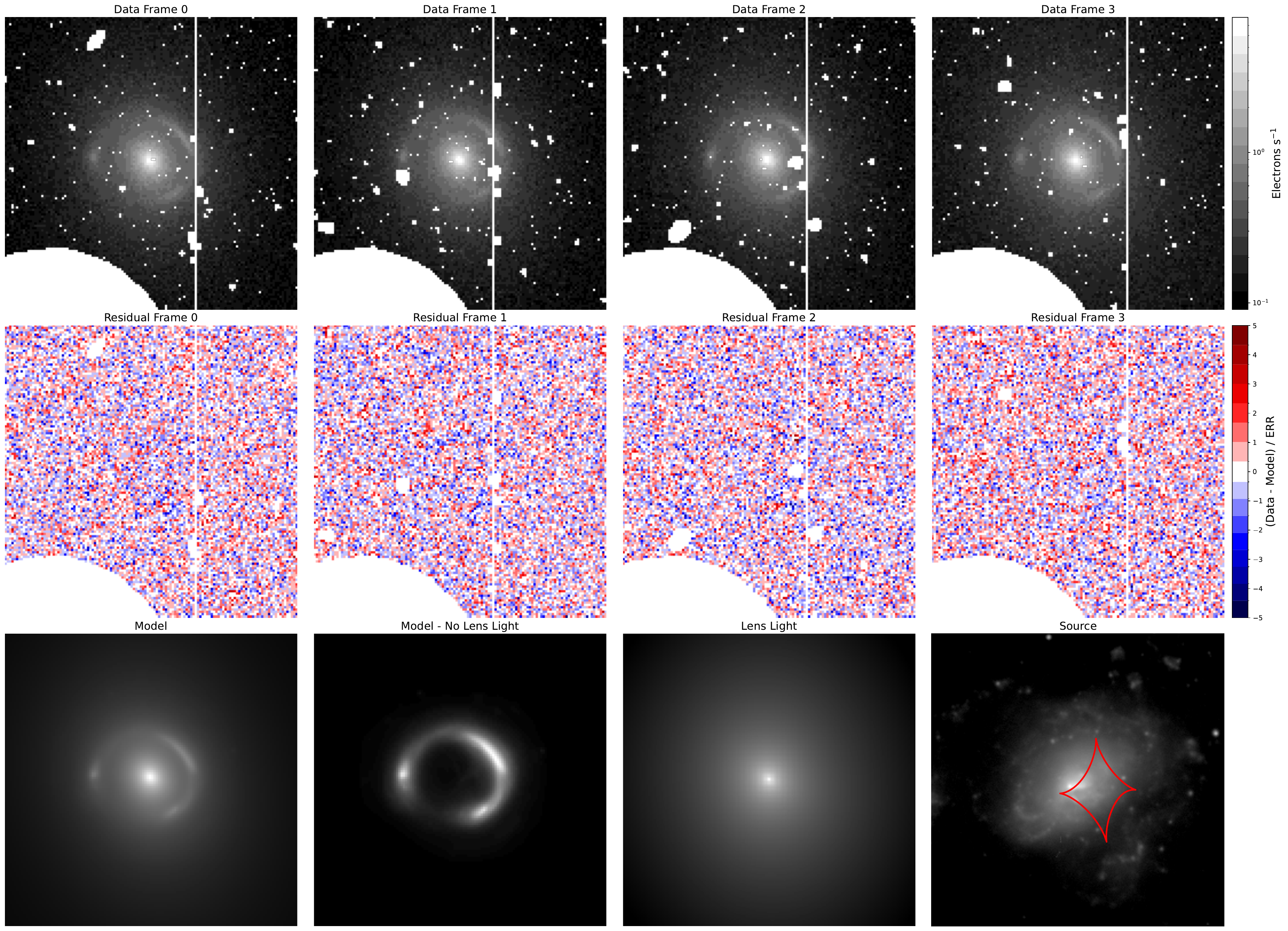}
    \caption{Same as Figure~\ref{fig:SDSSJ1430+4105_bestfit}, but for SDSSJ0737+3216.}
    \label{fig:stage1_SDSSJ0737+3216}
\end{figure*}

\textit{SDSSJ0737+3216} Background source galaxy ($z_s = 0.581$) with possible spiral arms as seen from posterior samples. Mild residuals near the arcs in Figure ~\ref{fig:stage1_SDSSJ0737+3216} indicate modest model mismatch. Foreground mass parameters show general agreement with \citet{Etherington2022} and \citet{Tan2024}.

\textit{SDSSJ0903+4116} High-redshift background source galaxy ($z_s = 1.065$) with recurring bright compact clumps of size on the order of $\sim$1000 pc visible in the posterior samples. The source model suggests an irregular galaxy morphology. Foreground mass parameters show general agreement with \citet{Bolton2008}, \citet{Etherington2022} and \citet{Tan2024}.

\begin{figure*}[t]
    \centering
    \includegraphics[width=\textwidth]{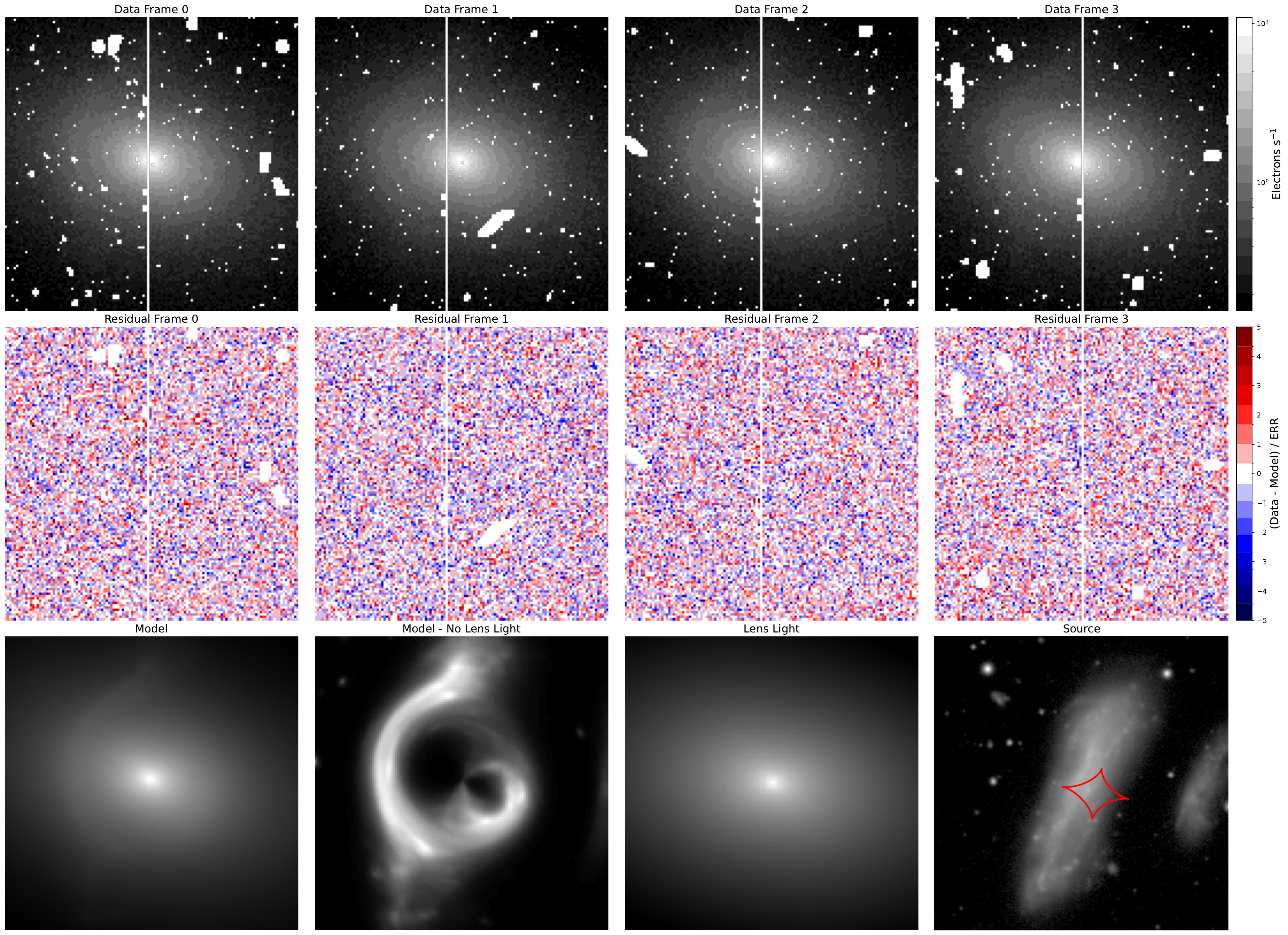}
    \caption{Same as Figure~\ref{fig:SDSSJ1430+4105_bestfit}, but for SDSSJ0912+0029.}
    \label{fig:stage1_SDSSJ0912+0029}
\end{figure*}

\textit{SDSSJ0912+0029} Low-redshift background source galaxy ($z_s = 0.324$) consistent with an edge-on disk morphology. Posterior sampling proved suboptimal, particularly due to poor convergence during the third Gibbs sampling stage (full non-annealed likelihood fine-tuning). In Table \ref{tab:macro_params_other}, we report the foreground mass parameter values used in the lens model featured in Figure \ref{fig:stage1_SDSSJ0912+0029}. 

\textit{SDSSJ0936+0913} Background source galaxy ($z_s = 0.588$) reconstructed as a likely pair of galaxies, indicative of a possible merger; the smaller component is strongly lensed into arcs. We report generally large uncertainties in the mass slope $t$ and external shear $\gamma_{\mathrm{ext}}$ relative to the works of \citet{Etherington2022} and \citet{Tan2024}.

\textit{SDSSJ0946+1006} Background source galaxy ($z_s = 0.609$) with a few compact, point-like features of size on the order $\sim$500 pc regularly seen within the lensing caustic in posterior samples. A fainter, separate lensing arc consistent with a second source at a different redshift was masked from the data. Residuals near the lensing arcs indicate model mismatch, potentially due to dark-matter substructure \citep{Vegetti2010, Quinn2021, Ballard2024, Enzi2025}. This system features one of the highest reported values for the third-order multipole amplitude $a_3$.

\begin{figure*}[t]
    \centering
    \includegraphics[width=\textwidth]{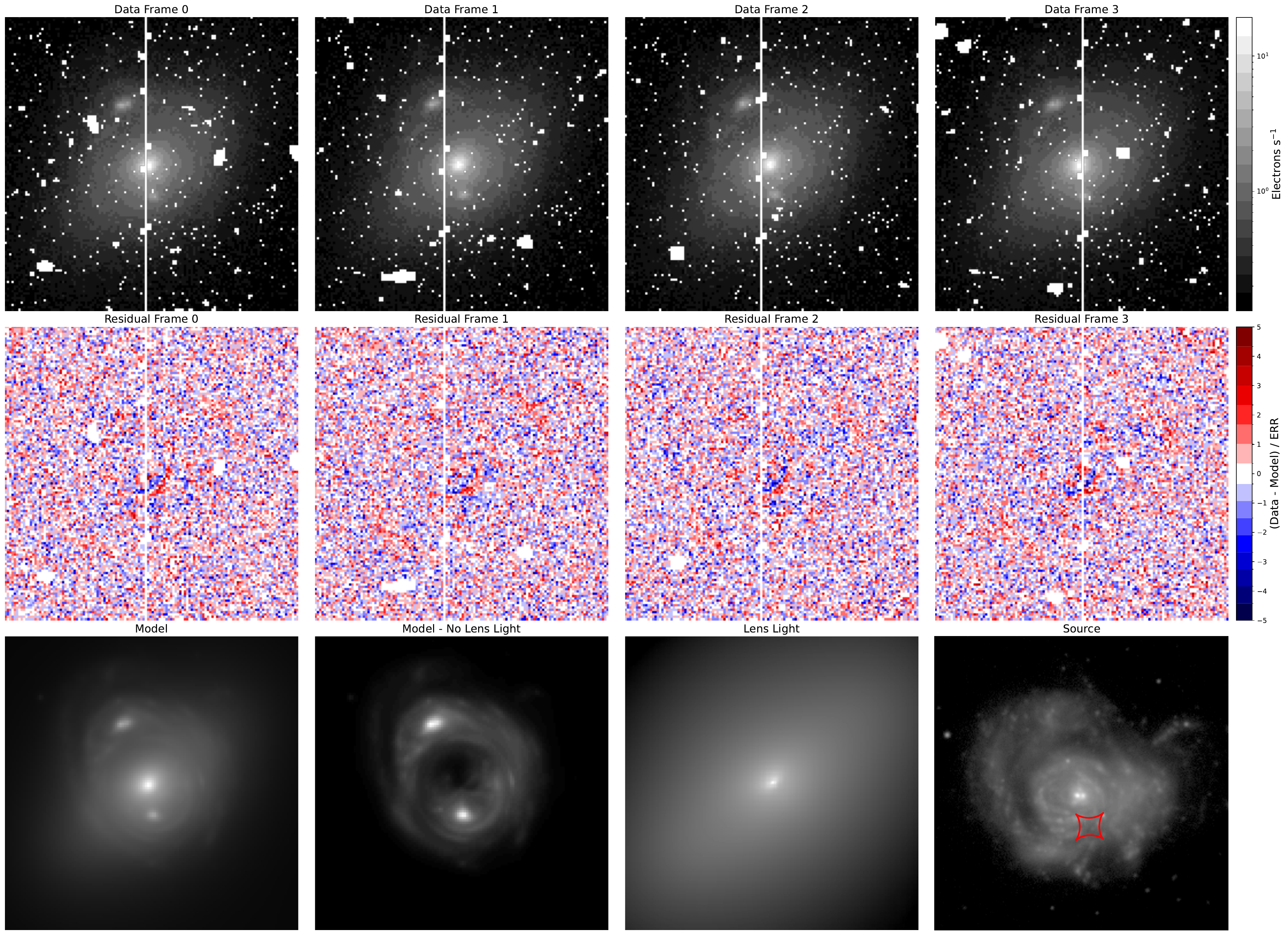}
    \caption{Same as Figure~\ref{fig:SDSSJ1430+4105_bestfit}, but for SDSSJ0959+0410.}
    \label{fig:stage1_SDSSJ0959+0410}
\end{figure*}

\textit{SDSSJ0959+0410} Background source galaxy ($z_s = 0.535$) with spiral arms and a bright core hosting two compact sources, as seen in the high-resolution ($512\times512$) reconstruction in Figure~\ref{fig:stage1_SDSSJ0959+0410}. However, the significance of these features cannot be confirmed due to the absence of proper posterior samples. Posterior sampling was unstable owing to significant residuals between our predicted foreground lens light model and the observed data.

\textit{SDSSJ1020+1122} Background source galaxy ($z_s = 0.553$) with a bright center surrounded by several compact bright point sources consistently seen in the majority of posterior samples. We report large error bars for the foreground external shear $\gamma_{\mathrm{ext}}$ relative to results from \citet{Etherington2022}.

\begin{figure*}[t]
    \centering
    \includegraphics[width=\textwidth]{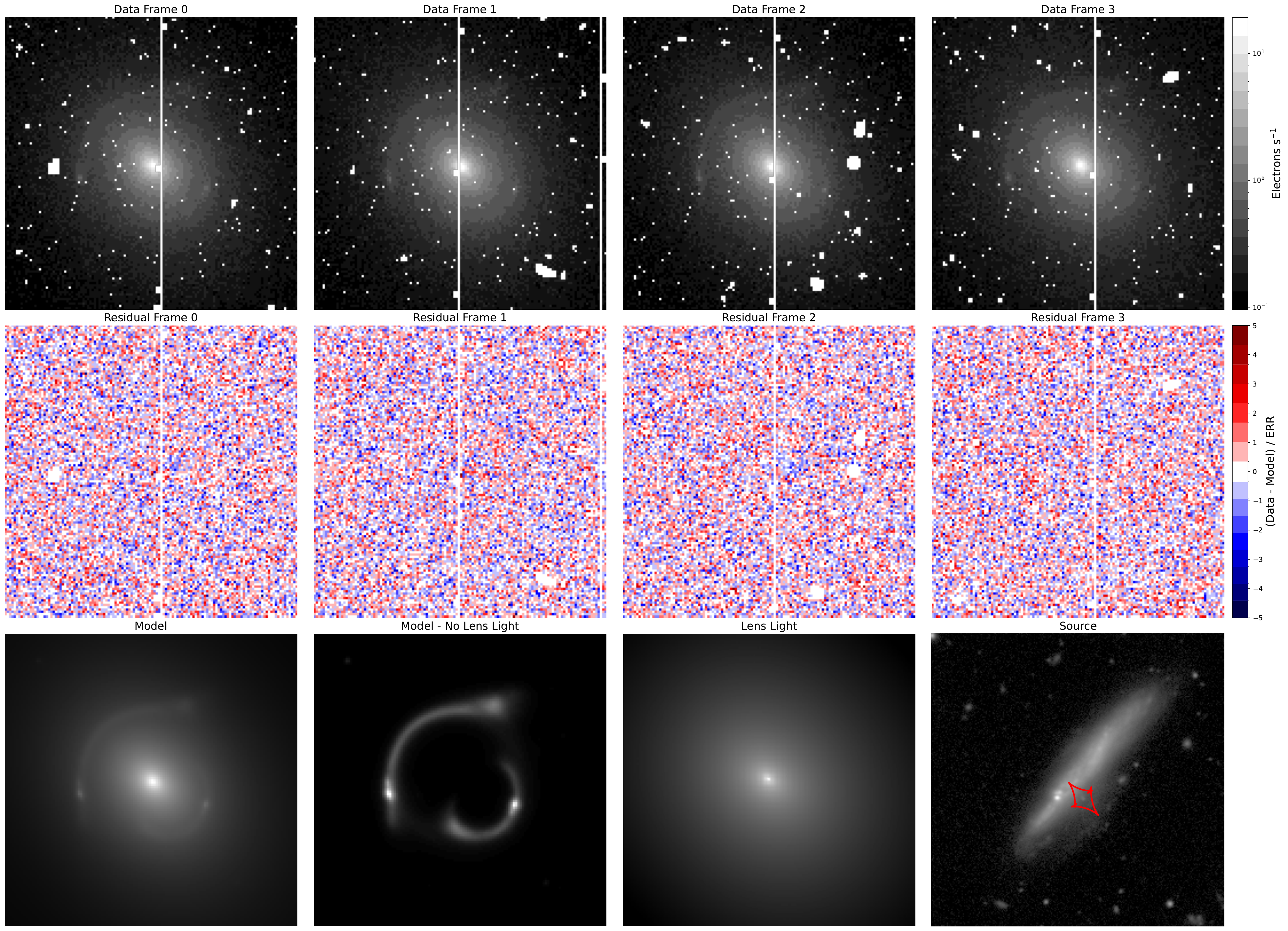}
    \caption{Same as Figure~\ref{fig:SDSSJ1430+4105_bestfit}, but for SDSSJ1023+4230.}
    \label{fig:stage1_SDSSJ1023+4230}
\end{figure*}

\textit{SDSSJ1023+4230} Background source galaxy ($z_s = 0.696$) possibly viewed edge-on, with two closely separated compact sources visible in the high-resolution ($512\times512$) posterior reconstruction in Figure~\ref{fig:stage1_SDSSJ1023+4230}. The lower-resolution ($256\times256$) posterior samples appear to have difficulty resolving this feature. Table~\ref{tab:macro_params} indicates a notably large fourth-order multipole amplitude $a_4$ from the posterior samples.

\begin{figure*}[t]
    \centering
    \includegraphics[width=\textwidth]{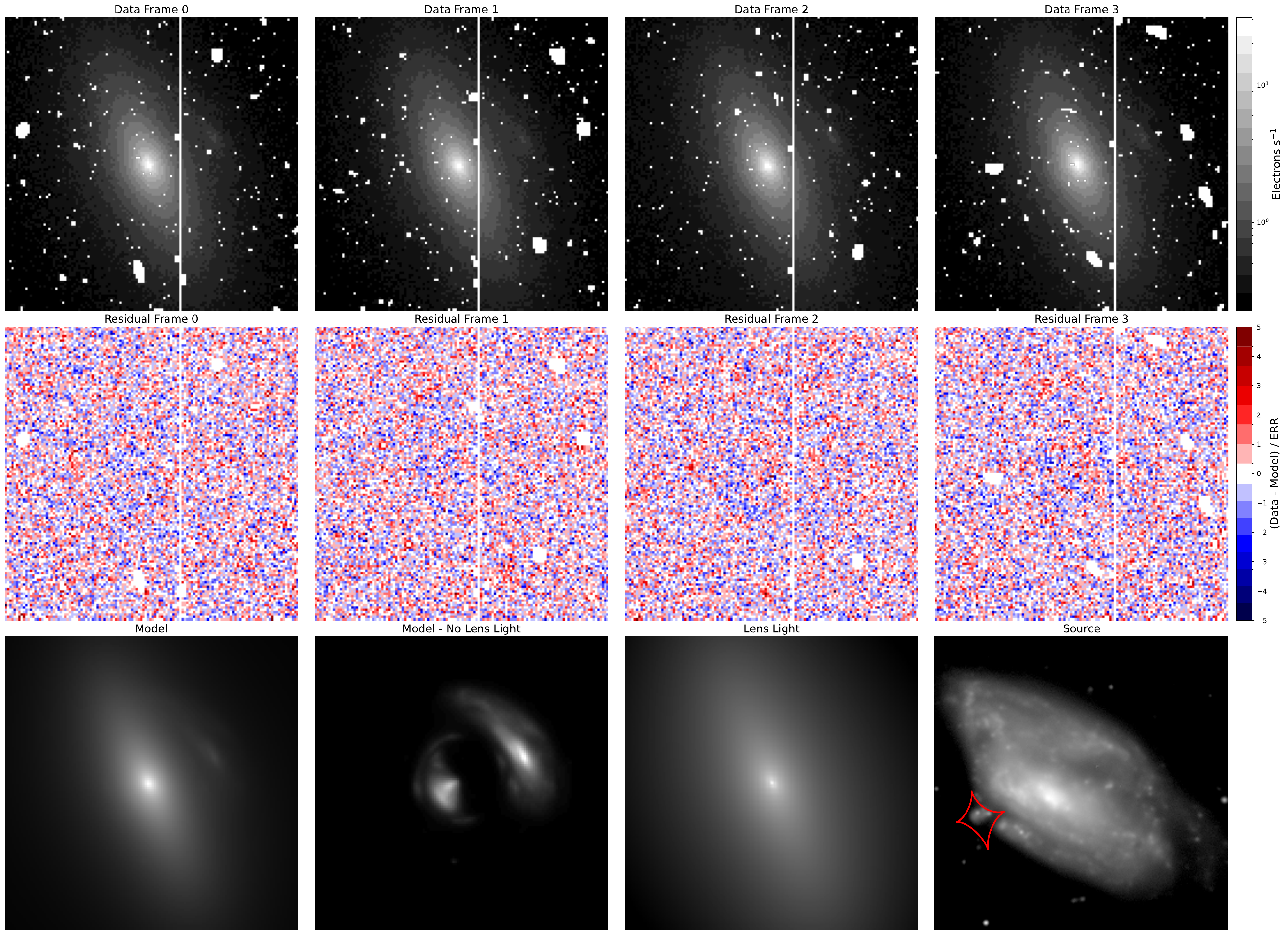}
    \caption{Same as Figure~\ref{fig:SDSSJ1430+4105_bestfit}, but for SDSSJ1029+0420.}
    \label{fig:stage1_SDSSJ1029+0420}
\end{figure*}

\textit{SDSSJ1029+0420} Background source galaxy ($z_s = 0.615$) with possibility of spiral arms seen in Figure~\ref{fig:stage1_SDSSJ1029+0420}. However, prominent artifacts in the high-resolution ($512\times512$) source reconstruction in Figure~\ref{fig:stage1_SDSSJ1029+0420}, resembling lens caustics, suggest sub-optimal foreground mass modeling. Posterior sampling was sub-optimal due to improperly converged chains during the third Gibbs sampling phase (full non-annealed likelihood fine-tuning). As such, we omit reports of parameter uncertainties and only report the foreground mass parameter values used in the lens model seen in Figure \ref{fig:stage1_SDSSJ1029+0420} in Table \ref{tab:macro_params_other}.

\begin{figure*}[t]
    \centering
    \includegraphics[width=\textwidth]{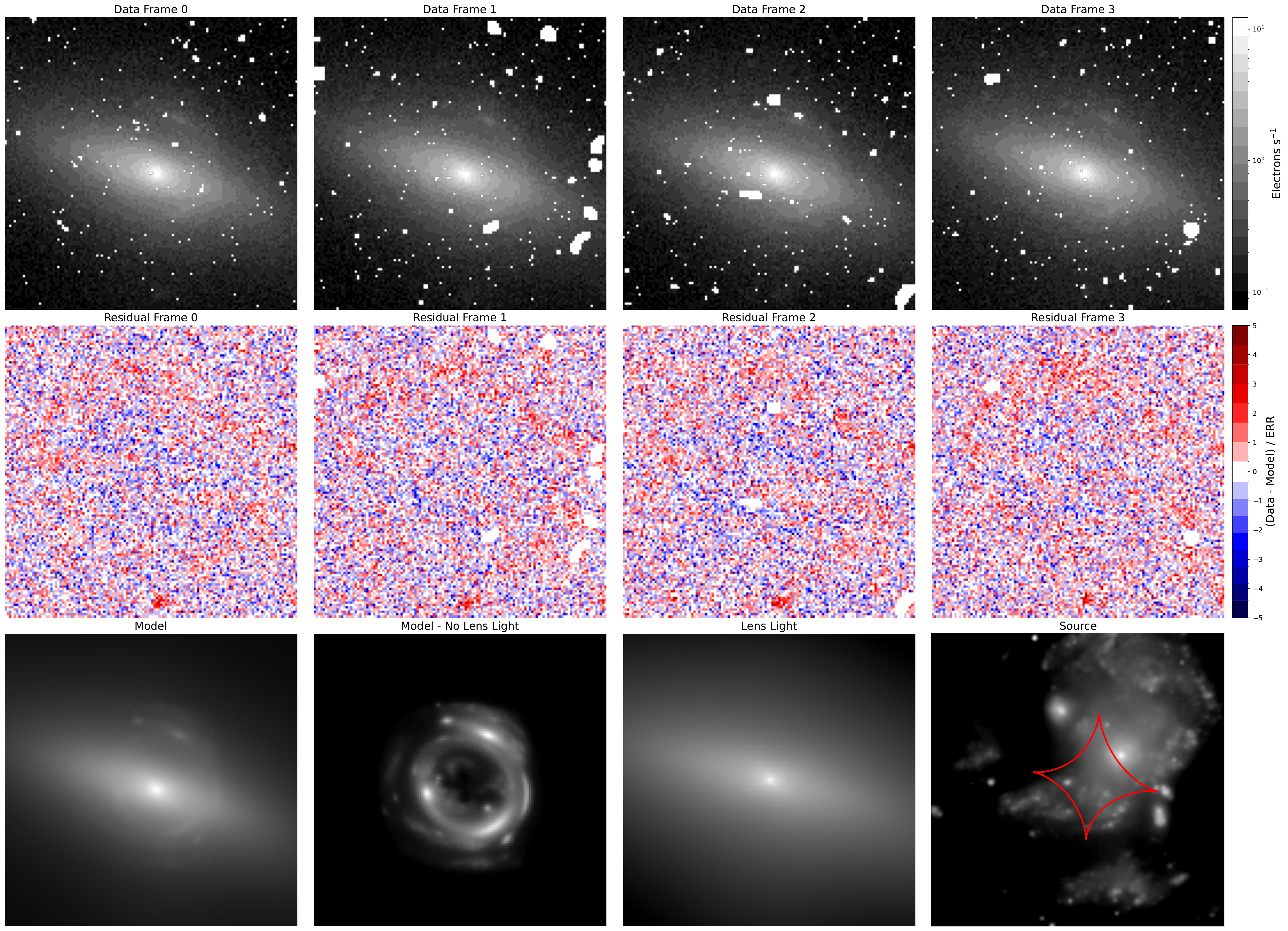}
    \caption{Same as Figure~\ref{fig:SDSSJ1430+4105_bestfit}, but for SDSSJ1103+5322.}
    \label{fig:stage1_SDSSJ1103+5322}
\end{figure*}

\textit{SDSSJ1103+5322} Background source model ($z_s = 0.735$) featuring many compact sources, though some bright neighbors may be foreground contaminants absorbed into the source model. Significant residuals in Figure~\ref{fig:stage1_SDSSJ1103+5322} and degeneracies between the lens light and source models led to unstable posterior sampling. We report the foreground mass parameter values used in the lens model seen in Figure \ref{fig:stage1_SDSSJ1103+5322} in Table \ref{tab:macro_params_other}.

\textit{SDSSJ1142+1001} Background source galaxy ($z_s = 0.504$) featuring a smooth brightness profile lying outside of the lensing caustic resulting in a doubly lensed image. Foreground mass parameters show general agreement with \citet{Bolton2008} and \citet{Etherington2022} and features large error bars for $\gamma_{\mathrm{ext}}$.

\begin{figure*}[t]
    \centering
    \includegraphics[width=\textwidth]{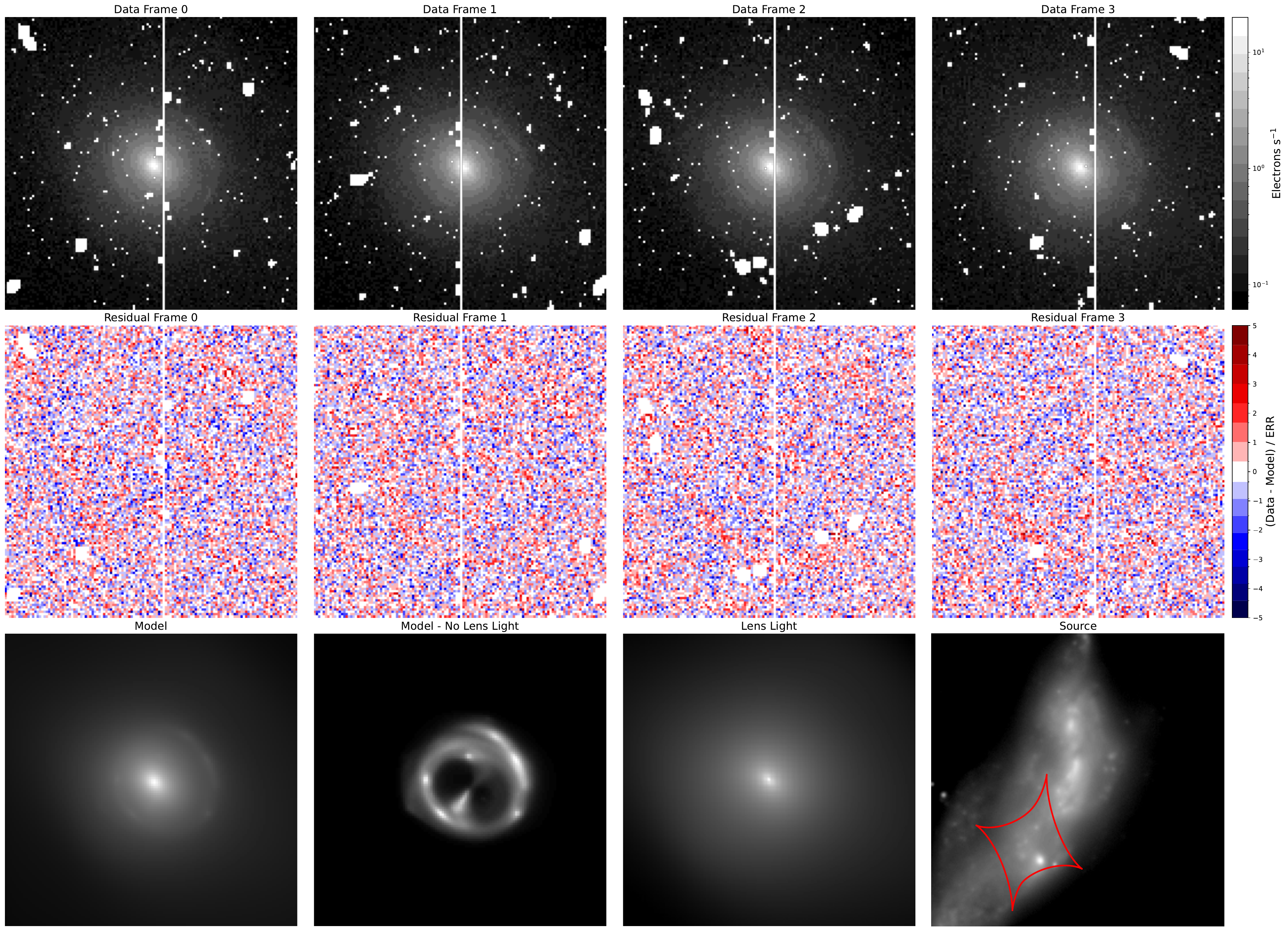}
    \caption{Same as Figure~\ref{fig:SDSSJ1430+4105_bestfit}, but for SDSSJ1153+4612.}
    \label{fig:stage1_SDSSJ1153+4612}
\end{figure*}

\textit{SDSSJ1153+4612} High-redshift background source galaxy ($z_s = 0.875$) with a compact, point-like source that is quadruply lensed. Residuals in Figure~\ref{fig:stage1_SDSSJ1153+4612} suggest a sub-optimal foreground lens light model. Our estimates of the foreground mass parameters are largely inconsistent with those of \citet{Bolton2008} and \citet{Etherington2022}.

\begin{figure*}[t]
    \centering
    \includegraphics[width=\textwidth]{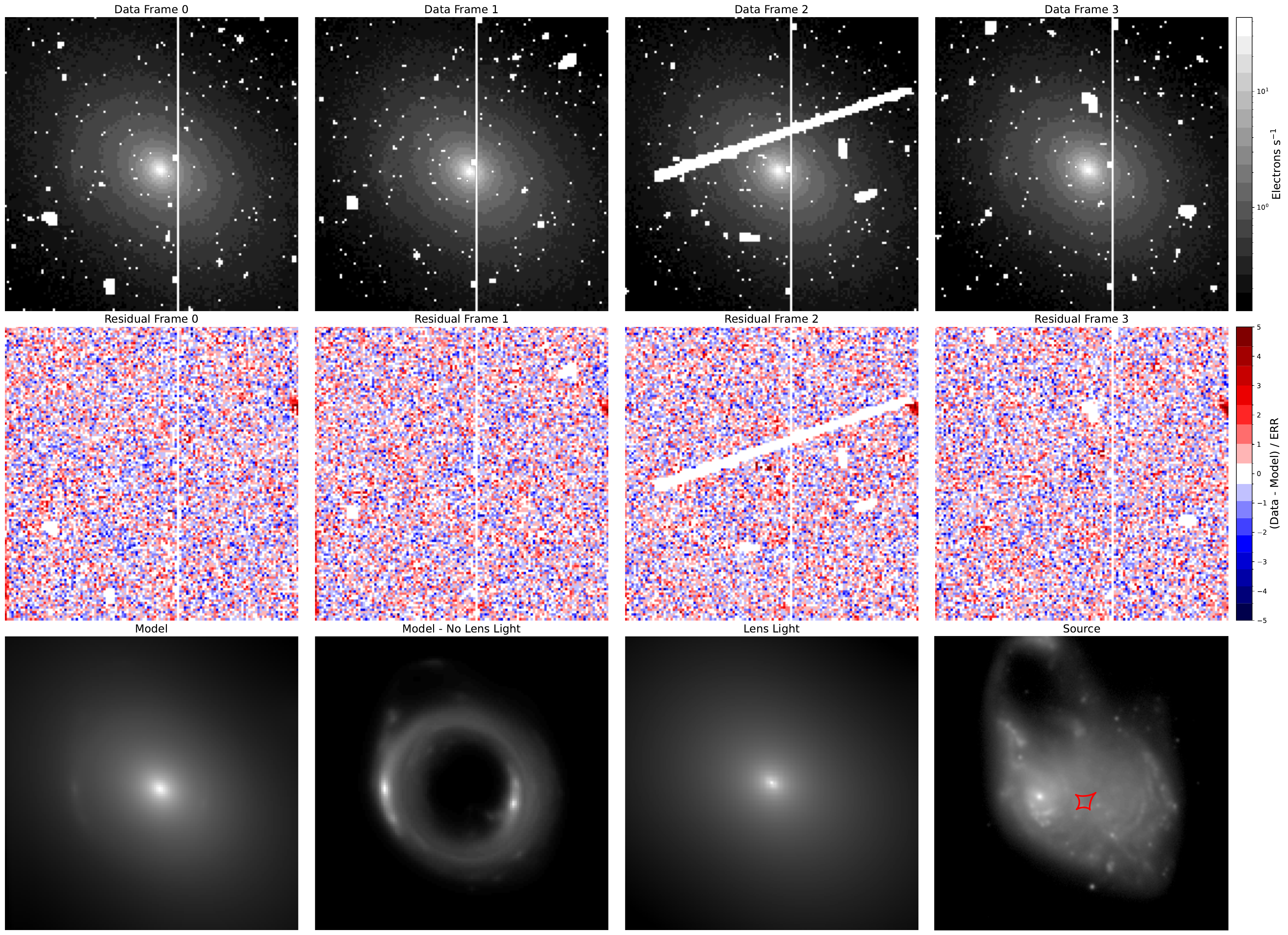}
    \caption{Same as Figure~\ref{fig:SDSSJ1430+4105_bestfit}, but for SDSSJ1213+6708.}
    \label{fig:stage1_SDSSJ1213+6708}
\end{figure*}

\textit{SDSSJ1213+6708} Lens system with a low signal-to-noise ratio for the lensed source galaxy ($z_s = 0.640$). Posterior sampling was unstable, as the annealed Gibbs sampler escaped from the initial high-posterior-probability region into unphysical lens model solutions. As such, we only report the foreground mass parameter values of the well-converged fit from Figure \ref{fig:stage1_SDSSJ1213+6708} in Table \ref{tab:macro_params_other}.

\begin{figure*}[t]
    \centering
    \includegraphics[width=\textwidth]{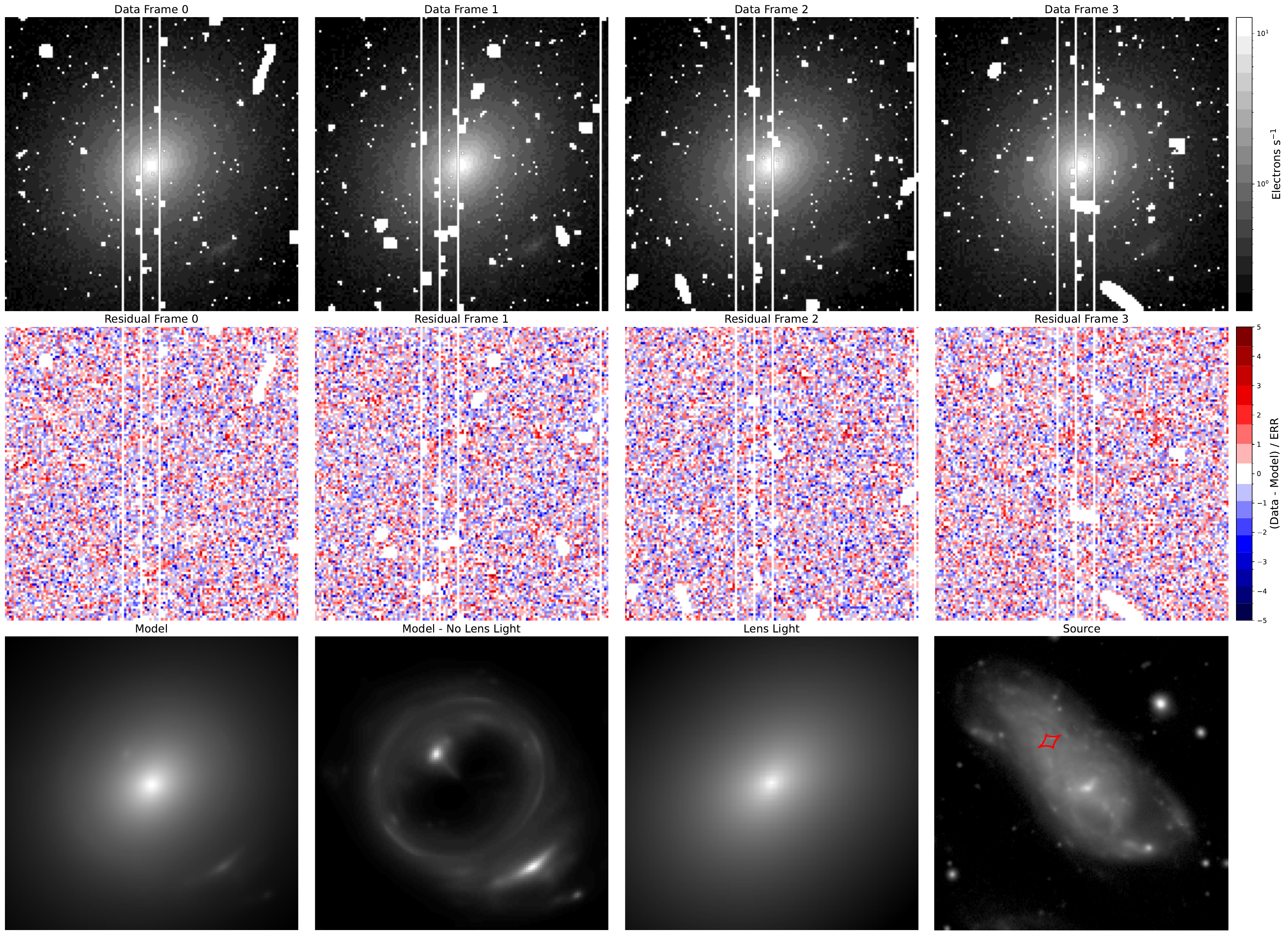}
    \caption{Same as Figure~\ref{fig:SDSSJ1430+4105_bestfit}, but for SDSSJ1218+0830.}
    \label{fig:stage1_SDSSJ1218+0830}
\end{figure*}

\textit{SDSSJ1218+0830} Background source galaxy ($z_s = 0.717$) with a central core and an additional source to the lower right of the source model (Figure~\ref{fig:stage1_SDSSJ1218+0830}). Posterior sampling was suboptimal due to improper convergence during the third stage of Gibbs sampling (full non-annealed likelihood fine-tuning). In Table \ref{tab:macro_params_other}, we report the foreground mass parameter values of the model fit featured in Figure \ref{fig:stage1_SDSSJ1213+6708}.

\textit{SDSSJ1250+0523} High-redshift background source galaxy ($z_s = 0.795$) and the largest reconstructed source in our sample, with visible spiral structure. The inner lensing caustic occupies only a small central region of the source. Foreground mass parameter values are generally in agreement with \citet{Bolton2008}, \citet{Etherington2022}, and \citet{Tan2024}, with the exception of a significant discrepancy in the mass slope $t$ compared to \citet{Etherington2022}.

\textit{SDSSJ1402+6321} Background source galaxy ($z_s = 0.481$) with three compact point-like sources of size on the order $\sim$500 pc aligned roughly vertically, regularly appearing in posterior samples; two lie within the caustic (quadruply imaged) and one outside (doubly imaged). Slight residuals near the center indicate a mismatch between the observed central peak brightness and our foreground lens light model. Foreground mass parameters agree well with \citet{Etherington2022}, while more significant discrepancies occur in comparison with \citet{Tan2024}.

\textit{SDSSJ1416+5136} High-redshift background source galaxy ($z_s = 0.811$) with irregular morphology, showing several bright clumpy features of size on the order $\sim$1000 pc consistent across posterior samples, possibly indicative of star-forming regions. Residuals near the lensing system center persist. In addition to the main deflector, a secondary foreground galaxy is present (masked out from the data) and may indicate an ongoing merger; assuming a single-deflector model may bias the inferred lens parameters. Foreground mass parameters are generally consistent with \citet{Bolton2008}, though a noticeable discrepancy is observed in the external shear magnitude $\gamma_{\mathrm{ext}}$ compared to \citet{Etherington2022}.

\begin{figure*}[t]
    \centering
    \includegraphics[width=\textwidth]{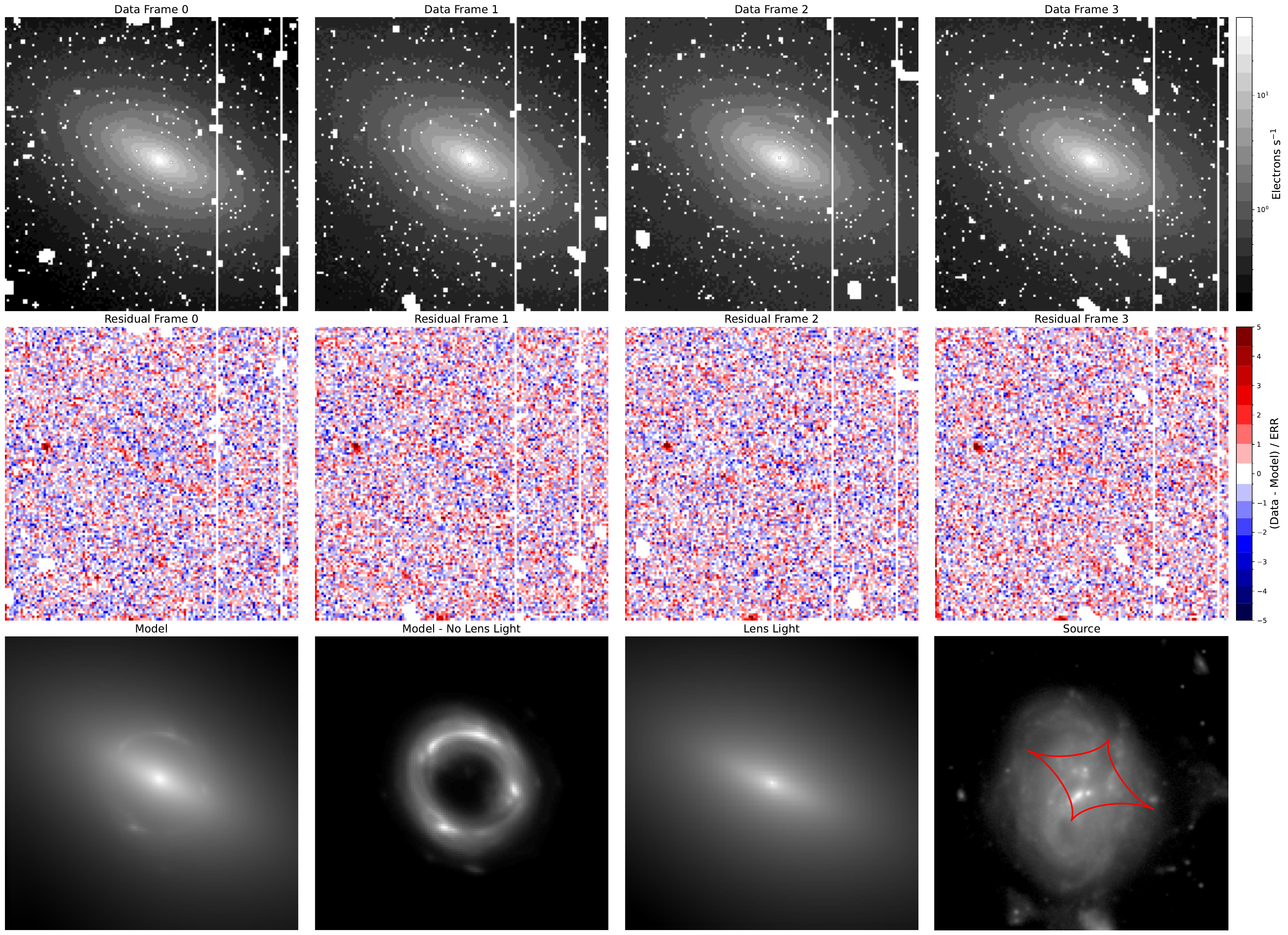}
    \caption{Same as Figure~\ref{fig:SDSSJ1430+4105_bestfit}, but for SDSSJ1420+6019.}
    \label{fig:stage1_SDSSJ1420+6019}
\end{figure*}

\textit{SDSSJ1420+6019} Background source galaxy ($z_s = 0.535$) with spiral-like morphology and multiple compact sources near the center, inside the inner caustic, as seen in the high-resolution ($512\times512$) reconstruction in Figure~\ref{fig:stage1_SDSSJ1420+6019}. Posterior sampling was suboptimal due to significant residuals arising from improper lens light modeling in the final retrieved posterior samples. Foreground mass parameter values used in the fit showcased in Figure \ref{fig:stage1_SDSSJ1420+6019} are presented in Table \ref{tab:macro_params_other} 

\textit{SDSSJ1430+4105} Background source galaxy ($z_s = 0.575$) with clearly resolved spiral arms and a bar-like central feature in the reconstructed galaxy, visible in both the high-resolution ($512\times512$) and posterior samples. The inferred foreground mass parameters exhibit among the smallest uncertainties in our sample of lenses, particularly for the Einstein radius $R_{\rm E}$ and the mass slope $t$. Foreground lens parameters are generally in good agreement with \citet{Bolton2008} and \citet{Etherington2022}, however, noticeable discrepancies in the external shear $\gamma_{\mathrm{ext}}$ is observed. Posterior samples reveal compact features as small as 1000 pc.

\textit{SDSSJ1627-0053} Background source galaxy ($z_s = 0.524$) producing a complete Einstein ring in the observed data. This system has one of the largest uncertainties in the foreground mass slope $t$ from our sample of lenses. Foreground lens parameters are overall in good agreement with \citet{Bolton2008}, \citet{Etherington2022} and \citet{Tan2024}.

\textit{SDSSJ1630+4520} High-redshift background source galaxy ($z_s = 0.793$) with multiple, seemingly separate source features. There is generally good agreement for the Einstein radius $R_{\rm E}$ compared to \citet{Bolton2008}, \citet{Etherington2022}, and \citet{Tan2024}, but more significant discrepancies are present for $t$ and $\gamma_{\mathrm{ext}}$.

\textit{SDSSJ2300+0022} Background source galaxy ($z_s = 0.463$) with a bright central core located just outside the inner caustic. Our results show a significant discrepancy in the Einstein radius $R_{\rm E}$ compared to \citet{Bolton2008}, \citet{Etherington2022}, and \citet{Tan2024}.

\textit{SDSSJ2303+1422} Background source galaxy ($z_s = 0.517$) consisting of a main structure plus two compact point-like sources near the lens caustic (one inside, one outside) of size on the order $\sim$500 pc. Foreground mass parameters appear significantly discrepant with the results of \citet{Etherington2022}, while they exhibit better consistency with \citet{Bolton2008} and \citet{Tan2024}.

\begin{figure*}[t]
    \centering
    \includegraphics[width=\textwidth]{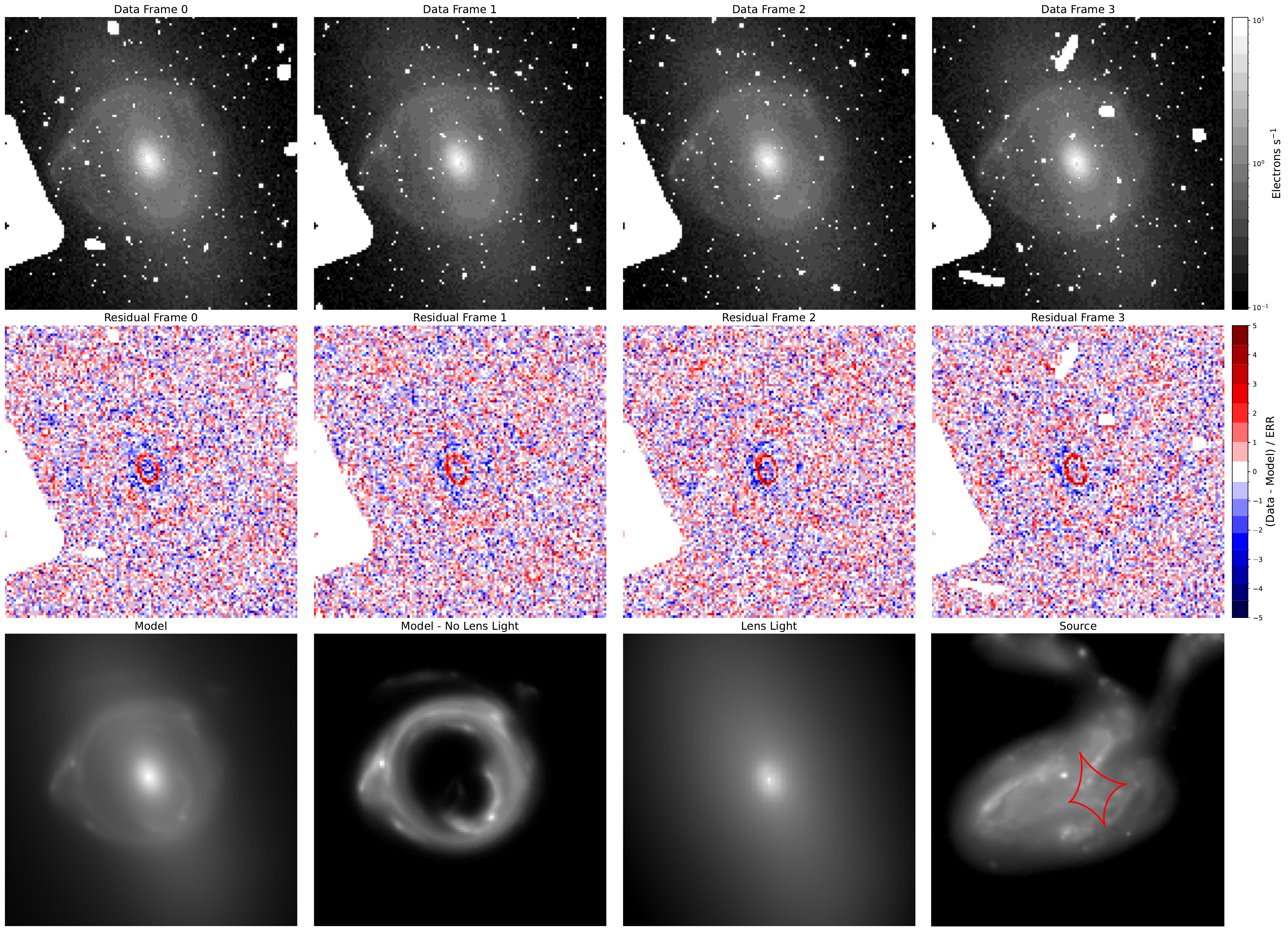}
    \caption{Same as Figure~\ref{fig:SDSSJ1430+4105_bestfit}, but for SDSSJ2341+0000.}
    \label{fig:stage1_SDSSJ2341+0000}
\end{figure*}

\textit{SDSSJ2341+0000} High-redshift background source galaxy ($z_s = 0.807$) with a compact, point-like source outside the caustic that appears doubly lensed in the image data. The source reconstruction shows evidence of foreground lens light contamination and significant residuals in Figure~\ref{fig:stage1_SDSSJ2341+0000} and in posterior samples. Because of this, we omit reporting uncertainties in Table~\ref{tab:macro_params} and only report foreground mass parameter values in Table \ref{tab:macro_params_other} from the fit showcased in Figure \ref{fig:stage1_SDSSJ2341+0000}.

\section{Additional Results} \label{sec:appendix_results}
\clearpage  

\begin{table*}
\centering
\begin{tabular*}{\textwidth}{l@{\extracolsep{\fill}}cccccccc}
\hline
System Name & $q$ & $\phi_{\mathrm{EPL}}$ & $R_{\mathrm{E}}$ & $t$ & $\gamma_{\mathrm{ext}}$ & $\phi_{\mathrm{ext}}$ & $a_3$ & $a_4$ \\
\hline
SDSSJ0912$+$0029 & $0.619$ & $1.413$ & $1.623$ & $0.898$ & $0.029$ & $1.545$ & $0.022$ & $0.004$ \\
SDSSJ0959$+$0410 & $0.658$ & $0.212$ & $0.982$ & $0.907$ & $0.070$ & $6.097$ & $0.001$ & $0.015$ \\
SDSSJ1029$+$0420 & $0.701$ & $6.095$ & $0.947$ & $0.548$ & $0.113$ & $0.005$ & $0.010$ & $0.008$ \\
SDSSJ1103$+$5322 & $0.562$ & $0.614$ & $1.074$ & $0.912$ & $0.070$ & $0.560$ & $0.016$ & $0.045$ \\
SDSSJ1213$+$6708 & $0.886$ & $1.468$ & $1.426$ & $1.211$ & $0.000$ & $4.885$ & $0.019$ & $0.018$ \\
SDSSJ1218$+$0830 & $0.536$ & $0.874$ & $1.730$ & $0.832$ & $0.191$ & $0.869$ & $0.001$ & $0.003$ \\
SDSSJ1420$+$6019 & $0.437$ & $5.891$ & $1.123$ & $0.920$ & $0.140$ & $5.871$ & $0.002$ & $0.003$ \\
SDSSJ2341$+$0000 & $0.875$ & $6.229$ & $1.382$ & $0.834$ & $0.028$ & $5.298$ & $0.020$ & $0.019$ \\
\hline
\end{tabular*}
\caption{Summary of macro lens parameters used in the high-resolution ($512 \times 512$) source model fit for lens systems for which posterior sampling failed (see Appendix \ref{sec:appendix_results} for data-model residuals).}
\label{tab:macro_params_other}
\end{table*}

\begin{figure*}[!t]
    \centering
    \includegraphics[width=\textwidth]{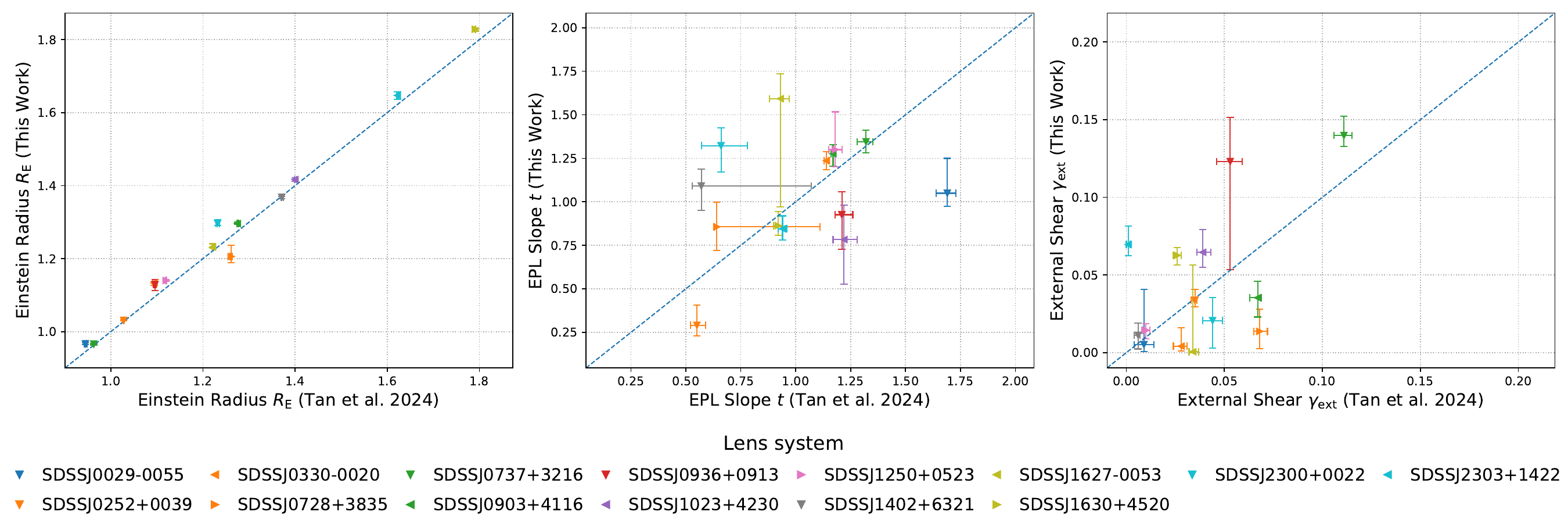}
   \caption{Same as Figure~\ref{fig:etherington_milex}, but comparing our results with those of \citet{Tan2024} taken from \url{https://www.projectdinos.com/dinos-i}. Their foreground mass model adopts a power-law profile with external shear but no multipoles. Therefore, exact one-to-one agreement is not guaranteed.}
    \label{fig:tan_milex}
\end{figure*}

\begin{figure*}[!t]
    \centering
    \includegraphics[width=\textwidth]{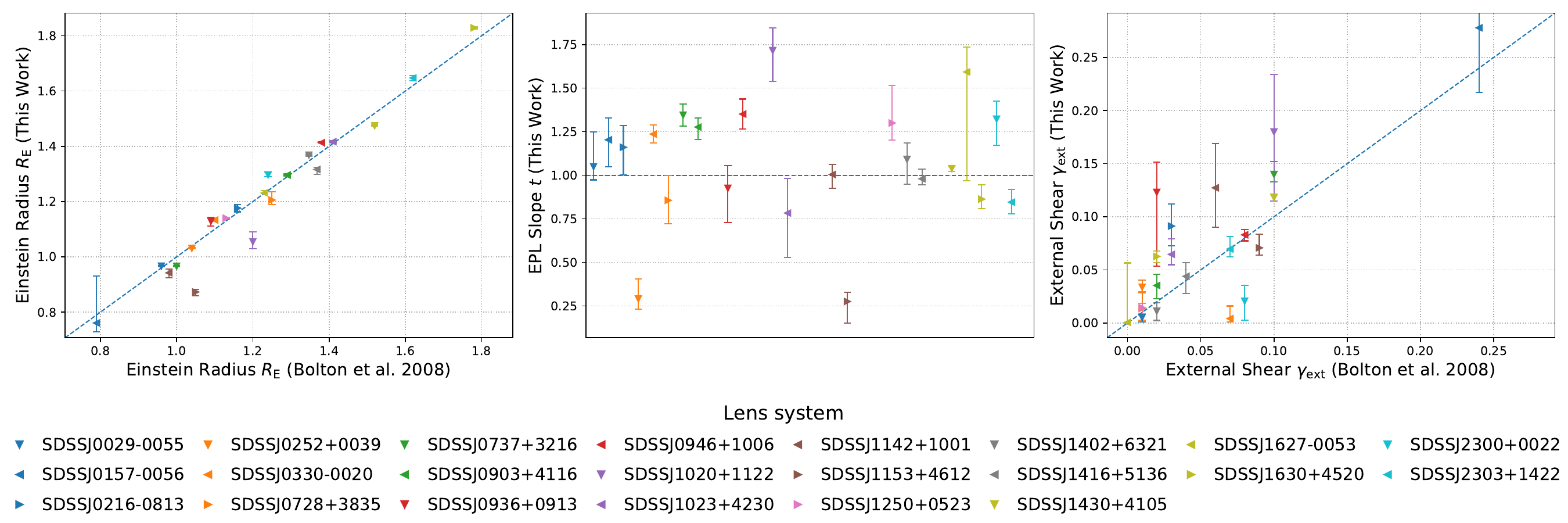}
    \caption{Same as Figure~\ref{fig:etherington_milex}, but comparing this work with \citet{Bolton2008}. Their analysis reports point estimates without uncertainties. Because \citet{Bolton2008} adopted an SIE mass profile, equivalent to an EPL slope \(t=1\), we include a horizontal reference line at $t=1$ in the $t$ panel to highlight departures from the SIE assumption.}
    \label{fig:bolton_milex}
\end{figure*}

\begin{figure*}[!t]
    \centering
    \includegraphics[width=\textwidth]{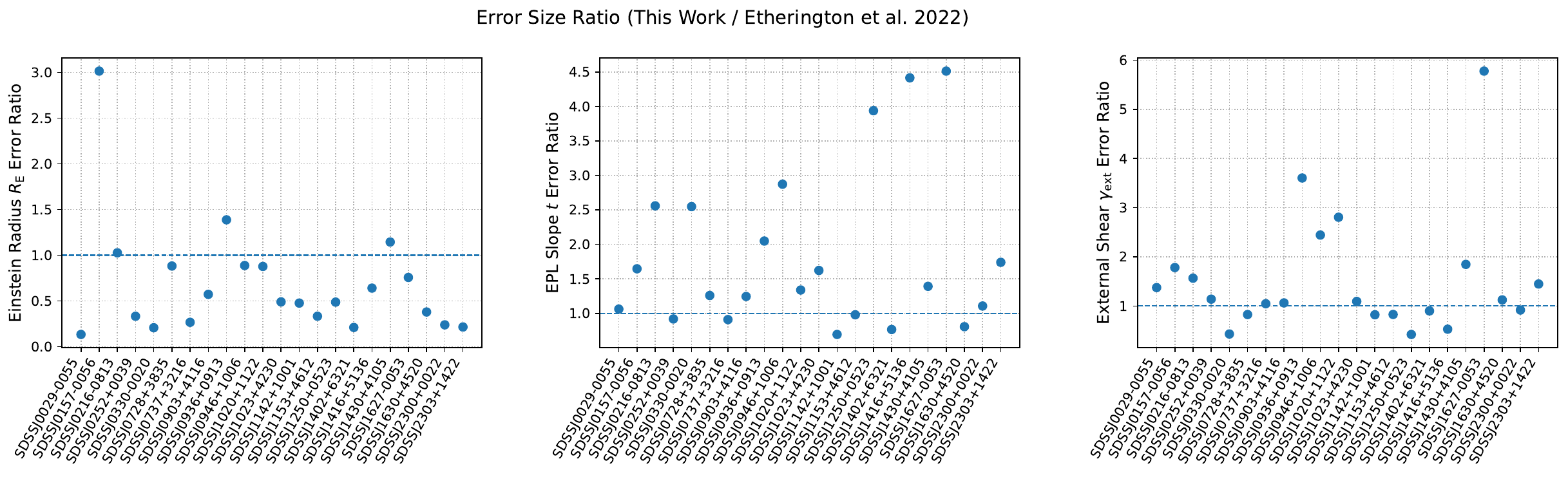}
    \caption{Ratio of \(68\%\) credible-interval widths (This Work)\,/\,\citealt{Etherington2022}) for the inferred foreground-mass parameters—Einstein radius \(R_{\mathrm{E}}\), EPL slope \(t\) and external shear \(\gamma_{\mathrm{ext}}\). Each point corresponds to a lens system along the x-axis; the y-axis shows the ratio of interval widths. The horizontal line at unity denotes equal \(68\%\) widths in the two works, with values above (below) one indicating broader (narrower) uncertainties in this work.}
    \label{fig:milex_etherington_error_width}
\end{figure*}

\begin{figure*}[!t]
    \centering
    \includegraphics[width=\textwidth]{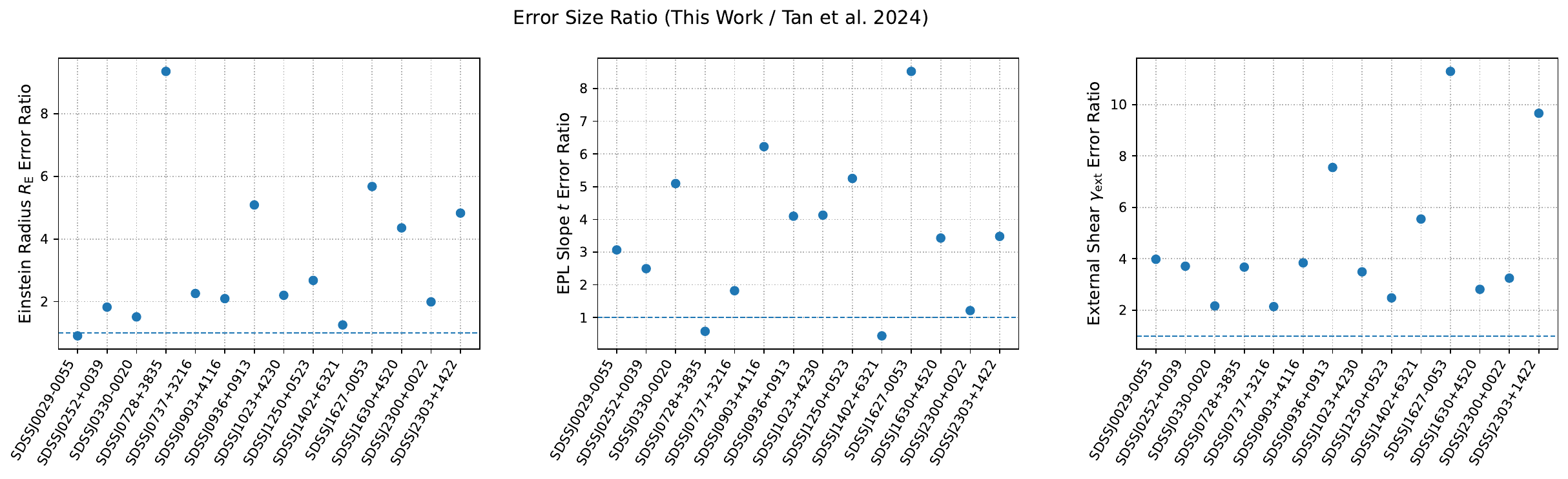}
    \caption{Same as Figure~\ref{fig:milex_etherington_error_width}, but comparing this work with \citet{Tan2024}.}
    \label{fig:milex_tan_error_width}
\end{figure*}




\begin{figure*}[t]
    \centering
    \includegraphics[width=\textwidth]{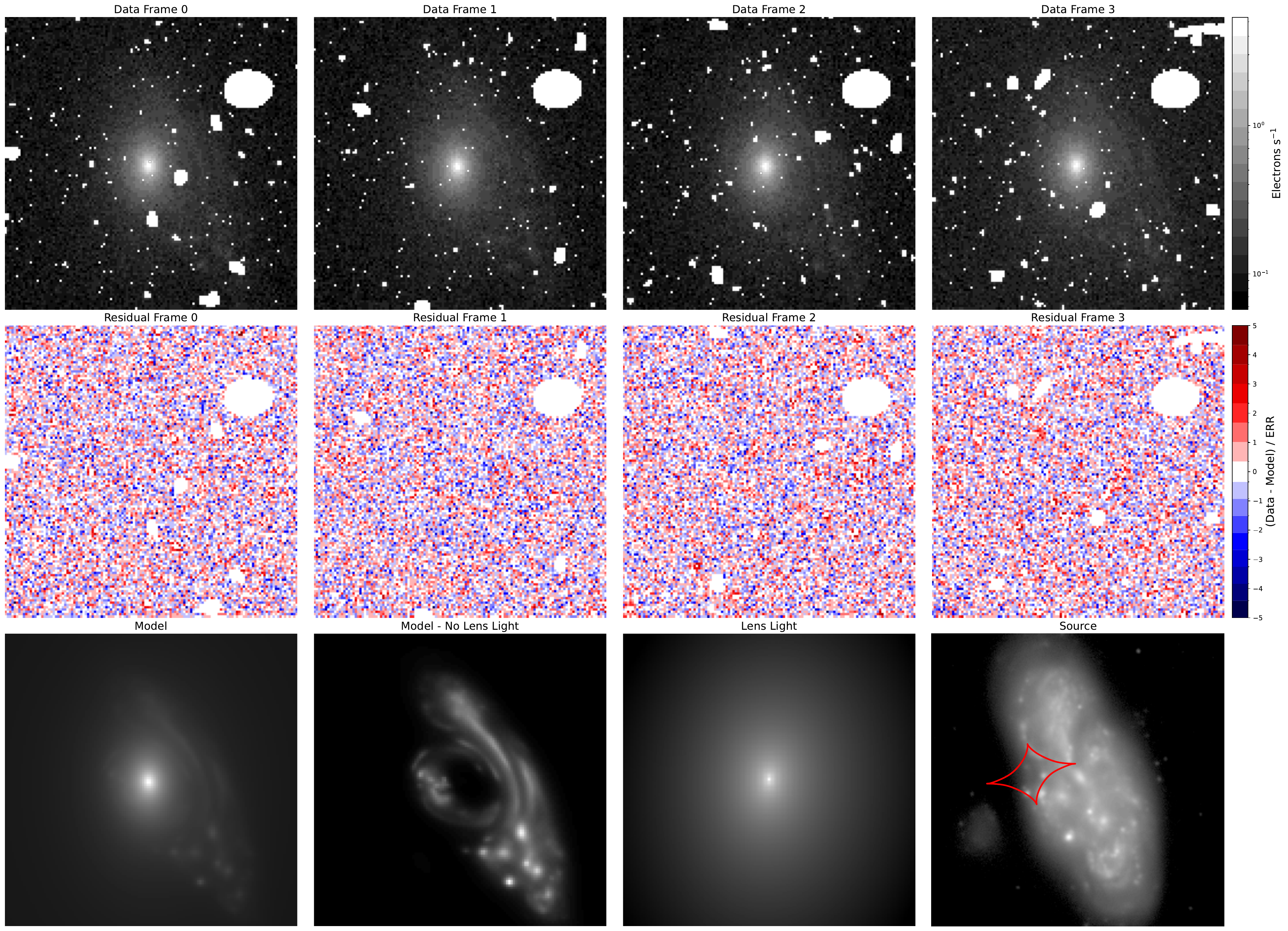}
    \caption{Same as Figure~\ref{fig:SDSSJ1430+4105_bestfit}, but for SDSSJ0157-0056.}
    \label{fig:stage1_SDSSJ0157-0056}
\end{figure*}

\begin{figure*}[t]
    \centering
    \includegraphics[width=\textwidth]{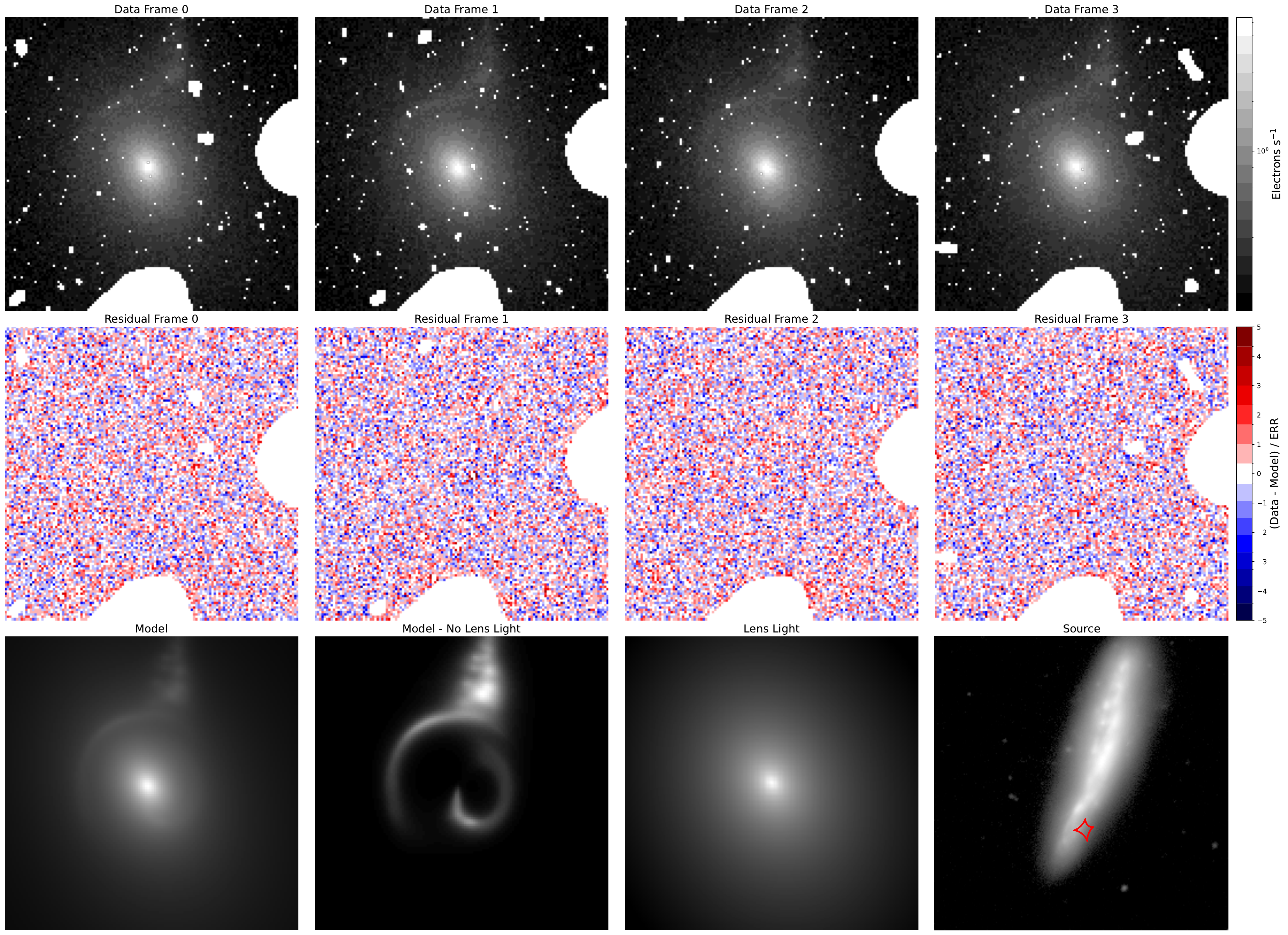}
    \caption{Same as Figure~\ref{fig:SDSSJ1430+4105_bestfit}, but for SDSSJ0216-0813.}
    \label{fig:stage1_SDSSJ0216-0813}
\end{figure*}

\begin{figure*}[t]
    \centering
    \includegraphics[width=\textwidth]{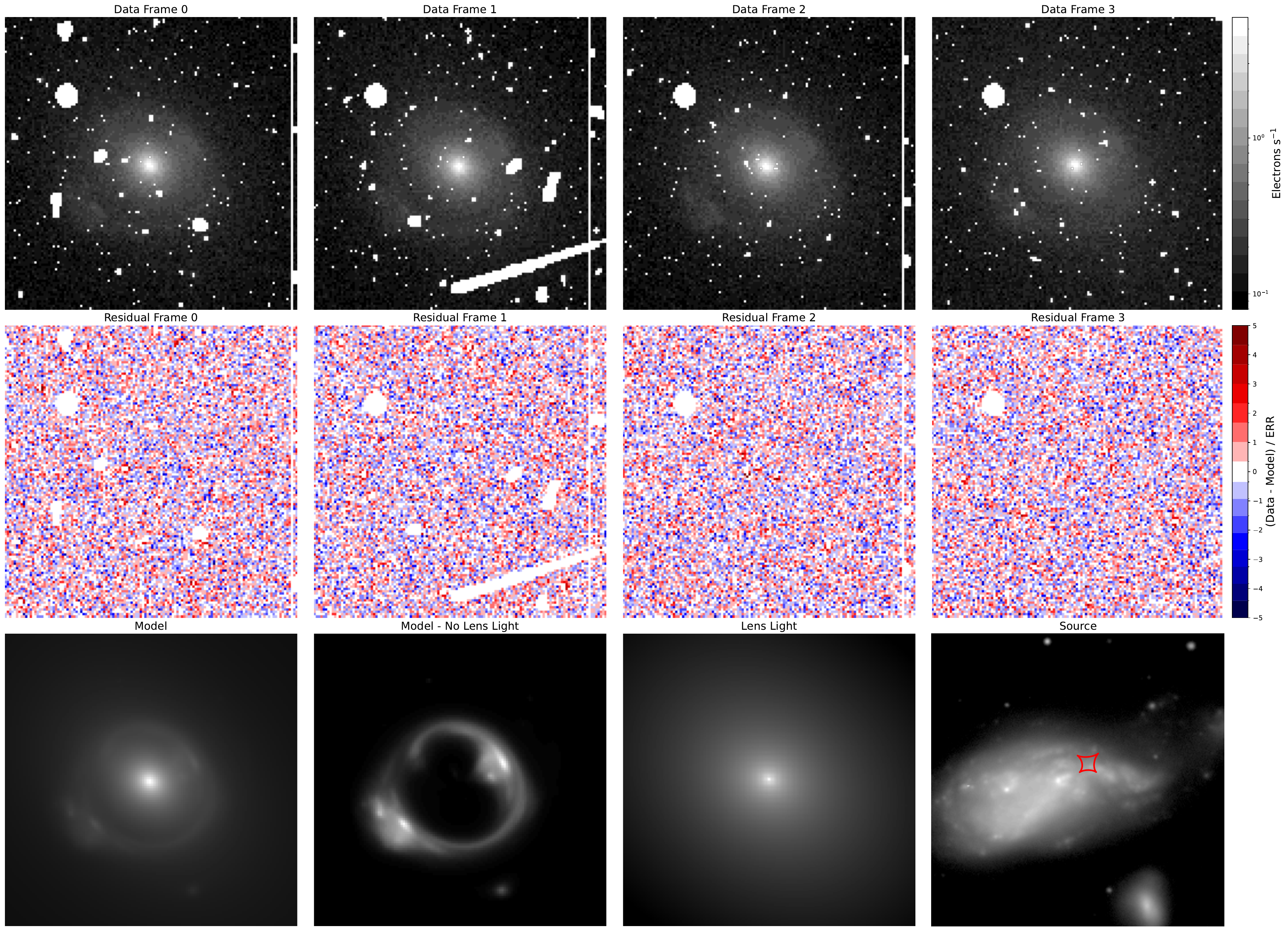}
    \caption{Same as Figure~\ref{fig:SDSSJ1430+4105_bestfit}, but for SDSSJ0903+4116.}
    \label{fig:stage1_SDSSJ0903+4116}
\end{figure*}

\begin{figure*}[t]
    \centering
    \includegraphics[width=\textwidth]{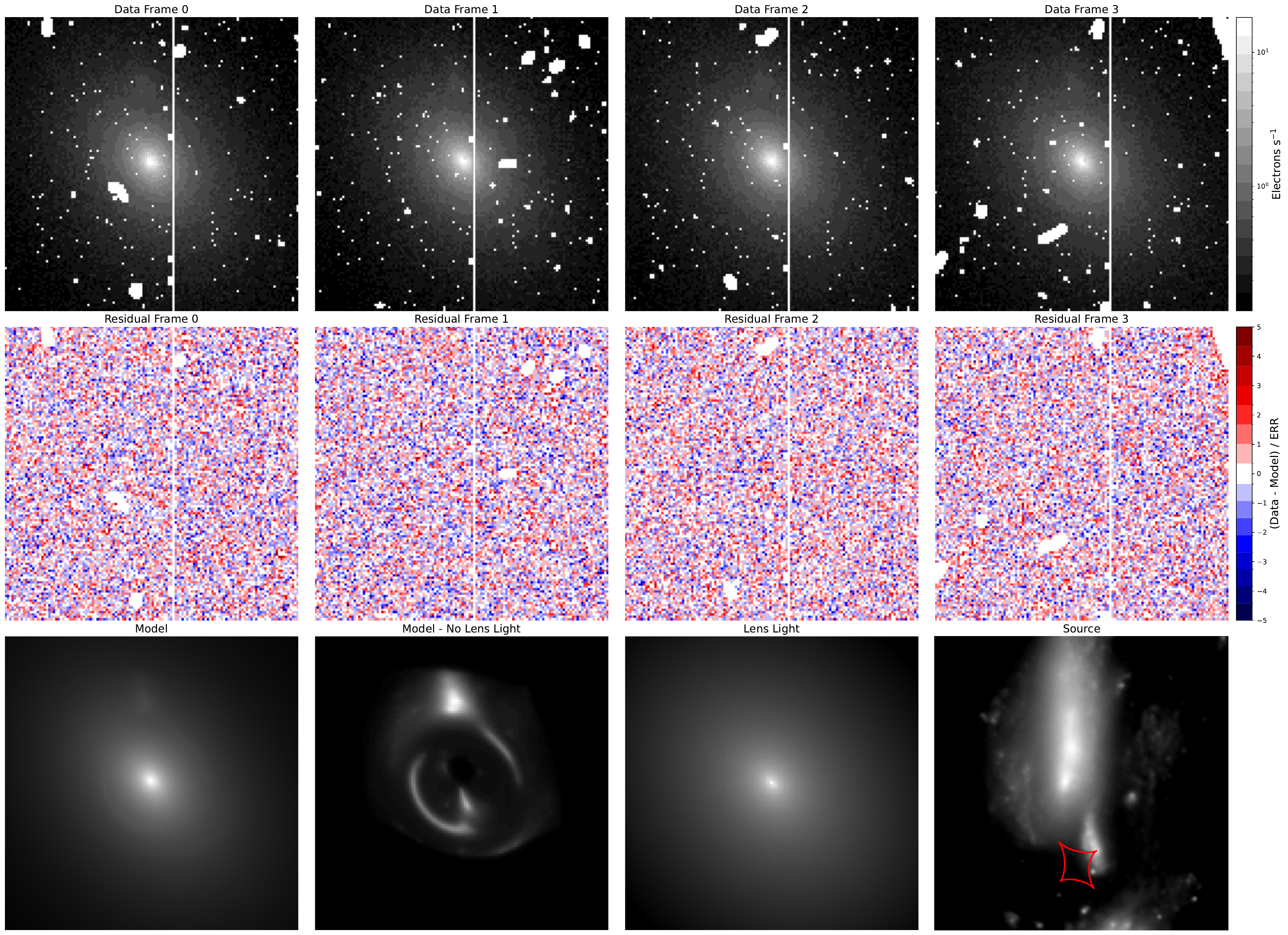}
    \caption{Same as Figure~\ref{fig:SDSSJ1430+4105_bestfit}, but for SDSSJ0936+0913.}
    \label{fig:stage1_SDSSJ0936+0913}
\end{figure*}

\begin{figure*}[t]
    \centering
    \includegraphics[width=\textwidth]{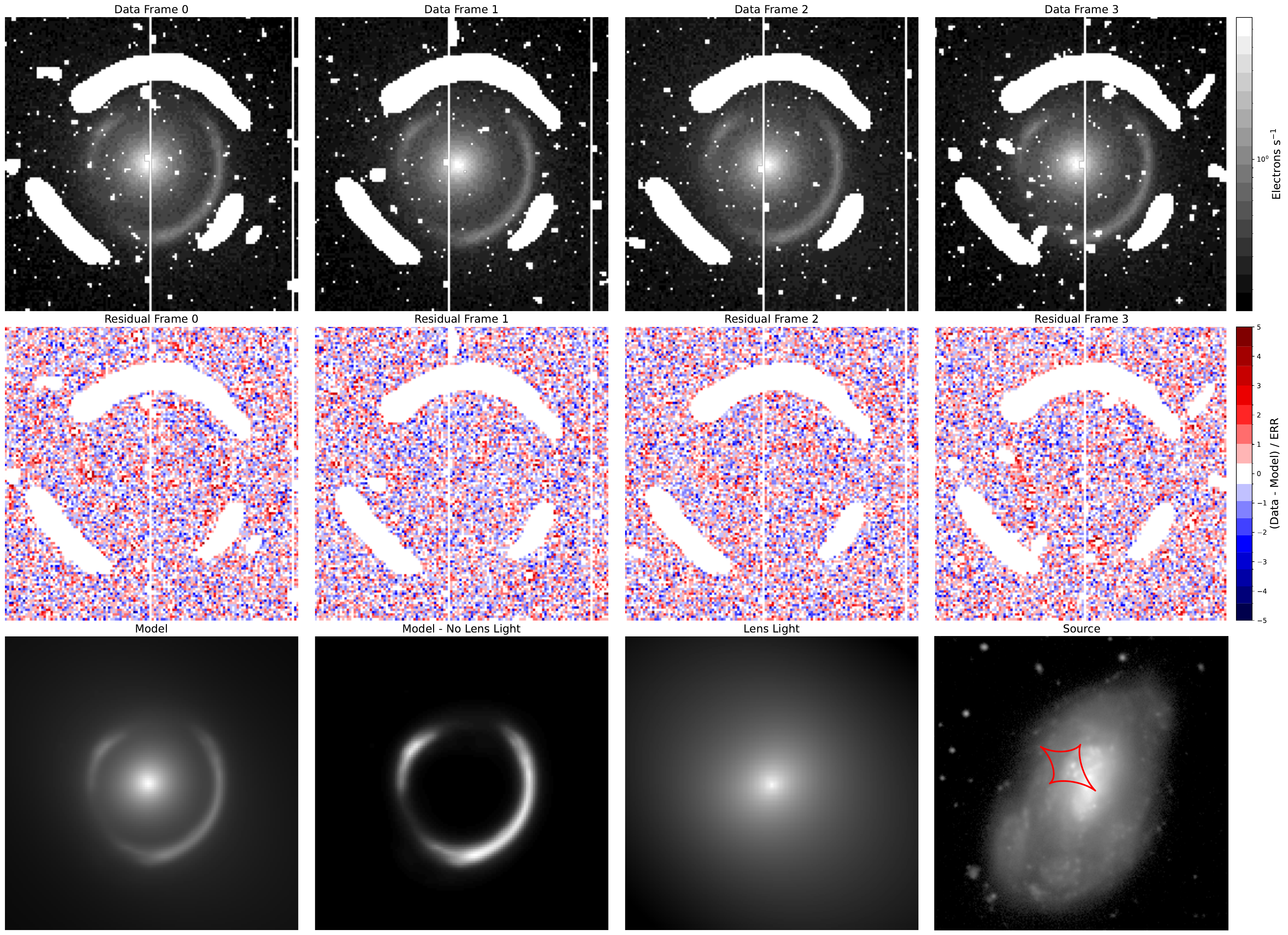}
    \caption{Same as Figure~\ref{fig:SDSSJ1430+4105_bestfit}, but for SDSSJ0946+1006.}
    \label{fig:stage1_SDSSJ0946+1006}
\end{figure*}

\begin{figure*}[t]
    \centering
    \includegraphics[width=\textwidth]{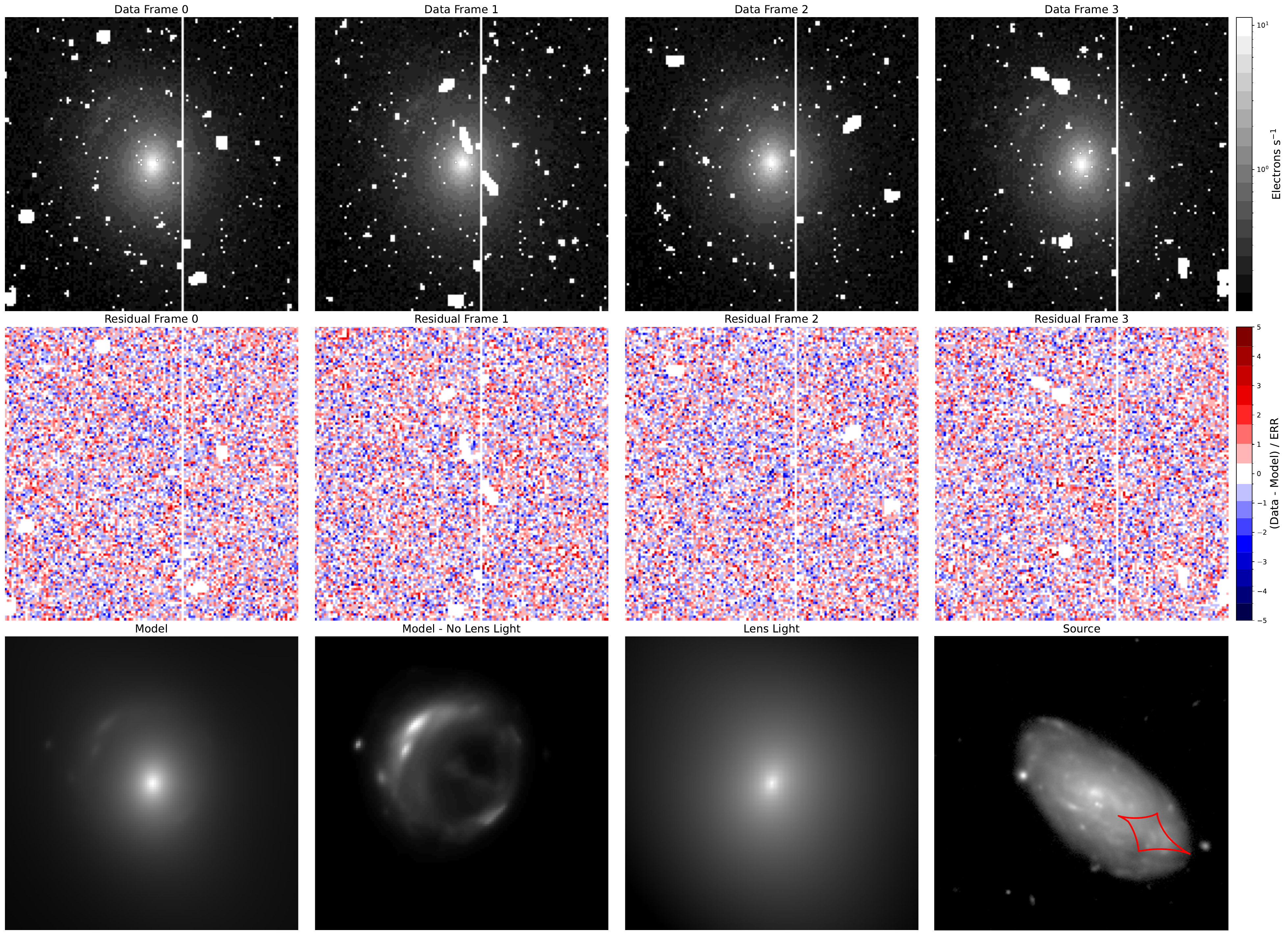}
    \caption{Same as Figure~\ref{fig:SDSSJ1430+4105_bestfit}, but for SDSSJ1020+1122.}
    \label{fig:stage1_SDSSJ1020+1122}
\end{figure*}

\begin{figure*}[t]
    \centering
    \includegraphics[width=\textwidth]{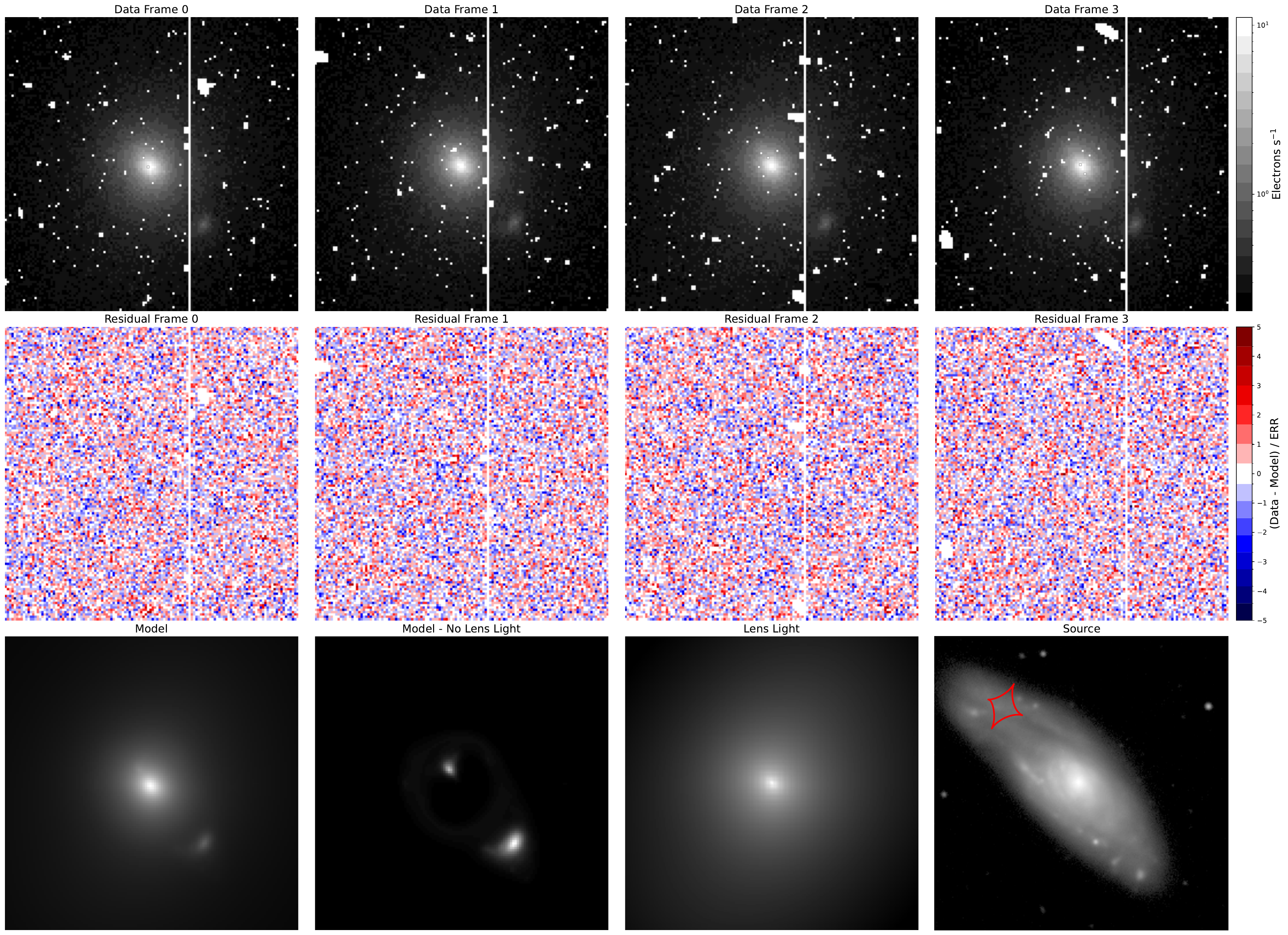}
    \caption{Same as Figure~\ref{fig:SDSSJ1430+4105_bestfit}, but for SDSSJ1142+1001.}
    \label{fig:stage1_SDSSJ1142+1001}
\end{figure*}

\begin{figure*}[t]
    \centering
    \includegraphics[width=\textwidth]{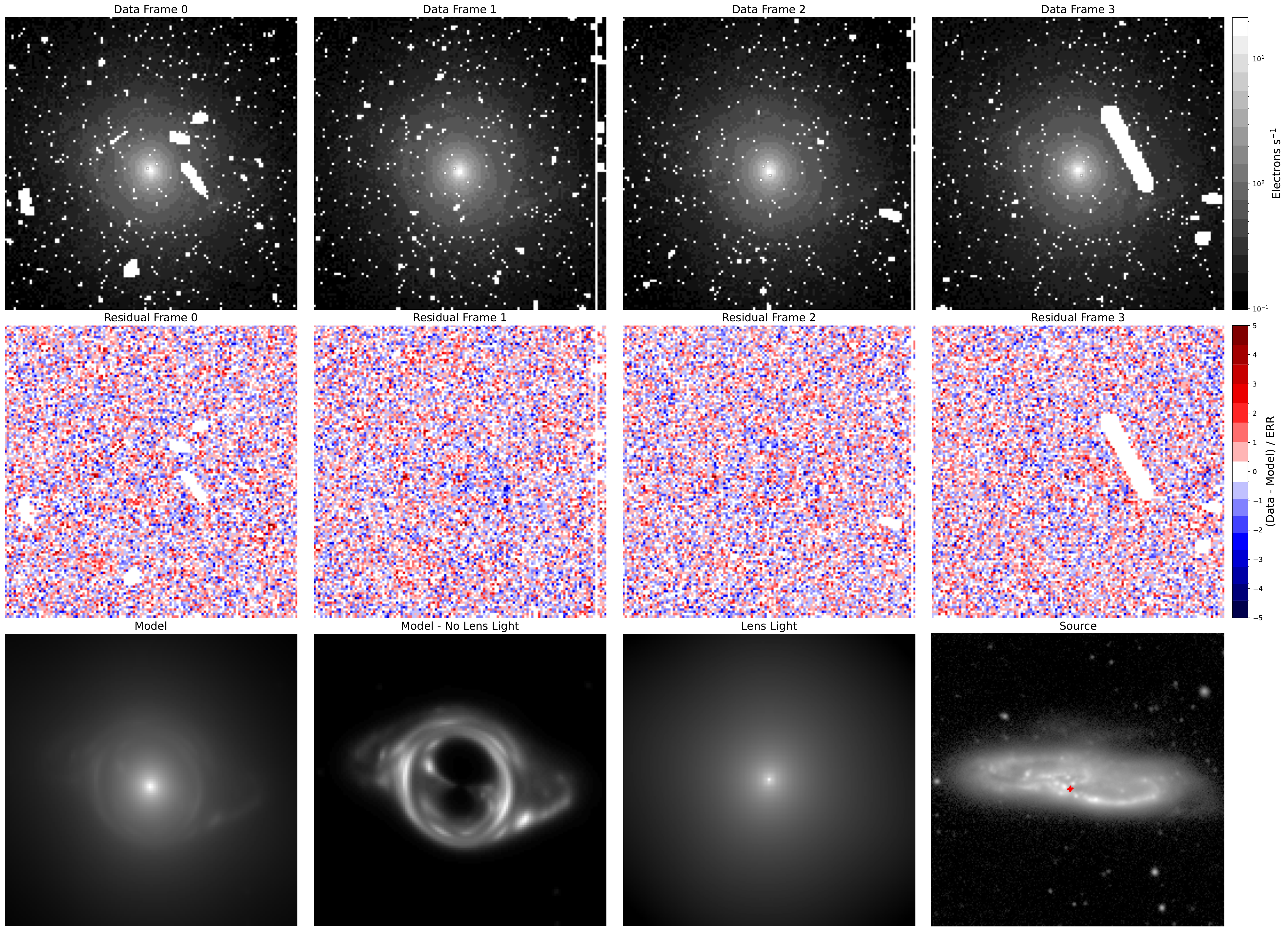}
    \caption{Same as Figure~\ref{fig:SDSSJ1430+4105_bestfit}, but for SDSSJ1250+0523.}
    \label{fig:stage1_SDSSJ1250+0523}
\end{figure*}

\begin{figure*}[t]
    \centering
    \includegraphics[width=\textwidth]{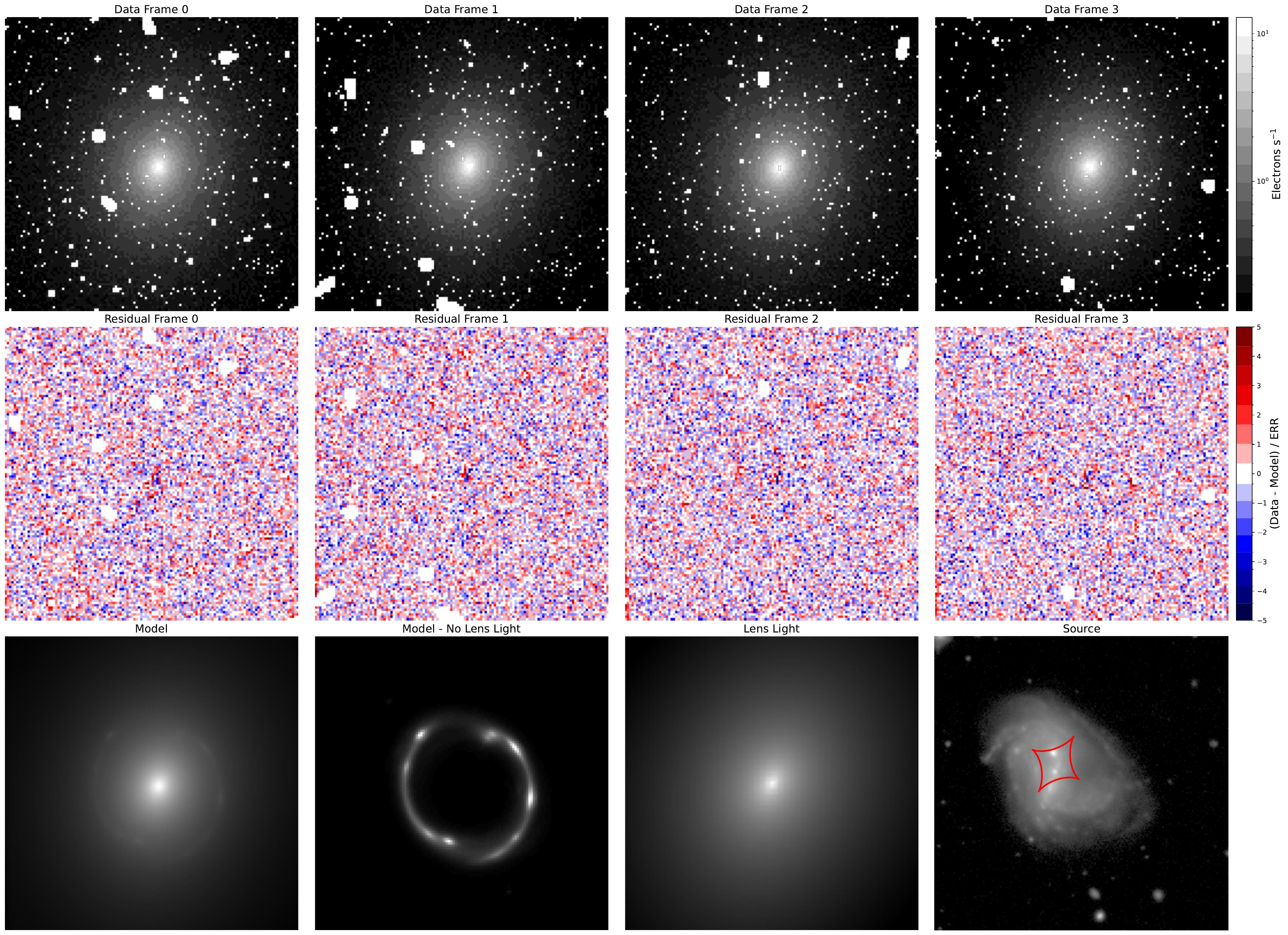}
    \caption{Same as Figure~\ref{fig:SDSSJ1430+4105_bestfit}, but for SDSSJ1402+6321.}
    \label{fig:stage1_SDSSJ1402+6321}
\end{figure*}

\begin{figure*}[t]
    \centering
    \includegraphics[width=\textwidth]{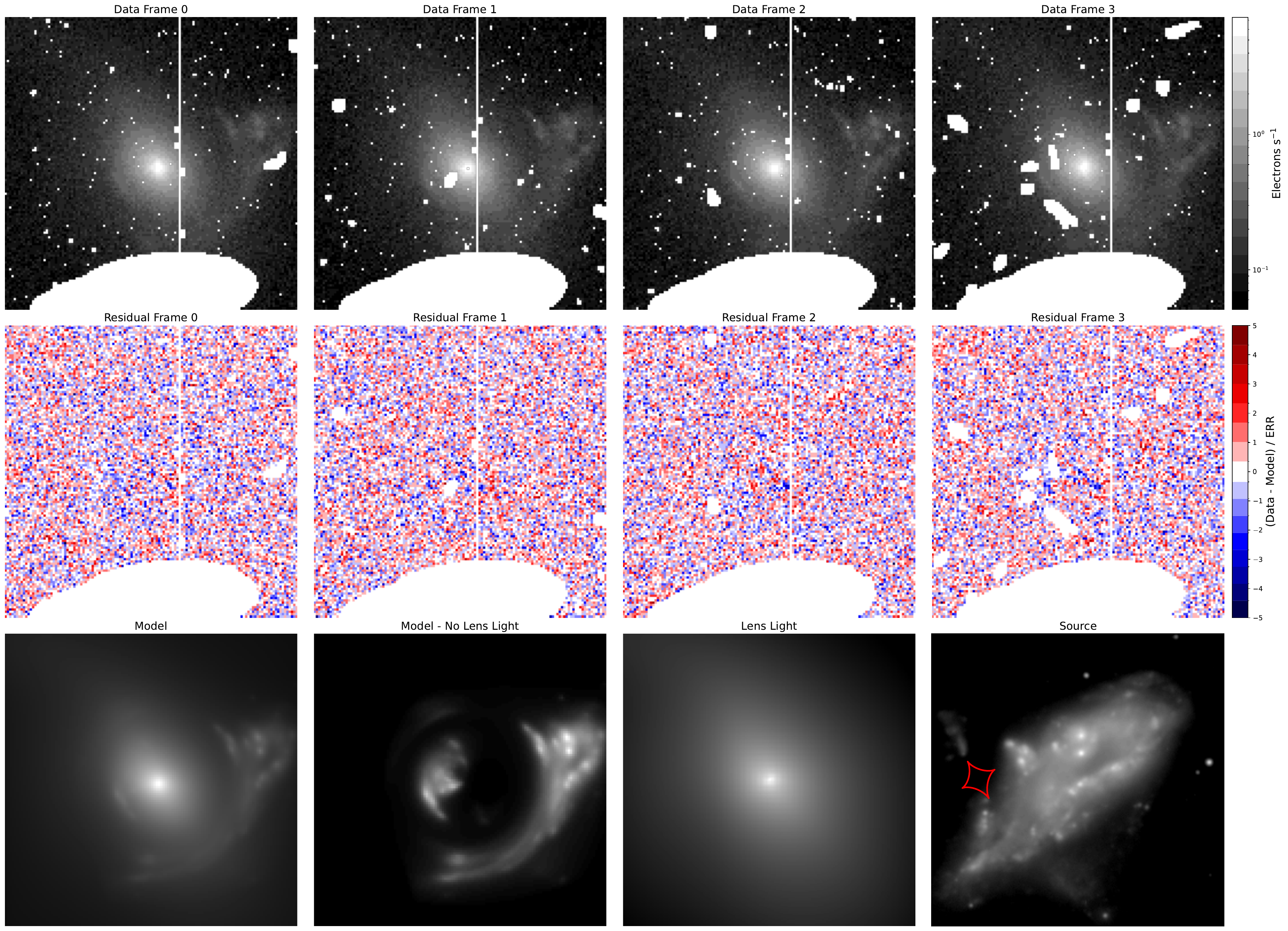}
    \caption{Same as Figure~\ref{fig:SDSSJ1430+4105_bestfit}, but for SDSSJ1416+5136.}
    \label{fig:stage1_SDSSJ1416+5136}
\end{figure*}

\begin{figure*}[t]
    \centering
    \includegraphics[width=\textwidth]{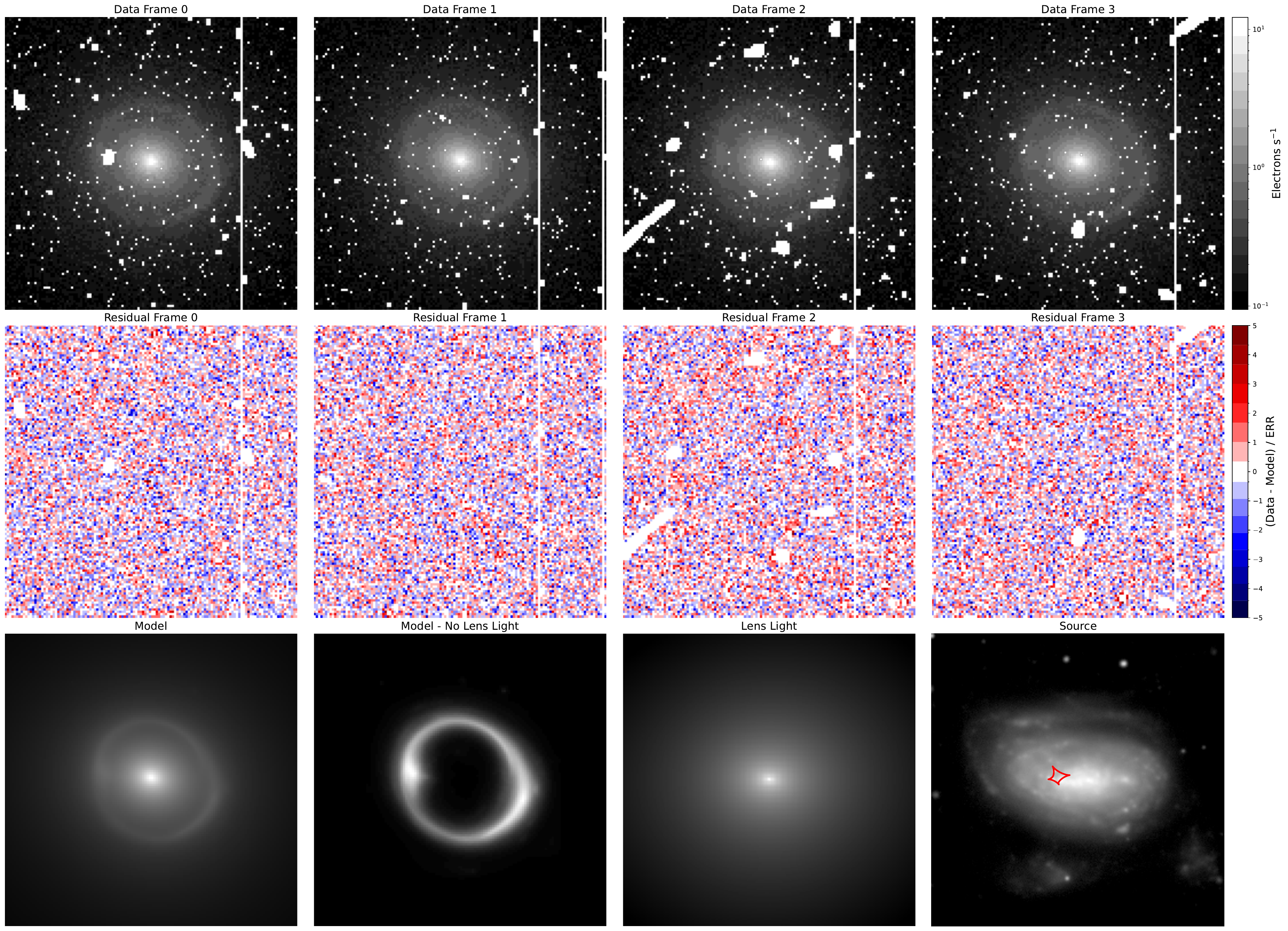}
    \caption{Same as Figure~\ref{fig:SDSSJ1430+4105_bestfit}, but for SDSSJ1627-0053.}
    \label{fig:stage1_SDSSJ1627-0053}
\end{figure*}

\begin{figure*}[t]
    \centering
    \includegraphics[width=\textwidth]{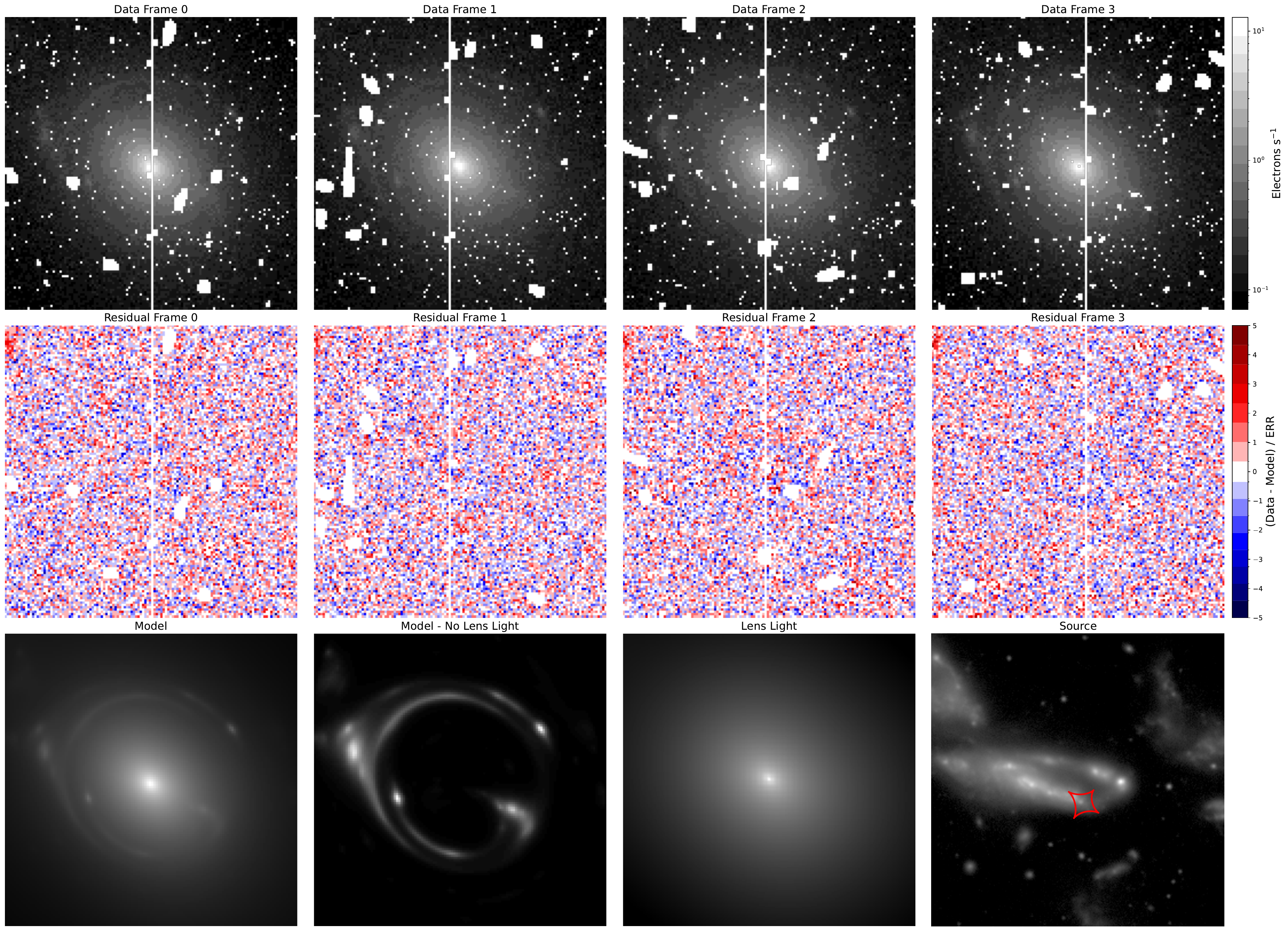}
    \caption{Same as Figure~\ref{fig:SDSSJ1430+4105_bestfit}, but for SDSSJ1630+4520.}
    \label{fig:stage1_SDSSJ1630+4520}
\end{figure*}

\begin{figure*}[t]
    \centering
    \includegraphics[width=\textwidth]{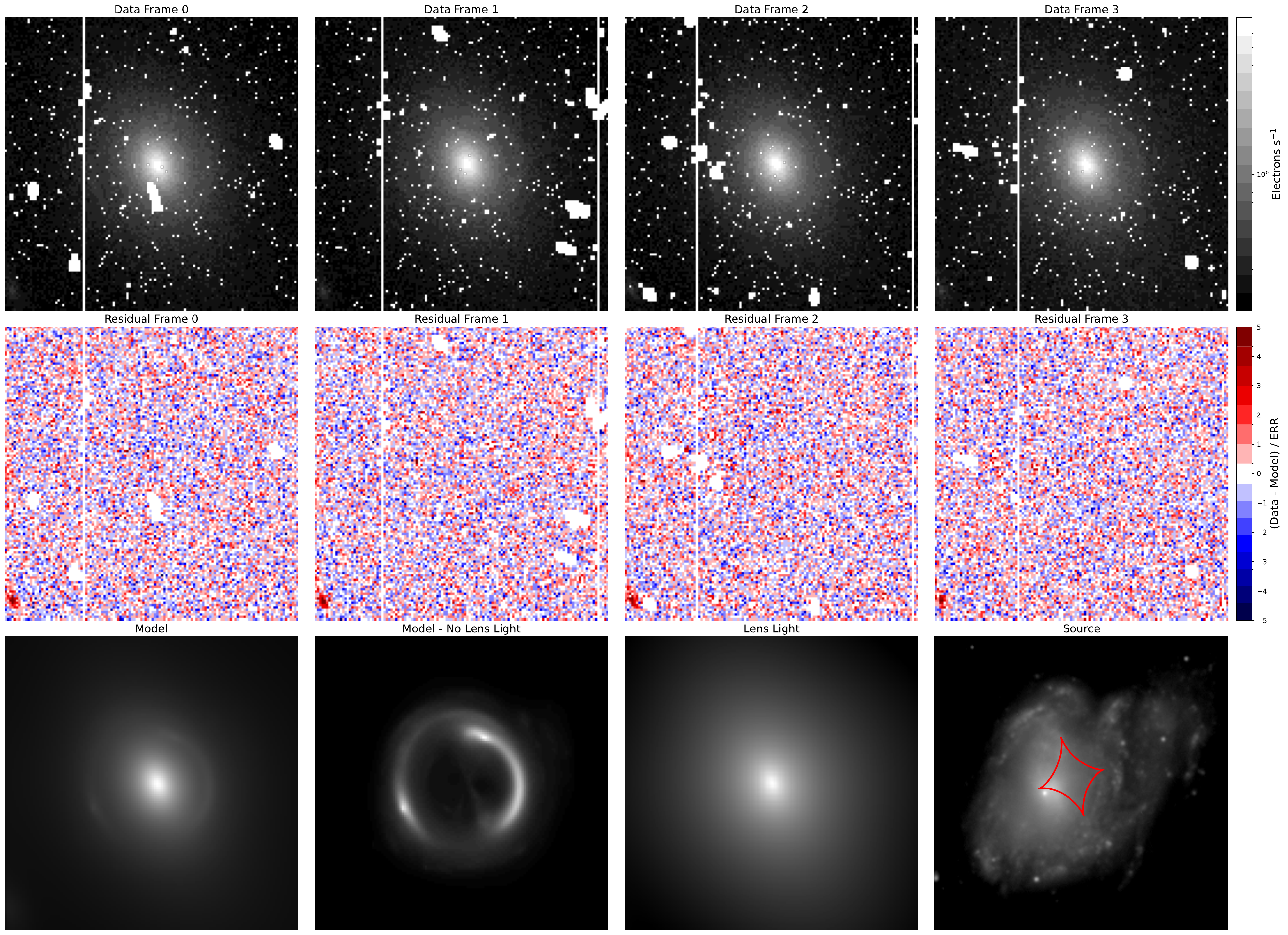}
    \caption{Same as Figure~\ref{fig:SDSSJ1430+4105_bestfit}, but for SDSSJ2300+0022.}
    \label{fig:stage1_SDSSJ2300+0022}
\end{figure*}

\begin{figure*}[t]
    \centering
    \includegraphics[width=\textwidth]{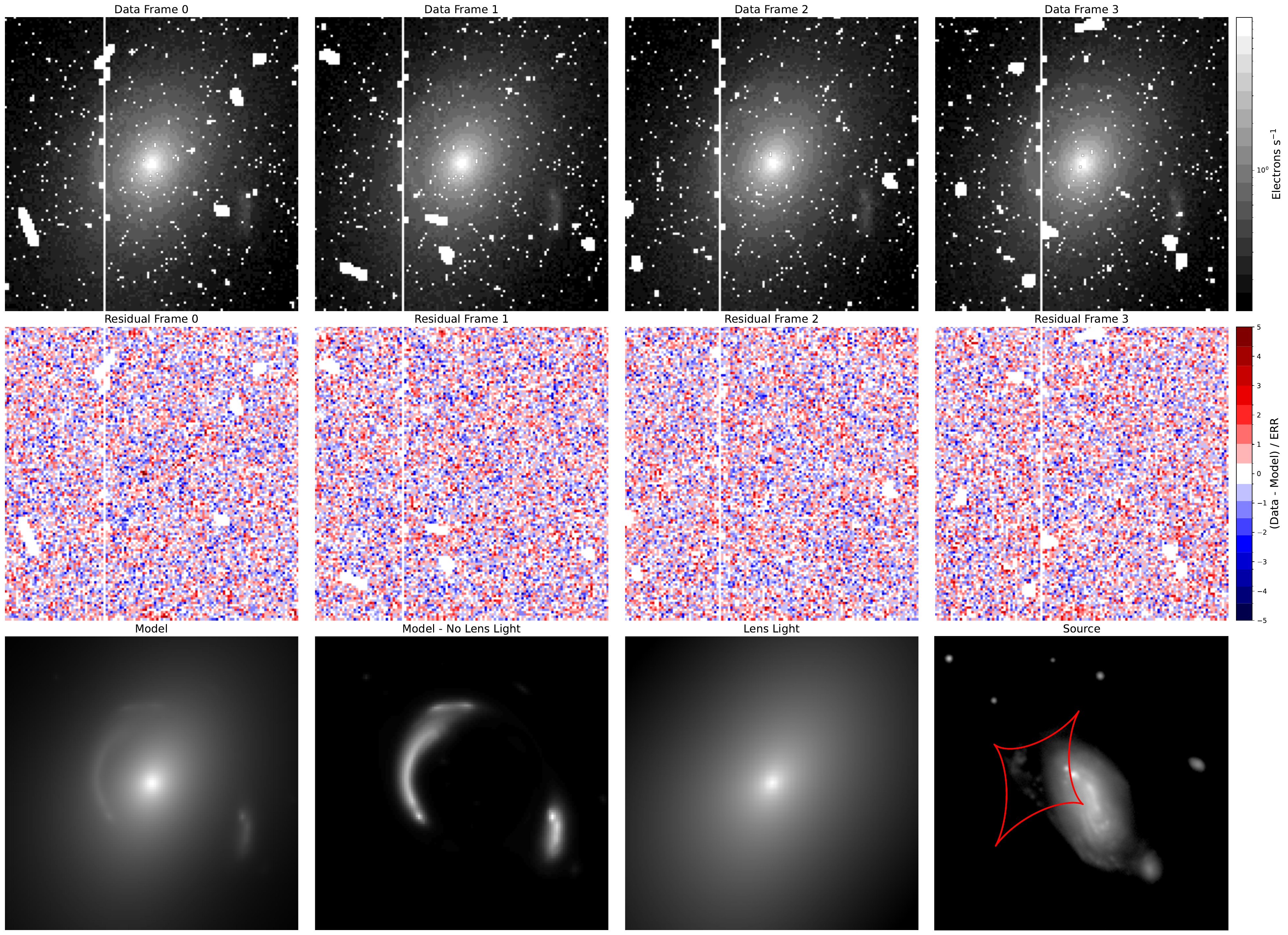}
    \caption{Same as Figure~\ref{fig:SDSSJ1430+4105_bestfit}, but for SDSSJ2303+1422.}
    \label{fig:stage1_SDSSJ2303+1422}
\end{figure*}

\end{document}